\documentclass[a4paper,11pt]{article}

\pdfoutput=1
\usepackage{setspace}
\usepackage{authblk}

\usepackage{fullpage}
\usepackage{parskip}
\usepackage{titlesec}
\usepackage[section]{placeins}
\usepackage[dvipsnames,table]{xcolor}

\usepackage{breakcites}
\usepackage{amssymb}
\usepackage{lineno}
\usepackage{hyphenat}
\usepackage[all]{nowidow}
\usepackage{bm}

\PassOptionsToPackage{hyphens}{url}
\usepackage[colorlinks = true]{hyperref} 

\hypersetup{
  linkcolor  = RedViolet,
  citecolor = Blue!90!black,
  urlcolor   = Aquamarine!80!black,
  colorlinks = true
}

\usepackage{etoolbox}

\renewenvironment{abstract}
  {{\bfseries\noindent{\abstractname}\par\nobreak}\footnotesize}
  {\bigskip}
\titlespacing{\section}{0pt}{*3}{*1}
\titlespacing{\subsection}{0pt}{*2}{*0.5}
\titlespacing{\subsubsection}{0pt}{*1.5}{0pt}

\usepackage{graphicx}
\usepackage{makecell}
\usepackage[space]{grffile}
\usepackage{latexsym}
\usepackage{textcomp}
\usepackage{longtable}
\usepackage{lscape}
\usepackage{tabularx}
\usepackage{tabulary}
\newcolumntype{W}[1]{>{\raggedright\arraybackslash}p{#1}}
\newcolumntype{R}[1]{>{\raggedleft\arraybackslash}p{#1}}
\usepackage{array,multirow}
\newsavebox\CBox
\def\textBF#1{\sbox\CBox{#1}\resizebox{\wd\CBox}{\ht\CBox}{\textbf{#1}}}
\usepackage{amsfonts,amsmath,amssymb,bbm}
\usepackage{subcaption}
\usepackage{calc}

\DeclareFontFamily{U}{MnSymbolD}{}
\DeclareFontShape{U}{MnSymbolD}{m}{n}{
  <-6> MnSymbolD5 <6-7> MnSymbolD6 <7-8> MnSymbolD7
  <8-9> MnSymbolD8 <9-10> MnSymbolD9 <10-12> MnSymbolD10 <12-> MnSymbolD12
}{}
\DeclareFontShape{U}{MnSymbolD}{b}{n}{
  <-6> MnSymbolD-Bold5 <6-7> MnSymbolD-Bold6 <7-8> MnSymbolD-Bold7 
  <8-9> MnSymbolD-Bold8 <9-10> MnSymbolD-Bold9 <10-12> MnSymbolD-Bold10  <12-> MnSymbolD-Bold12
}{}
\DeclareSymbolFont{MnSyD}{U}{MnSymbolD}{m}{n}
\SetSymbolFont{MnSyD}{bold}{U}{MnSymbolD}{b}{n}
\DeclareMathSymbol{\mnsfrown}{\mathrel}{MnSyD}{25}
\makeatletter
\DeclareRobustCommand{\invbreve}[1]{\mathpalette\invbreve@{#1}}
\newcommand{\invbreve@}[2]{
  \mathord{#2\mkern-6.7mu \raise1.57ex\hbox{$#1\scriptstyle\mnsfrown$}}%
}
\makeatother
\newif\iflatexml\latexmlfalse

\AtBeginDocument{\DeclareGraphicsExtensions{.pdf,.PDF,.eps,.EPS,.png,.PNG,.tif,.TIF,.jpg,.JPG,.jpeg,.JPEG}}

\usepackage[utf8]{inputenc}
\usepackage[english]{babel}
\usepackage{float}
\usepackage{titlesec}
\titleformat*{\section}{\Large\bfseries}
\titleformat*{\subsection}{\large\bfseries}
\usepackage{csquotes}

\usepackage[natbib=true, backend=biber, sorting=nyt, style=apa, apamaxprtauth=20]{biblatex}
\defcitealias{UCIrepo}{Kelly et al., n.d.}

\newcommand{\beginsupplement}{
        \setcounter{table}{0}
        \renewcommand{\thetable}{S\arabic{table}}
        \setcounter{figure}{0}
        \renewcommand{\thefigure}{S\arabic{figure}}
     }

\usepackage{orcidlink}
\usetikzlibrary{positioning,
                calc, 
                chains}

\newcommand*\samethanks[1][\value{footnote}]{\footnotemark[#1]}
 \usepackage{enumitem}
\newlist{tabitemize}{itemize}{1}
\setlist[tabitemize]{label=\textbf{\textperiodcentered},leftmargin=*,topsep=1ex,
parsep=0pt,
                  after=\vspace{-\baselineskip},
                  before=\vspace{-0.75\baselineskip}
                  }  

\setlist[itemize]{nosep,leftmargin=5.5mm} %remove distance in itemize and reduce left margin

\newcommand{\ds}{\mathcal{D}} 
\newcommand{\dsind}{r} 
\newcommand{\dsnum}{R} 
\newcommand{\dswithind}{{\ds}^{(\dsind)}}
\newcommand{\dsall}{\boldsymbol{\ds}}

\newcommand{\assumption}{Z} 

\newcommand{\dgm}{\mathcal{G}} 
\newcommand{\dgmdtrue}{\dgm^{*}_{\ds}}
\newcommand{\dgmdtrueind}{\dgm^{*}_{\ds^{(\dsind)}}} 
\newcommand{\struc}{\mathcal{M}}
\newcommand{\strucdtrueind}{\struc^{*}_{\ds^{(\dsind)}}}
\newcommand{\rdbparam}{\theta} 
\newcommand{\thetaall}{\boldsymbol{\rdbparam}} 
\newcommand{\thetadesign}{\boldsymbol{\rdbparam}_{\text{known}}} 
\newcommand{\thetaestim}{\boldsymbol{\rdbparam}_{\text{unknown}}} 
\newcommand{\thetaestimtarget}{\boldsymbol{\rdbparam}_{\text{unknown,target}}} 
\newcommand{\thetaestimother}{\boldsymbol{\rdbparam}_{\text{unknown,other}}} 
\newcommand{\paramind}{q} 
\newcommand{\paramnum}{Q}

\newcommand{\resparam}{\lambda} 
\newcommand{\lambdaall}{\boldsymbol{\resparam}} 
\newcommand{\lambdadesign}{\boldsymbol{\resparam}_{\text{known}}} 
\newcommand{\lambdaestim}{\boldsymbol{\resparam}_{\text{unknown}}} 
\newcommand{\lambdaestimtarget}{\boldsymbol{\resparam}_{\text{unknown,target}}} 
\newcommand{\lambdaestimother}{\boldsymbol{\resparam}_{\text{unknown,other}}} 

\newcommand{\realizedrdbparam}{\hat{\rdbparam}} 
\newcommand{\aggregatedrdbparam}{\tilde{\rdbparam}}
\newcommand{\aggregatednum}{A_{\paramind}}
\newcommand{\setofvaluesforparam}{\boldsymbol{\rdbparam}'} %set of inferred values for a given parameter
\newcommand{\consideredparamvector}{\boldsymbol{\invbreve{\rdbparam}}} %vector with a single value for each parameter in \thetaall
\newcommand{\paraminconsideredparamvector}{\invbreve{\rdbparam}}
\newcommand{\setofconsideredparamvectors}{\Theta} %set of vectors with a single value for each parameter in \thetaall
\newcommand{\consideredparamvectorind}{l} 
\newcommand{\consideredparamvectornum}{L}

\newcommand{\exordinal}{\hyperref[ex:ordinal]{\textit{Example-Ordinal}}}
\newcommand{\exsurv}{\hyperref[ex:survival]{\textit{Example-Survival}}}
\newcommand{\exde}{\hyperref[ex:de]{\textit{Example-DE-Analysis}}}
\newcommand{\exmeta}{\hyperref[ex:meta]{\textit{Example-Meta-Analysis}}}
\newcommand{\exordinalshort}{Ordinal}
\newcommand{\exsurvshort}{Survival}
\newcommand{\exdeshort}{DE-Analysis}
\newcommand{\exmetashort}{Meta-Analysis}

\newcommand{\nrep}{n_{\text{sim}}} 

\newcommand{\groupind}{k} 
\newcommand{\groupnum}{K} 
\newcommand{\obsind}{i} 
\newcommand{\obsnum}{n} 
\newcommand{\ordind}{m} 
\newcommand{\ordnum}{M} 
\newcommand{\varind}{j} 
\newcommand{\varnum}{p} 
\newcommand{\groupone}{1} %treatment group
\newcommand{\grouptwo}{2} %control group
\newcommand{\metaeffect}{\delta} 
\newcommand{\nstudy}{n_{\text{study}}}
\newcommand{\studyind}{i}
\newcommand{\nindindex}{n_{\text{obs}_{i}}}
\newcommand{\propDE}{p_{\text{DE}}} %also used in plots, so those would need to be updated if we change "p" to something else
\newcommand{\propupreg}{p_{\text{up}}}
\newcommand{\eventrate}{\eta}

\newcommand{\dsaim}{(D2)} 
\newcommand{\dsextract}{(D3)} 
\newcommand{\dsrepro}{(D1)} 

\newcommand{\releff}{RE} 
\newcommand{\baik}{B20}  % could also be set to \cite{..}

\usepackage[paper=a4paper,left=25mm,right=25mm,top=25mm,bottom=25mm]{geometry}  

\usepackage{marginnote}
\AtBeginEnvironment {table}{}
\AtBeginEnvironment {figure}{}
\begin{document}
\title{Statistical parametric simulation studies based on real data}

\def\correspondingauthor{\footnote{Corresponding author, e-mail: \href{mailto:christina.sauer@stat.uni-muenchen.de}{christina.sauer@stat.uni-muenchen.de}.}}
\author[1,2,3]{Christina Sauer\thanks{These authors contributed equally to this work.}\correspondingauthor\orcidlink{0000-0003-2425-7858}}
\author[2,3]{F. Julian D. Lange\samethanks[1]\orcidlink{0009-0000-2000-1527}}
\author[4,5]{Maria Thurow}
\author[4]{Ina Dormuth}
\author[2,3]{Anne-Laure Boulesteix\orcidlink{0000-0002-2729-0947}}
\affil[1]{Department of Statistics, LMU Munich, Munich, Germany}
\affil[2]{Institute for Medical Information Processing, Biometry and Epidemiology, Faculty of Medicine, LMU Munich, Munich, Germany}
\affil[3]{Munich Center for Machine Learning (MCML), Munich, Germany}
\affil[4]{Department of Statistics, TU Dortmund University, Dortmund, Germany}
\affil[5]{Research Center Trustworthy Data Science and Security, UA Ruhr, Dortmund, Germany}
\vspace{-1em}
  \date{November 7, 2025}
\maketitle

%%%%%%%%%%%%%%%%%%%%%%%%%%%%%%%%%%%%%%%%%%%%%%%%%%%%
%%%%%%%%%%%%%%%%%%%% Abstract %%%%%%%%%%%%%%%%%%%%%%
%%%%%%%%%%%%%%%%%%%%%%%%%%%%%%%%%%%%%%%%%%%%%%%%%%%%
\begin{abstract}

Simulation studies are indispensable for evaluating statistical methods and ubiquitous in statistical research. The most common simulation approach is parametric simulation, where the data-generating mechanism (DGM) corresponds to a parametric model from which observations are drawn. 
While many simulation studies aim to give practical guidance on method suitability, parametric simulations in particular are often criticized for being unrealistic.
To overcome this drawback, it is sensible to employ real data for constructing the parametric DGMs. 
However, although real-data-based parametric DGMs are widely used, the specific ways in which DGM components are inferred vary, and their implications may not be fully understood.
Additionally, researchers often rely on a limited selection of real datasets, with the rationale for their selection being unclear.
This paper reviews and formally discusses how components of parametric DGMs can be inferred from real data and how dataset selection can be performed more systematically.  
By doing so, we aim to support researchers in conducting simulation studies with a lower risk of overgeneralization and misinterpretation. We illustrate the construction of parametric DGMs based on a systematically selected set of real datasets using two examples: one on ordinal outcomes in randomized controlled trials and one on differential gene expression analysis.

\end{abstract}

\vspace{-1em}

\sloppy

%%%%%%%%%%%%%%%%%%%%%%%%%%%%%%%%%%%%%%%%%%%%%%%%%%%%
%%%%%%%%%%%%%%%%%%% Introduction %%%%%%%%%%%%%%%%%%%
%%%%%%%%%%%%%%%%%%%%%%%%%%%%%%%%%%%%%%%%%%%%%%%%%%%%
\section{Introduction}
\label{sec:1_introduction}
In substantive disciplines applying statistical methods, researchers are faced with an ever-growing number of methods to choose from. To aid them in these decisions and provide well-founded practical recommendations regarding the suitability of a method for a given application, empirical
methodological studies, i.e.\ studies that empirically evaluate and compare statistical methods, are indispensable. In these studies, the data to which the statistical methods are applied can be divided into two main categories: simulated data and real data. Accordingly, we refer to these studies as simulation studies and real-data studies, respectively, although they may be presented within the same publication, often in combination with the introduction of a new method.\\
The first key distinction between the two study types lies in how the data-generating mechanism (DGM) is determined: While in simulation studies, the DGM(s) must be explicitly specified by the researcher, in real-data studies, the DGM underlying each real dataset is inherently determined by real-world processes \citep{hothornDesignAnalysisBenchmark2005}. For real-data studies, this shifts the focus from specifying a DGM to the careful selection of appropriate datasets for the study.
Although simulation studies may employ semi-parametric approaches, where part of the DGM involves resampling from a real dataset, the most common approach is parametric simulation, where a predefined parametric model is used to draw observations \citep{morris2019using, siepe2024simulation}. Focusing for now on parametric simulations, another critical distinction between simulation studies and real-data studies emerges:  Unlike real-data studies, where the true DGMs remain unknown and reflect complex real-world processes, simulation studies operate with DGMs that are fully known, as they are explicitly constructed by the researcher. \\
Based on these two distinctions, simulation studies offer two key advantages over real-data studies. First, the full knowledge of the DGM (often referred to as access to the \enquote{ground truth}) enables researchers to evaluate the performance of statistical methods with respect to essentially any target of interest, such as a true effect size or the validity of a null hypothesis \citep{boulesteix2020introduction,friedrich2024ontherole}. In contrast, real datasets typically provide only a limited set of targets for which the truth is known, with prediction tasks (where the target corresponds to the true outcome of an observation) being a notable exception. 
Second, the full control over the DGM in simulation studies allows researchers to investigate statistical methods under virtually any scenario they wish to explore (e.g., varying parameter values and distributions) and to generate an unlimited number of datasets from the same DGM. In stark contrast to real-data studies, the amount of available data in simulation studies is essentially only limited by the available computational resources \citep{boulesteix2020introduction}.\\
Importantly, however, having full control over the DGM, while offering clear advantages, also places a substantial responsibility on the researcher, making it something of a mixed blessing. While this applies to all decisions in the design and execution of a study---commonly referred to as researcher degrees of freedom (RDFs; \citealp{simmons2011false-positive})---DGM-related RDFs are particularly impactful, as the choice of the DGM(s) can strongly influence the results of a simulation study and, as a consequence, the recommendations derived from them \citep{metcalfe2006importance,pateras2018data-generating,kulinskaya2021exploring,jansen2023random-effects,astivia2015cautionary,fairchild2024many}. To illustrate the mixed blessing of having full control over the DGM, consider a researcher conducting a simulation study to evaluate methods with respect to a specific target in a clinical trial with two treatment groups, a continuous outcome, and under the scenario where a specific assumption~$\assumption$ is violated. While it is advantageous that the DGM can be easily tailored to match the researcher’s specific interests (e.g., continuous outcome, two treatment groups, violation of $\assumption$), several parameters (e.g., effect size, sample size) and parts of\enlargethispage{\baselineskip} the model structure (e.g., outcome distribution, presence of covariates, extent of $\assumption$’s violation) still need to be defined, which is often a challenging process that requires careful consideration.

\paragraph{Practical relevance}
For simplicity, assume for now that only a single numerical parameter $\theta$ remains to be specified.
To enhance the generalizability of the simulation findings to its domain of interest, a reasonable approach is to choose values for $\theta$ that are practically relevant for that domain. Here, the \textit{domain of interest} refers to the (hypothetical) population of true real-world DGMs to which the simulation study’s results and recommendations should apply, and the set of chosen values for $\theta$ is \textit{practically relevant} if its distribution closely aligns with that of $\theta$ in the real-world DGMs within the domain of interest; this concept can similarly be extended to all components of the DGM. Importantly, having \textit{realistic} DGMs---defined as reflecting \textit{any} real-world DGM---does not automatically ensure practical relevance. Using again the example of a specific parameter $\theta$, practical relevance requires that the distribution of its values aligns with the distribution observed in the DGMs that actually belong to the study’s domain of interest. Thus, realistic DGMs are a necessary but insufficient condition for practical relevance. The concept of specifying practically relevant DGMs (or at least realistic ones) aligns with recommendations in the literature, including formal guidelines and related discussions (\citealp{burton2006design, boulesteix2020introduction, chipman2022lets, paxton2001monte, white2023importance, harwell2017experimental}). Of course, simulations using intentionally simplistic or unrealistic DGMs also serve important purposes, such as identifying a method’s breaking point (\citealp{heinze2024phases,morris2019using}). However, we argue that for many simulations, researchers implicitly aim for practical relevance, or at least this is what readers are likely to assume unless explicitly stated otherwise.\\
Assuming an aim of practical relevance, two main issues arise: First, achieving practically relevant DGMs for a given domain of interest is challenging. This is supported by several reviews identifying discrepancies between the DGMs used in simulations and real datasets. For example, when reviewing simulation studies on meta-analyses, \citet{langan2017comparative} and \citet{fernandez-castilla2020application} found that the number of simulated individual studies often exceeded the number typically observed in existing meta-analyses. Similar discrepancies were found in the context of recurrent events data \citep{penichoux2015simulating}, missing data \citep{morel2022dealing}, fMRI data \citep{welvaert2014review}, or the health, educational, and social sciences in general \citep{bono2017non-normal}. Since these reviews treat the discrepancies as criticisms, it is reasonable to assume that the studies aimed for practical relevance. These discrepancies may arise from arbitrary or non-neutral choices, with the latter reflecting often well-intentioned but potentially biased decisions that favor specific outcomes (e.g., demonstrating the superiority of a particular method; \citealp{pawel2024pitfalls}) and thus exploit the RDFs associated with the DGM. 
Regardless of the reason for the discrepancies, if a simulation study gives the impression of using practically relevant DGMs when, in fact, they are not---even though it is widely understood that such studies inevitably involve simplifications---this can lead to misinterpretation of the findings by readers and possibly even the researchers themselves. \\
Second, the definition of practical relevance given above inherently relies on a precise specification of the domain of interest, which is usually not clear from the context of the study. For instance, in the example above, assume the researcher has selected a set of DGMs; the domain of interest to which the results are intended to generalize could then lie anywhere between the real-world DGMs exactly corresponding to those considered in the study and all real-world DGMs with a continuous outcome, two treatment groups, and a violation of assumption~$\assumption$. While \citet{strobl2024against} rightly argue that it is nearly infeasible to formally and unambiguously define a domain of interest, failing to specify it entirely is not a better alternative. Without such a specification, the domain of interest may be assumed to be broader than it actually is, which can increase the risk of overgeneralizing the results \citep{niessl2024explaining}.\\
To address these issues, a reasonable approach in the analysis of results is to focus on analyzing the relationship between DGM characteristics and method performance (\citealp{strobl2024against}), as solely considering overall performance can be misleading if the DGMs lack practical relevance and is generally hard to interpret without a clearly defined domain of interest. However, this approach is limited if, for example, most of the values selected for a specific parameter in the DGM do not reflect any real-world DGMs. 
As a complementary perspective, attention may thus also be directed toward specifying the DGMs themselves.

\paragraph{Real-data-based parametric simulations} To ensure that the DGMs reflect any real-world DGM (a necessary condition for practical relevance, as noted above), a natural approach is to base them on real datasets. This can be done in a direct manner by resampling parts of the simulated data from real datasets, which, however, requires transitioning from parametric to semi-parametric simulation approaches, such as Plasmode simulation \citep{franklin2014plasmode,schreck2024plasmode}. If one wishes to remain within the framework of parametric simulations, specific parameters or parts of the model structure could still be derived from real datasets (as suggested, e.g., by \citealp{burton2006design}). This approach has been adopted in a number of simulation studies (see reviews by \citealp{morris2019using}, and \citealp{siepe2024simulation}), but the concept is typically implemented differently---both in terms of which parts of the DGM are informed by real data \citep{friedrich2024ontherole} and how directly the data inform these parts, which inherently affects the degree of realism achieved.
Thus, while real-data-based simulations reduce RDFs associated with the direct specification of DGMs, they introduce new RDFs related to the process of specifying the DGM based on the real dataset. To our knowledge, in contrast to Plasmode simulation, there is no literature systematically discussing the rationale or implications for different implementations of real-data-based parametric simulation.\\
In addition, basing simulations on real datasets also creates new RDFs related to the selection of these datasets.  In practice, very few datasets are typically used for this purpose. For example, in the reviews by \citet{morris2019using} and \citet{siepe2024simulation}, the real-data-based parametric simulation studies almost always rely on just one or two datasets. 
The selection of these datasets is rarely justified, often appearing to be one of convenience, and it is usually unclear which domain of interest they are meant to represent. Consequently, while the resulting DGMs might be realistic, they are not necessarily practically relevant---at least not beyond the specific DGMs underlying the selected datasets. In principle, the criticisms regarding dataset selection are similar to those raised for real-data studies (see, e.g., \citealp{herrmann2024position}, and references therein). A promising strategy for addressing these issues in real-data-based simulation studies could thus be to adopt what has already been suggested for real-data studies: systematically and\enlargethispage{\baselineskip} transparently selecting datasets by specifying a database and clear eligibility criteria for dataset inclusion (see, e.g., \citealp{boulesteix2017towards}). While this approach does not fully resolve the challenge of formally defining the domain of interest, the selection process can serve as a proxy, providing more clarity to readers about the practical settings in which a simulation study's findings are expected to hold. Additionally, shifting RDFs from the selection of individual datasets to defining a systematic selection process can facilitate more meaningful and neutral decisions, thereby enhancing the practical relevance of the considered DGMs.

\paragraph{The scope of this paper} Overall, while the concept of real-data-based simulations is widely recognized,  its specific implementation for parametric simulations has not yet been thoroughly addressed, and the process of selecting real datasets is often not conducted systematically. This paper aims to address these gaps by reviewing and discussing the possibilities, rationale, and implications of all steps of real-data-based simulations. While our focus is primarily on parametric simulations, the insights provided also apply to the parametric element of semi-parametric simulations. Additionally, the discussion on dataset selection is also relevant for the resampling element of semi-parametric simulations.\\
The paper is organized as follows. In \hyperref[sec:2_preliminaries]{Section~\ref*{sec:2_preliminaries}}, we cover all necessary preliminaries, including the types of DGMs used in simulations and key distinctions in the components of parametric DGMs. We then review and formalize the construction of real-data-based simulations, considering both the inference of DGMs from real datasets in \hyperref[sec:3_implementation]{Section~\ref*{sec:3_implementation}} and the systematic selection of these datasets in \hyperref[sec:3.3_selecting_datasets]{Section~\ref*{sec:3.3_selecting_datasets}}. In \hyperref[sec:4_results]{Section~\ref*{sec:4_results}}, we present two empirical examples of parametric simulations based on a systematically selected set of datasets, demonstrating that considering only purely researcher-specified DGMs or relying on a single dataset can lead to an incomplete picture of a method and results that do not generalize well. In \hyperref[sec:5_workflow]{Section~\ref*{sec:5_workflow}}, we provide a structured step-by-step workflow, and we conclude our paper in \hyperref[sec:6_conclusion]{Section~\ref*{sec:6_conclusion}}.

%%%%%%%%%%%%%%%%%%%%%%%%%%%%%%%%%%%%%%%%%%%%%%%%%%%%
%%%%%%%%%%%%%%%%%%%% Section 2 %%%%%%%%%%%%%%%%%%%%%
%%%%%%%%%%%%%%%%%%%%%%%%%%%%%%%%%%%%%%%%%%%%%%%%%%%%
\section{Preliminaries}
\label{sec:2_preliminaries}
\subsection{Parametric DGMs and other options}
\paragraph{Parametric DGMs} 
We refer to the DGMs employed in parametric simulations, which are the focus of this paper, as \textit{parametric DGMs}. Such DGMs correspond to parametric stochastic models that can be represented in closed form \citep{schreck2024plasmode,morris2019using}. A given parametric DGM is fully specified by (i) its model structure, which outlines how data is simulated from the parametric model and consists of various parts specifying the relationships among and the statistical distributions assigned to variables, and (ii) the numerical parameters that provide the specific values required to fully define the model. For example, if one is interested in multivariable regression models with a binary outcome variable, a logistic regression model may be assumed as the model structure. In this case, the set of parameters includes the number of covariates, the regression coefficients, and the sample size. As obvious from this example, the model structure inherently determines the set of parameters by outlining the distributions, relationships, and other aspects that require numerical values for full definition. We refer to the model structure (or its parts) and the parameters as the components of the DGM. Further examples of parametric DGMs are discussed in more detail in \hyperref[sec:2.4_examples]{Section~\ref*{sec:2.4_examples}}.

\paragraph{Semi-parametric DGMs}
As the name suggests, this type of DGM is not fully parametric but includes a non-parametric element in the form of resampling from a real dataset \citep{schreck2024plasmode}. Examples of resampling schemes are simple resampling of observations (with or without replacement) or more advanced methods such as sampling from a smoothed empirical distribution of the dataset estimated via kernel density estimation \citep[see][for further options]{stolte2024simulation}. A specific implementation of semi-parametric DGMs has become known as Plasmode simulation \citep{franklin2014plasmode, schreck2024plasmode}, which combines resampling of covariate information from a real dataset (non-parametric element) with an outcome-generating model specified by the researcher (parametric element).

\paragraph{Resampling-only approach} 
Building on the description of semi-parametric DGMs, one might also consider DGMs that are entirely based on resampling without any parametric element. While studies that rely solely on resampling without incorporating any parametric element are commonly classified as real-data studies, particularly in the context of prediction tasks \citep{hothornDesignAnalysisBenchmark2005}, generating data by resampling from an existing dataset can also be regarded as an approach for simulations (see, e.g., \citealp{morris2019using}). This is why we briefly address this entirely non-parametric approach here as well, even though we categorize resampling-only studies as real-data studies. Note that although the term \enquote{non-parametric} might suggest the absence of parameters, it refers only to the absence of a predefined parametric form. Researchers still need to specify parameters for the resampling scheme, such as the number of observations to be drawn and whether with or without replacement.

\paragraph{Comparison of the different approaches in terms of realism}
As noted in the introduction, parametric DGMs are widely used, likely because of two key advantages: access to the ground truth and full control over the DGM. However, when the aim is to utilize real datasets to improve practical relevance, other options may, at first glance, appear more suitable for this purpose than parametric DGMs. More formally, let $\ds$ be a real dataset that is considered given for now, and let $\dgm$ denote the DGM we aim to specify to closely approximate the true but unknown DGM of $\ds$, denoted as $\dgmdtrue$. While it is then possible to make the parametric DGM real-data-based by deriving it from $\ds$ (a process we intentionally leave vague for now but will elaborate on in the remainder of this paper), generating simulated data by resampling from $\ds$ might initially seem like a natural choice: it is intrinsically real-data-based, not constrained by a parametric model, and therefore generally expected to yield a DGM $\dgm$ that aligns closely with $\dgmdtrue$.
At the same time, as already outlined in the introduction, real-data studies face the critical limitation that the underlying DGM $\dgmdtrue$ remains (largely) unknown. As a result, the set of known targets available for evaluating methods by comparing their results to the truth is inherently restricted. In this respect, semi-parametric DGMs represent a promising compromise.
For instance, in the case of Plasmode simulations, the non-parametric resampling element allows to preserve complex covariate structures present in the real dataset, while the parametric element offers knowledge of the truth for specific aspects of the DGM, such as the relationship between the covariates and the outcome \citep{schreck2024plasmode}.

Based on these considerations, one might conclude that semi-parametric DGMs should generally be preferred over parametric DGMs for specifying realistic DGMs. However, this conclusion is not universally valid, as it depends on the characteristics of the real dataset $\ds$ and the specific procedure used to derive the parametric or semi-parametric DGMs from $\ds$. Moreover, there are general arguments in favor of (real-data-based) parametric DGMs over semi-parametric DGMs. First, the non-parametric element of semi-parametric DGMs lacks a closed-form representation, making it more difficult to comprehensively describe or evaluate its plausibility. If undesirable characteristics of the dataset $\ds$ (e.g., spurious correlations) are inadvertently incorporated into the DGM, these issues are more likely to go unnoticed in semi-parametric DGMs than in parametric DGMs. Additionally, when multiple real datasets are considered, comparing relevant differences between the resulting DGMs can be more challenging in the semi-parametric case. Second, semi-parametric DGMs face practical limitations related to the accessibility of the real dataset. While it is generally recommended to use openly available datasets, in cases where the dataset $\ds$ cannot be disclosed (e.g., due to data protection issues), a parametric DGM based on $\ds$ can still be shared. In contrast, the reproducibility of semi-parametric DGMs hinges on the availability of the complete dataset. Third, real-data-based parametric DGMs may in some cases be constructed without accessing the original dataset at all. For example, relevant parameters can often be derived from summary tables or similar sources, a convenience that semi-parametric DGMs generally lack. This will be discussed in more detail in \hyperref[sec:3.3_selecting_datasets]{Section~\ref*{sec:3.3_selecting_datasets}}.

Although this paper focuses on parametric DGMs, the discussion on deriving parametric DGMs from real datasets (\hyperref[sec:3_implementation]{Section~\ref*{sec:3_implementation}}) is equally applicable to the parametric element of semi-parametric DGMs, and the considerations for dataset selection (\hyperref[sec:3.3_selecting_datasets]{Section~\ref*{sec:3.3_selecting_datasets}}) are partially also relevant for the resampling element of semi-parametric DGMs.

\subsection{Differentiating components of parametric DGMs}
Before discussing how parametric DGMs can be inferred from real data, we first need to examine their components in more detail. For this purpose, we consider a given parametric DGM $\dgm$, of which some components were inferred from a real dataset $\ds$. Given this setup, we introduce additional differentiations beyond the distinction between model structure and parameters. Specifically, we consider two differentiations: one regarding how the components of $\dgm$ were specified (\hyperref[sec:2.2.1_spec_diff]{Section~\ref*{sec:2.2.1_spec_diff}}) and another regarding our knowledge of their form or value in the true DGM underlying $\ds$ (\hyperref[sec:2.2.2_know_diff]{Section~\ref{sec:2.2.2_know_diff}}). These distinctions apply to both individual parts of the model structure of $\dgm$ and individual parameters of $\dgm$.

\subsubsection{Specification-based differentiation: researcher-specified vs.\ real-data-based components}
\label{sec:2.2.1_spec_diff}
When defining a parametric DGM $\dgm$, some components may be specified by the researcher, while others may be based on real data. This differentiation, which should be made by the researcher when planning the study, is essential because it determines how real datasets (here: a single dataset $\ds$) are selected and which components are based on real data. Each component of $\dgm$ (i.e.\ each specific part of the model structure and each parameter) falls into one of the following three categories:

\begin{enumerate}[label=\roman*.]
    \item \textbf{Researcher-specified components of interest}: These are components of $\dgm$ that are explicitly specified by the researcher based on their research question. 
    Conceptually, these components anchor the domain of interest and thus determine the selection of real datasets (together with any constraints on real-data-based components, see the third category). 
    For example, the hypothetical researcher in the introduction was interested in a setting with two treatment groups, a continuous outcome, and the violation of a specific assumption~$\assumption$. 
    In this case, the researcher-specified components of interest in $\dgm$ include the type of outcome variable (continuous) and the status of assumption~$\assumption$ (violated), both of which are parts of the model structure, as well as the number of treatment groups ($2$), which is a parameter.
    \item \textbf{Researcher-specified components of convenience}: These are components of $\dgm$ that were specified by the researcher, but rather for reasons of convenience and not because they are directly aligned with the domain of interest. In the example above, such a component could be the distribution chosen for the continuous outcome, for example, a normal distribution. 
    While it may be argued that all components that are not of specific interest should be based on real data (see the next category), some might still be reasonably specified by the researcher for practical reasons. This may be because they are assumed to have negligible effects on simulation results, are expected to hold by default, or are impractical to infer from real data. 
    Importantly, the choice to directly specify components of the DGM that are not of primary interest is a delicate one and should be justified while considering potential unintended consequences for simulation results.
    \item \textbf{Real-data-based components}: These are components of $\dgm$ that are not specified by the researcher but instead are inferred from real data. During study planning, these components are not yet specified, but researchers might still impose explicit or implicit constraints on them based on their research question.  These constraints, along with the researcher-specified components of interest, help define the domain of interest and, for this reason, are also relevant for dataset selection. 
    For example, consider an effect size parameter. Researchers may explicitly specify a range of interest, directly restricting its possible values. Alternatively, they may restrict the simulation study to a specific substantive context, such as a certain type of disease, which will implicitly constrain the possible values of the effect size parameter when inferred from datasets about that disease type.
\end{enumerate}

\paragraph{Why is this distinction important?}
Researcher-specified components of interest will be a criterion for the selection of the real datasets from which real-data-based components will be inferred. For example, if the researcher is interested in methods for the analysis of data with $\groupnum = 2$ treatment groups (i.e.\ the researcher-specified parameter for the number of treatment groups is set to $\groupnum = 2$), the datasets they will infer real-data-based components from must have $\groupnum = 2$ groups. In contrast, researcher-specified components of convenience will not interact with the selection of the real datasets.

\subsubsection{Knowledge-based differentiation: known vs.\ unknown components}
\label{sec:2.2.2_know_diff}
The second differentiation of the components of $\dgm$ considers whether their true form or value (depending on whether the component is a part of the model structure or a parameter) in the true DGM $\dgmdtrue$ underlying the real dataset $\ds$ is known or unknown. While we provide examples for both categories below, more detailed examples will be given in \hyperref[sec:2.4_examples]{ Section~\ref*{sec:2.4_examples}}.

\begin{enumerate}[label=\roman*.]
    \item \textbf{Known components}: These are components of $\dgm$ whose true form (for parts of the model structure) or value (for parameters) in $\dgmdtrue$ is known, either because it can be directly determined from $\ds$ or because it is established by external knowledge about the application where $\ds$ originates (e.g., study design information). Typically, only a limited number of components belong to this category. 
    For parts of the model structure, examples of known components include variable types (e.g., whether the outcome variable is categorical or continuous). For parameters, examples include the total number of covariates and the number of observations, as these can be observed directly from $\ds$.
    \item \textbf{Unknown components}: These are components of $\dgm$ whose true form or value in $\dgmdtrue$ is unknown and can only be inferred from $\ds$ with uncertainty.  For example, if $\dgm$ includes a relationship between an outcome variable and covariates, unknown components include the functional form of this relationship (e.g., whether the relationship is linear or not), which is a part of the model structure, and the corresponding effect sizes, which are parameters. Within unknown DGM components, a further distinction can be made between those that explicitly or implicitly appear in the formulation of the simulation target and those that do not. 
    For instance, if, as in the previous example, $\dgm$ includes a relationship between an outcome and covariates, and the simulation aims to compare methods for estimating the effect size, this parameter is directly relevant to the simulation target.
\end{enumerate}

\paragraph{Why is this distinction important?}
This differentiation is relevant because unknown components introduce uncertainty that affects both the selection of real datasets and the inference of components from them. More specifically, for components of $\dgm$ that are intended to be real-data-based but are unknown, inference can only be conducted with uncertainty (e.g., a corresponding parameter can only be estimated rather than known exactly). If $\dgm$'s researcher-specified components of interest, which guide dataset selection, are unknown, it remains uncertain whether their specified form or value truly matches that in $\dgmdtrue$.

Also note that the distinction between known and unknown components is relevant for both individual parts of the model structure of $\dgm$ and individual parameters of $\dgm$. Since parameters in any parametric DGM are generally defined conditionally on the model structure, applying this differentiation to a parameter of $\dgm$ requires assuming that at least the relevant part of its\enlargethispage{\baselineskip} model structure matches that of $\dgmdtrue$. Otherwise, the corresponding parameter (and its true value) would not be meaningfully defined in $\dgmdtrue$.

\subsection{Notation and simplifying assumptions}
\label{sec:2.3_notation}
When describing the inference of real-data-based DGMs' components in \hyperref[sec:3_implementation]{ Section~\ref*{sec:3_implementation}}, 
we will start with the researcher-specified components already set, while the real-data-based components remain unspecified. In this section, we introduce the corresponding notation needed for this process. To simplify the notation, we make the following two assumptions.

\paragraph{Assumption 1: Researcher-specified components fixed}
In many cases, researcher-specified components of interest are not set to a single option or value. 
Returning to the hypothetical researcher from the introduction, we have, up to this point, suggested that they are only interested in a DGM with two treatment groups, a continuous outcome, and the violation of a specific assumption~$\assumption$. However, they may also be interested in DGMs with three or four treatment groups or with a dichotomous outcome instead of a continuous one. This would result in up to six possible combinations. To simplify the discussion, we assume that the researcher-specified components of interest are fixed to a single option or value at a time. In the example, this means that only one of these six combinations would be considered within a given process of constructing real-data-based DGMs. Accordingly, in practice, this process would need to be repeated separately for each combination. This is because the researcher-specified components of interest determine which real datasets are eligible to be used for constructing a real-data-based parametric DGM.\\
In contrast to the researcher-specified components of interest, the researcher-specified components of convenience are more likely to be set to a single option or value, particularly when their influence is assumed to be negligible or they are expected to hold by default. However, if such a component is assigned multiple options or values (e.g., because it is impractical to infer from real data but still important to vary), this does not require repeating the full process of constructing real-data-based DGMs. Instead, it only requires deciding how these components with multiple options or values will be combined with the later inferred real-data-based components. For simplicity, we assume that all researcher-specified components of convenience are also fixed to a single option or value in \hyperref[sec:3_implementation]{ Section~\ref*{sec:3_implementation}}, though we encounter the non-simplified case in the empirical illustrations in  \hyperref[sec:4_results]{ Section~\ref*{sec:4_results}}.

\paragraph{Assumption 2: Only parameter inference as the base scenario}
While both parts of the model structure and individual parameters can, in principle, be inferred from real data, this paper primarily focuses on parameter inference.
Consequently, we assume as a base scenario that only parameters are inferred from real data, meaning that all parts of the model structure are specified by the researcher, either as components of interest or convenience, and regard the case where parts of the model structure are also based on real data as an extended scenario. If, in addition, parts of the model structure are inferred from real data, this would lead to multiple possible model structures.

\paragraph{Notation for the base scenario}
Based on these simplifying assumptions, we introduce notation that will be used in the remainder of this paper (see \hyperref[tab:notation_sec2]{Table~\ref*{tab:notation_sec2}} for an overview). 
In the base scenario where the model structure is not inferred from real data but fully specified by the researcher, we denote this model structure as $\struc$. 
For parameters, which are inherently determined by $\struc$, we denote the set of parameters that are intended to be real-data-based as $\thetaall$, while the set of parameters that are researcher-specified, either as components of interest or as components of convenience, is denoted as $\lambdaall$. 
Because researcher-specified components are assumed to be fixed to a single option or value, each parameter in $\lambdaall$ is set to a single value. Consequently, given $\struc$ and $\lambdaall$, if a single value were inferred for each parameter in $\thetaall$, this would fully specify a single DGM.

To also integrate the knowledge-based differentiation specifically for parameters into the notation, we denote $\thetadesign$ and $\lambdadesign$ as the parameters whose true values in the true DGM underlying the real dataset are known, while $\thetaestim$ and $\lambdaestim$ are those whose true values are unknown. This differentiation is relevant for all parameters, regardless of whether they belong to $\thetaall$ (real-data-based) or $\lambdaall$ (researcher-specified). 
Within the category of unknown parameters, we further refine the notation to reflect the distinction between parameters that explicitly or implicitly appear in the formulation of the simulation target and those that do not. Accordingly, we write $\thetaestimtarget$ and $\lambdaestimtarget$ for the former and $\thetaestimother$ and $\lambdaestimother$ for the latter. 

%%%%% Begin Table: tab_notation_sec2 %%%%%----------------------
\begin{table}[ht]
\small
\centering
\caption{Overview of notation for the base scenario, where parameters of a parametric DGM $\dgm$ are inferred from a real dataset $\ds$ with the underlying DGM $\dgmdtrue$. }
{\renewcommand{\arraystretch}{1.2}
\begin{tabular}{ll}
\hline
 &  \\ \hline
$\struc$ & Model structure of $\dgm$  \\
$\thetaall=(\thetadesign^\top,\thetaestim^\top)^\top$ & Real-data-based  parameters \\
$\lambdaall=(\lambdadesign^\top,\lambdaestim^\top)^\top$ & Researcher-specified  parameters   \\ \hline
$\thetadesign$ & Real-data-based parameters whose true values in $\dgmdtrue$ are known  \\
$\lambdadesign$ & Researcher-specified parameters whose true values in $\dgmdtrue$ are known  \\ \hline
$\thetaestim$ & Real-data-based parameters whose true values in $\dgmdtrue$ are unknown   \\
\hspace{0.3cm} $\thetaestimtarget$ & \dots\ and that are related to the simulation target  \\
\hspace{0.3cm} $\thetaestimother$ & \dots\ and that are not related to the simulation target  \\
$\lambdaestim$ & Researcher-specified parameters whose true values in $\dgmdtrue$ are unknown  \\
\hspace{0.3cm} $\lambdaestimtarget$ & \dots\ and that are related to the simulation target  \\
\hspace{0.3cm} $\lambdaestimother$ & \dots\ and that are not related to the simulation target \\
\hline
\end{tabular}
}
\label{tab:notation_sec2}
\end{table}
%%%%% End Table: tab_notation_sec2 %%%%%----------------------

\subsection{Examples}
\label{sec:2.4_examples}
To illustrate the introduced notation, we now present descriptions of four simulation studies, which will also serve as recurring examples throughout the paper. These examples do not specify full study designs but focus on the aim, the model structure of the DGM, and the target/estimand, aligning with the \enquote{A}, \enquote{D}, and \enquote{E} aspects of the ADEMP framework for simulation studies proposed by \citet{morris2019using}.
Also, while examples 2–4 are based on actual published simulation studies, only a single aim, target, and model structure were selected per study (if multiple were present), and in some cases, these were slightly modified or simplified. Additionally, the notation was adjusted to ensure consistency across the examples. A summary of the parameters occurring in the examples is provided in \hyperref[tab:parameter_overview]{Table~\ref*{tab:parameter_overview}}.\\
Since the knowledge-based differentiation of parameters may require additional explanation, we explicitly highlight it in these examples. We assume all parameters to be based on real data and thus initially left unspecified, meaning that they are all contained in $\thetaall$, except for the number of groups ($\groupnum$), which we consider as a researcher-specified parameter of interest for practical reasons. 
Since the true value of $\groupnum$ can be known in the true DGM underlying a dataset $\ds$, we have $\lambdadesign = \{\groupnum\}$ (while $\lambdaestim = \emptyset$, as there are no other researcher-specified parameters).

\paragraph{Example 1} (\textit{Example-\exordinalshort})
\label{ex:ordinal}
\begin{itemize}
    \item \textbf{Aim}: Evaluation of methods testing the null hypothesis $H_0$ of no treatment differences in two-arm (i.e.\ $\groupnum = 2$) randomized controlled trials with ordinal outcomes having $\ordnum$ categories, in settings where $H_0$ is false
    \item \textbf{Model structure $\struc$ of the DGM}: Each simulated dataset is an $\obsnum \times 2$ matrix containing the ordinal outcome $y_{\obsind} \in \{1, \dots, \ordnum\}$ and the treatment assignment $x_{\obsind} \in \{\groupone, \grouptwo\}$ for $\obsnum$ individuals, $\obsind = 1, \dots, \obsnum$. Half of the individuals ($\obsnum/2$) are assigned to each treatment group. For each individual in group $\groupind \in \{1,2\}$, the outcome is generated by drawing from  $\operatorname{Multinomial}(1, \bm{\pi}_{\groupind})$, where $\bm{\pi}_{\groupind} = (\pi_{1,{\groupind}},\dots,\pi_{\ordnum,\groupind})$, with $\pi_{\ordind,\groupind} = P(Y = \ordind \mid X = \groupind)$, $\ordind = 1, \dots, \ordnum$, and $\sum_{\ordind=1}^\ordnum \pi_{\ordind,\groupind} = 1$. In addition, since $H_0$ is false, $\pi_{\ordind,\groupone} \neq \pi_{\ordind,\grouptwo}$ for at least one $\ordind \in \{1, \dots, \ordnum\}$.
    \item \textbf{Estimand/Target}: The null hypothesis $H_0: \pi_{\ordind,\groupone} = \pi_{\ordind,\grouptwo}$ for all $\ordind \in \{1, \dots, \ordnum\}$
\end{itemize} 
The parameters in $\thetaall$ in {\exordinal} are the number of individuals, $\obsnum$, the number of ordinal categories, $\ordnum$, and the outcome probabilities for each group, $\bm{\pi}_1$ and $\bm{\pi}_2$. The true values of $\obsnum$ and $\ordnum$ in $\dgmdtrue$ are known, making them part of $\thetadesign$, i.e.\ $\thetadesign = \{\obsnum,\ordnum\}$. In contrast, the true probabilities $\bm{\pi}_1$ and $\bm{\pi}_2$ in $\dgmdtrue$  cannot be known. Since the simulation target is to evaluate the null hypothesis $H_0: \bm{\pi}_1 = \bm{\pi}_2$, it follows that $\thetaestimtarget = \{\bm{\pi}_1, \bm{\pi}_2\}$. In this example, there are no additional parameters, so $\thetaestimother = \emptyset$.

\paragraph{Example 2} (\textit{Example-\exsurvshort}, based on the study by \citealp{dormuth2023comparative})
\label{ex:survival}
\begin{itemize}
    \item \textbf{Aim}: Evaluation of methods testing the null hypothesis $H_0$ of no differences in two-arm (i.e.\ $\groupnum = 2$) clinical trials with survival outcome, in settings where $H_0$ is false
    \item \textbf{Model structure $\struc$ of the DGM}: Each simulated dataset is an $\obsnum \times 3$ matrix containing the (uncensored or right-censored) survival time $y_{\obsind} \in \mathbb{R}^+$, the censoring indicator $d_{\obsind} \in \{0, 1\}$ (with $d_{\obsind} = 1 $ if the event was observed and $d_{\obsind} = 0$ otherwise), and the treatment assignment $x_{\obsind} \in \{\groupone, \grouptwo\}$ for $\obsnum$ individuals, $\obsind = 1, \dots, \obsnum$. Each treatment group contains $\obsnum/2$ individuals. For each individual in group $\groupind \in \{1,2\}$, the observed survival time and the censoring indicator are generated as $y = \min(t, c)$ and $d = \mathbbm{1}(t \leq c)$, respectively, with theoretically observable survival time $t$ and censoring time $c$ being drawn independently from $ \operatorname{Exp}(\eventrate_{\groupind})$ and  $\operatorname{Unif}(0, u)$, respectively. Since $H_0$ is false, $\eventrate_\groupone \neq \eventrate_\grouptwo$.
    \item \textbf{Estimand/Target}: The null hypothesis $H_0: S_{\groupone}(t) = S_{\grouptwo}(t)$ for all $t \in \mathbb{R}^+$, where $S_{\groupone}(t)$ and $S_{\grouptwo}(t)$ are the survival functions of groups $\groupone$ and $\grouptwo$
\end{itemize}
In {\exsurv}, the parameters in $\thetaall$ are the number of individuals, $\obsnum$, the event rate parameters $\eventrate_\groupone$ and $\eventrate_\grouptwo$, and the upper bound of the censoring distribution, $u$. Similar to the first example, the true value of $\obsnum$ is known, i.e.\ $\thetadesign = \{\obsnum\}$. This is in contrast to the parameters $\eventrate_\groupone$, $\eventrate_\grouptwo$, and $u$, whose true values cannot be known (unless $u$ is explicitly determined by the\enlargethispage{\baselineskip} study design). Since the target considers the survival function, which for group $\groupind$ under the exponential distribution is given by $S_\groupind(t) = \exp(-\eventrate_\groupind t)$, it follows that $\thetaestimtarget = \{\eventrate_\groupone, \eventrate_\grouptwo\}$, while $\thetaestimother = \{u\}$.

%%%%% Begin Table: tab_parametertypes %%%%%----------------------
\begin{table}[ht]
\small
\centering
\caption{Summary of all parameters in $\thetaall$, categorized according to the knowledge-based differentiation of components, in the example simulation studies. Except for the number of groups, which is researcher-specified and set to $\groupnum=2$ in all examples, these constitute the full set of parameters defined by the corresponding model structure $\struc$.}
{\renewcommand{\arraystretch}{1.2}
\begin{tabularx}{\textwidth}{|l|X|p{2.07cm}|p{2.9cm}|p{3.28cm}|}
\hline
\textbf{Example} & {\raggedright \textbf{Aim: Evaluate methods for \dots}} & $\thetadesign$ & $\thetaestimtarget$ & $\thetaestimother$ \\ \hline
\hyperref[ex:ordinal]{\textit{\exordinalshort}} & \dots\ testing $H_0$ of no treatment differences in two-arm randomized controlled trials with ordinal outcomes 
& 
{\raggedright
\begin{tabitemize}
    \item $\obsnum$: No. of individuals 
    \item $\ordnum$: No. of outcome categories
\end{tabitemize}
} 
& 
{\raggedright
\begin{tabitemize}
    \item $\bm{\pi}_1, \bm{\pi}_2$: Outcome probabilities per group
\end{tabitemize}
} 
& 
{\raggedright
    --
} \\ \hline
\hyperref[ex:survival]{\textit{\exsurvshort}} & \dots\ testing $H_0$ of no differences in two-arm trials with survival outcomes 
& 
{\raggedright
\begin{tabitemize}
    \item $\obsnum$: No. of individuals
\end{tabitemize}
} 
& 
{\raggedright
\begin{tabitemize}
    \item $\eventrate_1, \eventrate_2$: Event rate per group
\end{tabitemize}
} 
& 
{\raggedright
\begin{tabitemize}
    \item $u$: Censoring upper bound
\end{tabitemize}
} \\ \hline
\hyperref[ex:meta]{\textit{\exmetashort}} & \dots\ estimating the variance of true effect sizes (between-study heterogeneity variance) 
& 
{\raggedright
\begin{tabitemize}
    \item $\nstudy$: No. of studies
\end{tabitemize}
} 
& 
{\raggedright
\begin{tabitemize}
    \item $\tau^2$: Between-study heterogeneity
\end{tabitemize}
} 
& 
{\raggedright
\begin{tabitemize}
    \item $\metaeffect$: Overall effect
    \item $u_{\text{min}}, u_{\text{max}}$: Range for sample size
    \item $\mu_{1,\studyind}$: Mean for group 1 (per study)
    \item $\sigma^2$: Within-group variance
\end{tabitemize}
} \\ \hline
\hyperref[ex:de]{\textit{\exdeshort}} & \dots\ identifying differentially expressed (DE) genes between two groups 
& 
{\raggedright
\begin{tabitemize}
    \item $\obsnum$: No. of samples
    \item $\varnum$: No. of genes
\end{tabitemize}
} 
& 
{\raggedright
\begin{tabitemize}
    \item $\operatorname{FC}_{\varind}$: Fold change
    \item $\propDE$: Proportion of DE genes
\end{tabitemize}
} 
& 
{\raggedright
\begin{tabitemize}
    \item $\mu_{\varind}, \phi_{\varind}$: Expression mean and dispersion
\end{tabitemize}
} \\ \hline
\end{tabularx}
}

\label{tab:parameter_overview}
\end{table}
%%%%% End Table: tab_parametertypes %%%%%----------------------

\paragraph{Example 3} (\textit{Example-\exmetashort}, based on the study by \citealp{langan2019comparison}) 
\label{ex:meta}
\begin{itemize}
    \item \textbf{Aim}: Evaluation of methods to estimate the variance of the true effect sizes (between-study heterogeneity variance) in meta-analyses of studies with two groups (i.e.\ $\groupnum = 2$) and continuous outcomes
    \item \textbf{Model structure $\struc$ of the DGM}:  Each simulated dataset represents a meta-analysis of $\nstudy$ studies. It is an $\nstudy \times 2$ matrix containing the estimated effect size $\hat{\metaeffect}_\studyind \in \mathbb{R}$ and its estimated within-study variance $\hat{\sigma}_{\studyind}^2 \in \mathbb{R}^+$ for the $\nstudy$ studies, $\studyind = 1,\dots,\nstudy$. 
    The evaluated methods are applied exclusively to the meta-analysis dataset. However, to generate this dataset, additional study-level data must be simulated for each of the $\nstudy$ studies. For each study $\studyind$, the true study effect $\metaeffect_\studyind$ is drawn from $\mathcal{N}(\metaeffect, \tau^2)$, where $\metaeffect$ is the true overall effect and $\tau^2$ is the between-study heterogeneity variance, and a study sample size  $\nindindex$ is drawn from $\operatorname{Unif}(u_{\text{min}}, u_{\text{max}})$ and then split evenly into two groups. Outcome values for the individuals in group 1 and 2 are drawn from $\mathcal{N}(\mu_{1,\studyind}, \sigma_{1,\studyind}^2)$ and $\mathcal{N}(\mu_{2,\studyind}, \sigma_{2,\studyind}^2)$, respectively, where  $\mu_{2,\studyind} - \mu_{1,\studyind} = \metaeffect_\studyind$ and $\sigma^2_{1,\studyind}=\sigma^2_{2,\studyind}=\sigma^2$. Based on the simulated data at the study level, the estimated effect size $\hat{\metaeffect}_\studyind$ and within-study variance $\hat{\sigma}_{\studyind}^2$ are calculated using Hedges' $g$ \citep{Hedges1981}.
    \item \textbf{Estimand/Target}: The between-study heterogeneity variance $(\tau^2)$
\end{itemize}
The parameters in $\thetaall$ in {\exmeta} can be grouped by study level: At the meta-analysis level, they include the number of studies, $\nstudy$, the overall effect $\metaeffect$, the between-study heterogeneity variance $\tau^2$, and the parameters $u_{\text{min}}$ and $u_{\text{max}}$, which define the range for the study sample sizes.  At the study level, the parameters are the mean for group 1 (for each study $\studyind$), $\mu_{1,\studyind}$, and the within-group variance $\sigma^2$. Accordingly, $\dgmdtrue$ essentially represents a two-level mechanism, specifying both the generation of study-level data and the meta-analysis dataset. If $\ds$ represents a meta-analysis dataset (where each row corresponds to a study), additional datasets for the $\nstudy$ studies summarized in $\ds$ would be needed to infer the study-level parameter values. Similar to the previous examples, the true value of $\nstudy$ in $\dgmdtrue$ can be known, i.e.\ $\thetadesign = \{\nstudy\}$, while the true values of the remaining parameters cannot.
Among the latter, $\tau^2$ represents the simulation target, i.e.\ $\thetaestimtarget = \{\tau^2\}$, while the remaining parameters are included in $\thetaestimother = \{\metaeffect, \sigma^2, u_{\text{min}}, u_{\text{max}}, \mu_{1,\studyind} \mid \studyind = 1, \dots, \nstudy\}$.

\paragraph{Example 4} (\textit{Example-\exdeshort}, based on the study by \citealp{baik2020benchmarking})
\label{ex:de}
\begin{itemize}
    \item \textbf{Aim}: 
    Evaluation of methods for differential gene expression analysis, i.e.\ methods that identify genes with differences in their RNA-Seq expression levels, in a two-group (i.e.\ $\groupnum = 2$) setting (e.g., cancer vs.\ normal) 
    \item \textbf{Model structure $\struc$ of the DGM}: Each simulated dataset is an $\obsnum \times (\varnum+1)$ matrix containing the RNA-Seq read count $r_{\obsind,\varind} \in \mathbb{Z}^{0+}$ for $\obsnum$ samples, $\obsind = 1, \dots, \obsnum$, and $\varnum$ genes, $\varind = 1, \dots, \varnum$, where the read count represents the gene expression level, with a larger count indicating higher expression. The matrix also includes the group indicator $x_{\obsind} \in \{\groupone, \grouptwo\}$, and each group contains $\obsnum/2$ samples. For sample $\obsind$ and gene $\varind$, the read count is generated by drawing from a negative binomial distribution, specifically $\operatorname{NB}(\mu_{\varind} \cdot \operatorname{FC}_{\varind}, \phi_{\varind})$, $\mu_{\varind}, \phi_{\varind} \geq 0$, if $x_{\obsind} = \groupone$, and  $\operatorname{NB}(\mu_{\varind}, \phi_{\varind})$, $\mu_{\varind}, \phi_{\varind} \geq 0$, if $x_{\obsind} = \grouptwo$. Here, the fold change $\operatorname{FC}_{\varind}$ quantifies the relative change in expression. Among all genes, a proportion $\propDE$ is simulated as differentially expressed (DE), with $\operatorname{FC}_{\varind} \neq 1$ for those genes, while $\operatorname{FC}_{\varind} = 1$ for non-differentially expressed genes.
    \item \textbf{Estimand/Target}: The null hypothesis $H_0: \operatorname{FC}_{\varind} = 1$ for all $\varind \in \{{1, \dots, \varnum}\}$
\end{itemize}   
In {\exde}, the parameters in $\thetaall$ are the number of samples, $\obsnum$, the number of genes, $\varnum$, the mean expression level $\mu_{\varind}$ and  the dispersion parameter $\phi_{\varind}$ for each gene $\varind$, $\varind = 1, \dots, \varnum$, the proportion of DE genes, $\propDE$, and the fold change $\operatorname{FC}_{\varind}$ (where $\operatorname{FC}_{\varind} \neq 1$ only for DE genes). 
While the true values of $\obsnum$ and $\varnum$ in $\dgmdtrue$ are known, i.e.\ $\thetadesign = \{\obsnum, \varnum\}$, the true values of $\mu_{\varind}$, $\phi_{\varind}$, $\propDE$, and $\operatorname{FC}_{\varind}$ are not. 
Since the simulation target is to evaluate the null hypothesis of no differential expression  ($\operatorname{FC}_{\varind} = 1$) for each gene, $\thetaestimtarget$ includes $\operatorname{FC}_{\varind}$ and $\propDE$, as $\propDE$ represents the proportion of genes where the null hypothesis does not hold. Accordingly, $\thetaestimtarget = \{\operatorname{FC}_{\varind} \mid \operatorname{FC}_{\varind} \neq 1\} \cup \{\propDE\}$. The remaining parameters, $\mu_{\varind}$ and $\phi_{\varind}$, are part of $\thetaestimother$, as they are required for generating the data but not directly related to the simulation target, i.e.\ $\thetaestimother = \{\mu_{\varind}, \phi_{\varind} \mid \varind = 1, \dots, \varnum\}$.

%%%%%%%%%%%%%%%%%%%%%%%%%%%%%%%%%%%%%%%%%%%%%%%%%%%%
%%%%%%%%%%%%%%%%%%%% Section 3 %%%%%%%%%%%%%%%%%%%%%
%%%%%%%%%%%%%%%%%%%%%%%%%%%%%%%%%%%%%%%%%%%%%%%%%%%%
\section{Inferring components of real-data-based parametric DGMs from a set of real datasets}
\label{sec:3_implementation}
In this section, we provide a detailed overview of the inference of real-data-based parametric DGMs' components, along with practical recommendations, while the systematic selection of the real datasets on which the DGMs are based will be addressed in  \hyperref[sec:3.3_selecting_datasets]{Section~\ref*{sec:3.3_selecting_datasets}}. 

We begin with the inference of DGM parameters in the base scenario with a single, fully researcher-specified model structure $\struc$, where only parameters ($\thetaall$) are inferred from real data (\hyperref[sec:3.1_inferring_params]{Section~\ref*{sec:3.1_inferring_params}}), and then address the extended scenario, where parts of the model structure are no longer specified by the researcher and instead also inferred from real data (\hyperref[sec:3.2_rdb_ms]{Section~\ref*{sec:3.2_rdb_ms}}). For both scenarios, we assume that any researcher-specified parameters $\lambdaall$ are fixed to a single value. 

\subsection{Inferring parameters from a set of real datasets} \label{sec:3.1_inferring_params}
\subsubsection{Overview}
Given a set of $\dsnum$ real datasets (the selection of which will be discussed in \hyperref[sec:3.3_selecting_datasets]{Section~\ref*{sec:3.3_selecting_datasets}}), $\dsall = \{\ds^{(1)},\dots,\ds^{(\dsnum)}\}$, a model structure $\struc$, and a set of $\paramnum$ parameters, $\thetaall$, which we now consider to be arranged as a vector, i.e.\ $\thetaall=(\rdbparam_1, \ldots, \rdbparam_\paramnum)$, there are different approaches for inferring the set of $\consideredparamvectornum$ considered parameter vectors, $\setofconsideredparamvectors = \{\consideredparamvector^{(1)}, \ldots, \consideredparamvector^{(\consideredparamvectornum)}\}$, from the real datasets. 
Each considered parameter vector $\consideredparamvector^{(\consideredparamvectorind)}$, $\consideredparamvectorind=1, \ldots, \consideredparamvectornum$, contains a single value for each parameter in $\thetaall$ (i.e.\ $\paramnum$ values in total). See \hyperref[tab:notation_sec3]{Table~\ref*{tab:notation_sec3}} for an overview of relevant notation introduced throughout \hyperref[sec:3.1_inferring_params]{Section~\ref*{sec:3.1_inferring_params}}.

%%%%% Begin Table: tab_notation_sec3 %%%%%----------------------
\begin{table}[ht]
\small
\centering
\caption{Overview of notation for inferring a set of considered parameter vectors $\setofconsideredparamvectors$ from a set of real datasets $\dsall$, where each considered parameter vector, together with a model structure $\struc$, defines a DGM. 
}
{\renewcommand{\arraystretch}{1.2}
\begin{tabularx}{1\textwidth}{W{2cm}X}
\hline
\\ \hline
$\dsall$ & Set of selected real datasets \\
$\dsnum$ & Number of datasets in $\dsall$ (i.e.\ number of selected real datasets) \\ 
$\dsind = 1, \ldots, \dsnum$ & Index for the datasets in $\dsall$ (i.e.\ index for the  selected real datasets) \\ 
$\dswithind$ & A given selected real dataset, $\dsind = 1, \ldots, \dsnum$ \\
\hline
$\thetaall$ & Vector of real-data-based parameters \\
$\paramnum$ & Number of parameters in $\thetaall$ (i.e.\ number of real-data-based parameters) \\
$\paramind = 1, \ldots, \paramnum$ & Index for the parameters in $\thetaall$ (i.e.\ index for the real-data-based parameters) \\
$\rdbparam_\paramind$ & A given real-data-based parameter, $\paramind = 1, \ldots, \paramnum$ \\
\hline
$\boldsymbol{\realizedrdbparam}_\paramind$ & Set of inferred parameter values if parameter $\rdbparam_\paramind$ is inferred via direct inference   \\
$\realizedrdbparam_\paramind^{(r)}$ & Parameter value (directly) inferred from $\dswithind$ for parameter $\rdbparam_\paramind$\\
$\boldsymbol{\aggregatedrdbparam}_\paramind$ & Set of inferred parameter values if parameter $\rdbparam_\paramind$ is inferred via aggregated inference    \\
\hline
$\setofconsideredparamvectors$ & Set of considered parameter vectors, each of which, together with $\struc$, defines one DGM \\
$\consideredparamvectornum$ & Number of parameter vectors in $\setofconsideredparamvectors$ (i.e.\ number of considered parameter vectors)\\
$\consideredparamvectorind = 1, \ldots, \consideredparamvectornum$ & Index for the parameter vectors in $\setofconsideredparamvectors$ (i.e.\ index for the considered parameter vectors) \\ 
$\consideredparamvector^{(\consideredparamvectorind)}$ & A given considered parameter vector, $\consideredparamvector^{(\consideredparamvectorind)} = (\paraminconsideredparamvector_1^{(\consideredparamvectorind)}, \ldots, \paraminconsideredparamvector_\paramnum^{(\consideredparamvectorind)})$, $\consideredparamvectorind = 1, \ldots, \consideredparamvectornum$ \\
\hline
\end{tabularx}
}
\label{tab:notation_sec3}
\end{table}
%%%%% End Table: tab_notation_sec3 %%%%%----------------------

The inference process can be broken down into two steps. In the first step, for each of the $\paramnum$ parameters in $\thetaall$, values are inferred from $\dsall$, resulting in a set of inferred parameter values for\enlargethispage{\baselineskip} each parameter. In the second step, those sets are mapped to the set of considered parameter vectors, $\setofconsideredparamvectors$. 

The approach that is expected to yield DGMs most closely approximating the true DGMs of the real datasets proceeds as follows:
In the first step, for each parameter, a value is inferred from each dataset (i.e.\ $\dsnum$ values are inferred from $\dsall$, and values are not aggregated across datasets). 
In the second step, the inferred values for each parameter are combined per dataset to form the set of considered parameter vectors, with each of the vectors containing the parameter values inferred from one dataset.
Since this approach essentially maps each dataset to one of the considered parameter vectors (and thus is equivalent to constructing one DGM per dataset, given $\struc$ and $\lambdaall$), we will refer to it as the one-to-one inference approach. However, one may also deviate from this approach and consider alternative strategies for the two outlined steps of the inference process. An overview is presented in \hyperref[tab:inferring_parameters]{Table~\ref*{tab:inferring_parameters}}.
In the following, we first discuss the one-to-one inference approach before exploring these alternatives. 
Note that regardless of the chosen inference approach, the resulting DGMs should be checked for plausibility. 

%%%%% Begin Table: tab_inferring_parameters %%%%%----------------
  \begin{table}[h]
    \caption{Overview of approaches for inferring parameters from a set of real datasets, $\dsall = \{\ds^{(1)},\dots,\ds^{(\dsnum)}\}$.}
    \small
    \centering
    {\renewcommand{\arraystretch}{1.3}
   \begin{tabularx}{1\textwidth}{|W{3.8cm}|W{5cm}|X|}\hline
%%%%%%%%%%%%%%%%%%%%%%%%%%%%%%%%%%%%%%%%%%%%%%%%%%%%%%%%%%%%%%%%%%%%%%%%%%%%%
       \textbf{Step of the inference process} &   \textbf{One-to-one  approach} & \textbf{Deviation}\\ \hline
1. Infer set of parameter values for each parameter $\rdbparam_\paramind$ in $\thetaall$. \newline $\rightarrow$ $\setofvaluesforparam_\paramind$  & Direct inference: Use value from each of the $\dsnum$ datasets directly.
\newline $\rightarrow$ $\setofvaluesforparam_\paramind \mathrel{\mathop:}=\boldsymbol{\realizedrdbparam}_\paramind = \{\realizedrdbparam_\paramind^{(1)}, \ldots, \realizedrdbparam_\paramind^{(\dsnum)}\}$ & Aggregated inference: Use information from $\dsnum$ datasets in aggregated form to generate $\aggregatednum$ values.
\newline $\rightarrow$ $\setofvaluesforparam_\paramind \mathrel{\mathop:}=\boldsymbol{\aggregatedrdbparam}_\paramind = \{\aggregatedrdbparam_\paramind^{(1)}, \ldots, \aggregatedrdbparam_\paramind^{(\aggregatednum)}\}$ \\ \hline
2. Map sets of inferred values for individual parameters, $\setofvaluesforparam_1, \ldots, \setofvaluesforparam_\paramnum$, to set of considered parameter vectors. \newline $\rightarrow$ $\setofconsideredparamvectors = \{\consideredparamvector^{(1)}, \ldots, \consideredparamvector^{(\consideredparamvectornum)}\}$ & Combine values of directly inferred parameters per dataset, i.e.\ $\boldsymbol{\realizedrdbparam}^{(\dsind)} = (\realizedrdbparam_1^{(\dsind)}, \ldots, \realizedrdbparam_\paramnum^{(\dsind)})$ for each $\dswithind$.
\newline $\rightarrow$ $\consideredparamvector^{(\consideredparamvectorind)}$ contains the values from one dataset ($\consideredparamvectornum = \dsnum$).
\newline $\rightarrow$ $\setofconsideredparamvectors  \mathrel{\mathop:}=\{\boldsymbol{\realizedrdbparam}^{(1)}, \ldots, \boldsymbol{\realizedrdbparam}^{(\dsnum)}\}$ & Combine values of directly inferred parameters across datasets and/or aggregately inferred parameters.
\newline $\rightarrow$ $\consideredparamvector^{(\consideredparamvectorind)}$ contains not only values from one dataset (typically $\consideredparamvectornum \neq \dsnum$).
\newline $\rightarrow$ $\setofconsideredparamvectors  \mathrel{\mathop:}=\setofvaluesforparam_1 \times \cdots \times \setofvaluesforparam_\paramnum$ (for fully factorial design), with $\setofvaluesforparam_\paramind \mathrel{\mathop:}=\boldsymbol{\realizedrdbparam}_\paramind$ or $\setofvaluesforparam_\paramind \mathrel{\mathop:}=\boldsymbol{\aggregatedrdbparam}_\paramind$ \\ \hline
    \end{tabularx} 
    }
\label{tab:inferring_parameters}
\end{table}
%%%%% End Table: tab_inferring_parameters %%%%%----------------

\subsubsection{One-to-one inference approach}\label{sec:3.1.1_rdb_param_direct}
More formally, in the one-to-one approach, each parameter $\rdbparam_\paramind$ in the parameter vector $\thetaall$, $\paramind = 1, \ldots, \paramnum$, is first directly inferred from each real dataset $\dswithind$, $\dsind = 1, \ldots, \dsnum$. For a given parameter $\rdbparam_\paramind$, this results in a set $\boldsymbol{\realizedrdbparam}_\paramind$, which contains $\dsnum$ values:
\begin{equation}
\label{eq:parameter_q_set}
\boldsymbol{\realizedrdbparam}_\paramind = \{\realizedrdbparam_\paramind^{(1)}, \ldots, \realizedrdbparam_\paramind^{(\dsnum)}\}.
\end{equation}
We refer to this way of inferring the values of a given parameter $\rdbparam_\paramind$ from $\dsall$ as direct inference.\\
To subsequently map the sets $\boldsymbol{\realizedrdbparam}_1, \dots, \boldsymbol{\realizedrdbparam}_\paramnum$ to the set of considered parameter vectors, $\setofconsideredparamvectors$, in the one-to-one approach, these directly inferred parameter values are combined per dataset, forming a full parameter vector  $\boldsymbol{\realizedrdbparam}^{(\dsind)} = (\realizedrdbparam_1^{(\dsind)}, \ldots, \realizedrdbparam_\paramnum^{(\dsind)})
$ for each dataset $\dswithind$. The set of considered parameter vectors therefore contains $\consideredparamvectornum = \dsnum$ considered parameter vectors:
\begin{equation}
\label{eq:considered_parameter_vectors_one-to-one}
    \setofconsideredparamvectors = \{\consideredparamvector^{(1)}, \ldots, \consideredparamvector^{(\consideredparamvectornum)}\} = \{\boldsymbol{\realizedrdbparam}^{(1)}, \ldots, \boldsymbol{\realizedrdbparam}^{(\dsnum)}\} =\mathrel{\mathop:} \setofconsideredparamvectors^{\text{one-to-one}}.
\end{equation}
This way of mapping the sets of inferred values for the individual parameters to the set of considered parameter vectors corresponds to a scattershot design where each $\boldsymbol{\realizedrdbparam}^{(\dsind)}$ contains potentially distinct values \citep{siepe2024simulation}.\\ Given a model structure $\struc$, the one-to-one approach results in $\dsnum$ DGMs, $\dgm^{(1)}, \dots, \dgm^{(\dsnum)}$, each corresponding to a specific dataset and individually parameterized by $\boldsymbol{\realizedrdbparam}^{(1)}, \dots, \boldsymbol{\realizedrdbparam}^{(\dsnum)}$ (and $\lambdaall$). 
Since each dataset is essentially treated individually in the one-to-one inference approach, the following considerations focus on the direct inference of parameters from a single real dataset $\dswithind$. For this dataset, we denote its true underlying DGM as $\dgmdtrueind$ and the corresponding true model structure as $\strucdtrueind$. To explore the direct inference of parameters $\thetaall$ from $\dswithind$, we apply the categorization of parameters introduced in \hyperref[sec:2.3_notation]{Section~\ref*{sec:2.3_notation}},  distinguishing between $\thetadesign$ and $\thetaestim$, with a further differentiation between $\thetaestimtarget$ and $\thetaestimother$.

\paragraph{Parameters in $\thetadesign$}
As illustrated by the four example simulation studies, parameters whose true values in $\dgmdtrueind$ can be known typically specify quantities such as the number of variables, the number of categories within a variable, or the number of observations in the DGM. 
While it is straightforward to infer their true value in $\dswithind$, these parameters often have the least need to be explicitly real-data-based (i.e.\ to be included in $\thetadesign$ rather than $\lambdadesign$), as researchers are likely to choose reasonably realistic values even without relying on specific datasets.
Additionally, parameters such as the number of observations are often chosen as rounded, tidy values (e.g., $n \in  \{10, 50, 100\}$). Using values that deviate from these conventions might make results harder to interpret and their reporting tedious. Nevertheless, when using researcher-specified values, unintentionally unrealistic values can arise. As mentioned in the introductory section, \citet{langan2017comparative} observed that the number of studies used in meta-analysis simulations was often unrealistically large compared to real-life meta-analyses. 
Therefore, having parameters in $\thetadesign$ can be reasonable, and while directly inferring their values from real datasets may not be necessary, they can still be based on aggregated information derived from the real datasets (see \hyperref[sec:3.1.2_rdb_param_indirect]{Section~\ref*{sec:3.1.2_rdb_param_indirect}}).

\paragraph{Parameters in $\thetaestim$}
As discussed in \hyperref[sec:2.3_notation]{Section~\ref*{sec:2.3_notation}}, the true values of $\thetaestim$  in $\dgmdtrueind$ are not known and can only be estimated from $\dswithind$. This estimation process introduces several challenges, which we now address in detail. We begin by discussing the general challenges of estimating parameters in $\thetaestim$, before highlighting the additional considerations specific to $\thetaestimtarget$. \\
A straightforward approach to estimate the parameters in $\thetaestim$ is to apply a maximum likelihood (ML) estimation method, using the given model structure $\struc$ as a basis. This approach relies on the assumption that $\struc$ and $\strucdtrueind$ align, at least for the parts relevant for estimating $\thetaestim$---an assumption that may not hold in practice, particularly if $\struc$ is overly simplistic. While perfect alignment between $\struc$ and $\strucdtrueind$ is likely rare, a close alignment may suffice in many cases. However, substantial mismatches can make the dataset $\dswithind$ unsuitable for parameter estimation, resulting in estimates that differ considerably from their true values or become effectively meaningless. To address this issue, estimation methods that account for and correct potential deviations between $\struc$ and $\strucdtrueind$ can be employed. 
For instance, in {\exmeta}, the model structure $\struc$ assumes a normal distribution for the study effect sizes ($\metaeffect_\studyind \sim \mathcal{N}(\metaeffect, \tau^2)$). However, in $\strucdtrueind$, the true distribution of effect sizes may deviate from normality, which can result in biased estimates of the between-study heterogeneity variance $\tau^2$. This issue can be mitigated by employing estimators that do not rely on the normality assumption (e.g., the Sidik-Jonkman estimator; \citealp{sidik2005simple}). In other cases, the mismatch between $\struc$ and $\strucdtrueind$ may be so pronounced that it cannot be reasonably addressed, leading to essentially meaningless parameter estimates. For example, in {\exsurv}, if the distribution of the (theoretically observable) survival times in $\strucdtrueind$ deviates substantially from an exponential distribution, the resulting parameters would fail to capture the characteristics of $\dgmdtrueind$ and no longer serve a meaningful purpose in the simulation study. 
Similarly, looking beyond the four examples from \hyperref[sec:2.4_examples]{Section~\ref*{sec:2.4_examples}}, estimating regression coefficients for $\varnum$ covariates ($\beta_1, \dots, \beta_{\varnum}$) in a linear regression model specified by $\struc$ becomes problematic if the relationship between the outcome and covariates is, for example, strongly non-linear. In such cases, it may be necessary to revise the model structure $\struc$ or select real datasets $\dsall$ based on criteria that ensure a better alignment with $\struc$.
While we treat both $\struc$ and $\dsall$ as fixed here, these considerations are addressed in Sections~\ref{sec:3.2_rdb_ms} and
\ref{sec:3.3_selecting_datasets}.\\
Even when $\struc$ and $\strucdtrueind$ align closely, additional challenges arise due to the finite nature of $\dswithind$, which represents a sample generated by $\dgmdtrueind$ and is thus subject to sampling variability. Consider, for example, {\exordinal}, where the true probabilities $\pi_{\ordind,\groupind}$ of the ordinal outcome variable in treatment group $\groupind$ can be reasonably estimated from $\dswithind$ via ML estimation (i.e.\ by calculating the proportion of individuals within treatment group $\groupind$ who fall into ordinal category $\ordind$). Note that even in this simple example, $\struc$ and $\strucdtrueind$ are not perfectly aligned: While $\struc$ assumes the outcome depends solely on treatment, other (observable and latent) factors also influence it in $\strucdtrueind$. Due to randomization, however, the outcome's distribution within treatment groups remains unaffected, allowing the probabilities to be estimated without bias.
Although the ML estimator seems appropriate in this context, it can exhibit substantial variance, leading to discrepancies between the estimated and true probabilities. This can cause practical challenges, such as categories with low (but non-zero) true probabilities having zero counts in $\dswithind$ due to sampling variability, resulting in estimated probabilities of zero. Consequently, simulated datasets derived from such estimates would lack certain outcome categories entirely. One potential solution is to impose a minimum sample size criterion when selecting the real datasets, though this pertains to dataset selection (see \hyperref[sec:3.3_selecting_datasets]{Section~\ref*{sec:3.3_selecting_datasets}}). 

Another issue for {\exordinal} arises because its target is a null hypothesis. Specifically, the estimated probabilities almost always exhibit small differences across treatment groups, even when the true probabilities in $\dgmdtrueind$ are equal (i.e.\ under $H_0$). As a result, in the DGM constructed from the estimated probabilities, $H_0$ will almost always be false, whether $H_0$ actually holds in $\dgmdtrueind$ or not. 
While this aligns with the model structure $\struc$ of {\exordinal}, which specifies unequal probabilities across treatment groups for at least one ordinal category, it raises concerns about how well the constructed DGM reflects reality if $H_0$ might plausibly hold in $\dgmdtrueind$. A pragmatic ad hoc approach to address this issue would be to incorporate the variance in parameter estimation through a statistical test (e.g., a Wilcoxon rank-sum test). The resulting $p$-values could then be used to decide whether $H_0$ should be assumed true or false for a given $\dswithind$. This could, for example, be achieved by applying a threshold (e.g., $p < 0.05$) or using $p$-values as sampling weights for determining $H_0$ status (a similar idea is applied by \citealp{benidt2015simseq}, who used $p$-values derived from the real dataset to sample genes for which the null hypothesis of no differential expression should hold in the resulting DGM). Only probabilities derived from datasets meeting these criteria would then be used. However, as with the sample size criterion discussed earlier, such procedures effectively act as additional exclusion criteria for dataset selection (see \hyperref[sec:3.3.1_rdb_select_database]{Section~\ref*{sec:3.3.1_rdb_select_database}}), which $\dswithind$ would already need to satisfy. 

So far, we have discussed the estimation of $\thetaestim$ in general without differentiating between $\thetaestimtarget$ and $\thetaestimother$. While the estimation tasks for both parameter types face similar challenges, the estimation of $\thetaestimtarget$ is particularly critical. 
This is because the estimates of $\thetaestimtarget$ determine the parameter values involved in the target that the methods being examined in the simulation study must recover, making the selection of an appropriate estimation method inherently more impactful. Moreover, when the statistical task of interest is estimation, the method used to estimate $\thetaestimtarget$ could itself be among the competing methods evaluated in the simulation study. This introduces an element of circularity and potential bias, as the chosen method might gain an unfair advantage. To address this, all competing methods could be applied to estimate $\thetaestimtarget$, with their results aggregated to provide a more balanced basis for the simulation study.

While the specific challenges associated with estimating $\thetaestim$ may differ, the provided examples illustrate the inherent complexity of the process. In principle, the challenge of inferring information from a real dataset that is not directly observable mirrors the issues encountered when analyzing data in real-world applications. For parameters related to the target, these challenges, as noted above, are closely tied to the very issues the simulation study aims to investigate or improve upon.

\subsubsection{Deviating from the one-to-one inference approach: Aggregated inference and factorial designs}
\label{sec:3.1.2_rdb_param_indirect}
While the one-to-one approach, i.e.\ the direct inference of values from $\dsall$ for every parameter and the subsequent mapping to $\setofconsideredparamvectors$ by combining the inferred values per dataset, is generally expected to produce DGMs that closely represent the true DGMs of the selected real datasets, there are alternative strategies worth considering for either step of the inference process from $\dsall$ to $\setofconsideredparamvectors$.%\\

\paragraph{Aggregated inference}
For the first step of inferring the values for each parameter in $\thetaall$, there are several reasons not to use direct inference for every parameter. 
As discussed in the previous section, for $\thetaestim$, the uncertainties associated with estimation may make it impractical to rely strictly on specific values inferred from the datasets. For $\thetadesign$, practical considerations---such as the preference for numerically tidy values---may influence the decision against direct inference.
In addition to these previously discussed issues, the number of datasets, even if adjustable during their selection (see \hyperref[sec:3.3.1_rdb_select_database]{Section~\ref*{sec:3.3.1_rdb_select_database}}), may prove insufficient or excessive, as the researcher might prefer to consider more or fewer than $\dsnum$ parameter values for a given parameter. In all these cases, a reasonable alternative may be to use the information from the real datasets in aggregated form to generate the values for a parameter. We refer to this alternative as aggregated inference. A simple implementation involves identifying the minimum and maximum values for each parameter across the datasets to approximate a reasonable range and then selecting a number of values, $\aggregatednum$, that are systematically distributed within this range (e.g., equidistant). Alternatively, these boundaries could define a uniform distribution, from which $\aggregatednum$ values are sampled. Moving beyond uniform distributions (and considering values beyond just the minimum and maximum), parameter values could also be used to fit and sample from other distributions (e.g., a normal distribution). 
Generally speaking, if $\aggregatednum$ values are generated for a given parameter $\rdbparam_\paramind$ through aggregated inference instead of direct inference, the result is not a set $\boldsymbol{\realizedrdbparam}_\paramind$ containing $\dsnum$ values (see \hyperref[eq:parameter_q_set]{Equation~\ref*{eq:parameter_q_set}}) but a set $\boldsymbol{\aggregatedrdbparam}_\paramind$ containing $\aggregatednum$ values:
\begin{equation}
\label{eq:aggregated_parameter_q_set}
\boldsymbol{\aggregatedrdbparam}_\paramind = \{\aggregatedrdbparam_\paramind^{(1)}, \ldots, \aggregatedrdbparam_\paramind^{(\aggregatednum)}\}.
\end{equation}
The described procedures are typically applied individually for each parameter, although in principle, one could also model a joint distribution and then draw the values for multiple parameters simultaneously. 
While inferring parameters via aggregated inference is unlikely to yield entirely unrealistic values, it may, depending on the chosen procedure for generating parameter values, still produce a (slightly) distorted representation of the true distribution of parameter values in the real-world DGMs targeted by the simulation study. 
Note that unless joint modeling is used, there is no natural way to map the sets of aggregated values for the individual parameters into realistic considered parameter vectors, as there is no inherent correspondence between the individual values. Accordingly, combining the values per dataset (as in the second step of the one-to-one approach) is not possible, and any other method of combination---including other scattershot designs and factorial designs (see below)---carries the risk of producing unrealistic parameter combinations.

\paragraph{Factorial designs}
For the second step of mapping the sets of inferred values for the individual parameters to the set of considered parameter vectors, we also want to discuss an alternative to the scattershot design that is implied in the one-to-one approach.
The combination of parameter values per dataset, therefore only considering $\dsnum$ specific (and potentially unique) combinations of values for all parameters, can complicate the analysis of the effects of individual parameters and their interactions on the performance of the methods being evaluated. An alternative is to construct a factorial design using the inferred parameter values. The implementation of factorial designs, which are employed in most simulation studies, involves combining parameter values independently---considering either all possible combinations (fully factorial) or a subset thereof (partially factorial) \citep{morris2019using, siepe2024simulation}. When a fully factorial design is used, the resulting set of considered parameter vectors is given by
\begin{equation}
\label{eq:factorial}
    \setofconsideredparamvectors = \{\consideredparamvector^{(1)}, \ldots, \consideredparamvector^{(\consideredparamvectornum)}\} = \setofvaluesforparam_1 \times \cdots \times \setofvaluesforparam_\paramnum =\mathrel{\mathop:} \setofconsideredparamvectors^{\text{factorial}},
\end{equation}where $\setofvaluesforparam_\paramind$ represents the set of inferred values for a given parameter $\rdbparam_\paramind$ and is the result of either direct inference ($\setofvaluesforparam_\paramind \mathrel{\mathop:}=\boldsymbol{\realizedrdbparam}_\paramind$, see \hyperref[eq:parameter_q_set]{Equation~\ref*{eq:parameter_q_set}}) or aggregated inference ($\setofvaluesforparam_\paramind \mathrel{\mathop:}=\boldsymbol{\aggregatedrdbparam}_\paramind$, see \hyperref[eq:aggregated_parameter_q_set]{Equation~\ref*{eq:aggregated_parameter_q_set}}).
If every parameter is inferred directly from the $\dsnum$ datasets (i.e.\ $\setofvaluesforparam_\paramind \mathrel{\mathop:}=\boldsymbol{\realizedrdbparam}_\paramind$ for all $\paramind \in \{1, \dots, \paramnum\}$), the set $\setofconsideredparamvectors^{\text{factorial}}$ (see \hyperref[eq:factorial]{Equation~\ref*{eq:factorial}}) includes the set of considered parameter vectors that would result from the second step of the one-to-one approach (see \hyperref[eq:considered_parameter_vectors_one-to-one]{Equation~\ref*{eq:considered_parameter_vectors_one-to-one}}; i.e.\ $\setofconsideredparamvectors^{\text{factorial}} \supsetneq \setofconsideredparamvectors^{\text{one-to-one}}$). 
If, additionally, each dataset provides unique values for every parameter, the fully factorial design results in a total of $\consideredparamvectornum = \lvert \setofconsideredparamvectors^{\text{factorial}} \rvert = \dsnum^\paramnum$ considered parameter vectors in $\setofconsideredparamvectors^{\text{factorial}}$. 
Importantly, when using direct inference, employing a factorial design instead of a scattershot design introduces the risk of producing unrealistic considered parameter vectors: while the individual parameter values inferred from real datasets may each be plausible, their combinations might not be, potentially resulting in DGMs that fail to represent any real-world DGM relevant to the simulation study. If aggregated inference is used instead, the same concern applies in principle---except that, as noted above, mapping sets of aggregated parameter values carries this risk regardless of the way in which the values are combined.

\subsection{Inferring parts of the model structure from a set of real datasets} \label{sec:3.2_rdb_ms}
In the previous section, we considered the scenario where only parameters are set to be real-data-based. However, researchers may also wish to infer specific parts of the model structure from the selected real datasets. For parts of the model structure whose true form in the underlying DGM of a real dataset can only be inferred with uncertainty (e.g., a distribution or functional relationship), similar issues as those discussed in \hyperref[sec:3.1.1_rdb_param_direct]{ Section~\ref*{sec:3.1.1_rdb_param_direct}} for $\thetaestim$ arise and must be taken into account---such as the impact of sampling variability---when choosing an inference method. As with parameter inference, the choice of method can substantially influence the results of the simulation study, particularly for parts of the model structure that explicitly appear in the formulation of the simulation target. In our four examples, the only parts of the model structure falling into this category are whether the null hypothesis (or hypotheses) holds in {\exordinal}, {\exsurv}, and {\exde}.\\
Unlike parameter inference, which involves estimating numerical values, inferring parts of the model structure typically requires categorical decisions, such as determining whether a null hypothesis holds or selecting an appropriate distribution for a variable. 
Often, this can be done using hypothesis tests, as already outlined for {\exordinal} in \hyperref[sec:3.1.1_rdb_param_direct]{ Section~\ref*{sec:3.1.1_rdb_param_direct}}, where we considered how the status of the null hypothesis of no treatment effect could be assessed via a test (albeit as a criterion for selecting the real datasets rather than for inference from them). Other examples include testing for the presence of an interaction effect or correlation. In some cases---such as selecting the distribution of a variable---inferring a specific part of the model structure via hypothesis testing additionally requires defining a set of plausible options, from which the best-fitting choice is then selected using an appropriate test. Importantly, this approach does not guarantee that any of the considered options closely approximate the true model structure of any of the selected real datasets. To illustrate the data-driven selection of distributions, consider {\exsurv}, for which we discussed in \hyperref[sec:3.1.1_rdb_param_direct]{ Section~\ref*{sec:3.1.1_rdb_param_direct}} that the true distribution of survival times may deviate from the exponential distribution assumed by $\struc$. This can be addressed by considering more flexible distribution options, such as Weibull, gamma, Gompertz, or mixture distributions, with the best-fitting distribution identified through goodness-of-fit tests like the Cramér--von Mises test. This procedure is implemented by \citet{thurow2024how} in a meta-scientific study on simulating realistic survival data. \\
If parts of the model structure are real-data-based and thus may vary across datasets, multiple model structures can emerge.
Accordingly, when inferring parts of the model structure from the set of real datasets, this procedure must be applied before inferring the parameters. Specifically, the procedure described in \hyperref[sec:3.1_inferring_params]{Section~\ref{sec:3.1_inferring_params}} must be applied separately for each subset of datasets corresponding to each resulting model structure derived from the inference of real-data-based parts. 
For example, in an extended version of {\exsurv}, suppose half of the real datasets suggest an exponential distribution, while the other half align more closely with a Weibull distribution. In this case, parameter inference must be conducted separately for the datasets associated with $\struc_{\text{Exp}}$ and $\struc_{\text{Weibull}}$.

%%%%%%%%%%%%%%%%%%%%%%%%%%%%%%%%%%%%%%%%%%%%%%%%%%%%
%%%%%%%%%%%%%%%%%%%% Section 4 %%%%%%%%%%%%%%%%%%%%%
%%%%%%%%%%%%%%%%%%%%%%%%%%%%%%%%%%%%%%%%%%%%%%%%%%%%
\section{Selecting a set of real datasets 
for real-data-based parametric DGMs}
\label{sec:3.3_selecting_datasets}
We propose three general requirements for real datasets used to construct real-data-based DGMs: {\dsrepro} they must be accessible to others, with a transparent and reproducible selection process; {\dsaim} their true DGMs must correspond to a representative subset of the simulation study’s domain of interest; and {\dsextract} they must provide the necessary information to both meaningfully infer the DGM components intended to be real-data-based and assess their eligibility.
In the following, we examine these requirements in more detail, including strategies to fulfill them and the challenges that may arise. Specifically, we discuss the identification of a database likely to contain eligible datasets as well as the specification of additional eligibility criteria to ensure that the final selection meets all requirements.

 \subsection{Database}\label{sec:3.3.1_rdb_select_database}
Regarding the choice of database, requirement {\dsrepro} excludes collections of datasets that are only accessible to the researchers conducting the simulation study. Public data repositories, therefore, represent a natural solution to fulfill {\dsrepro}. For example, in the context of {\exde}, the open data repository of The Cancer Genome Atlas (TCGA) program (\url{https://www.cancer.gov/tcga}) could be utilized, offering genomic, epigenomic, transcriptomic, and proteomic data for 33 cancer types. Similarly, publicly accessible platforms like OpenML \citep{vanschoren2014openml, bischl2025OpenML} or the UCI Machine Learning Repository \citepalias{UCIrepo}, which offer datasets across various research domains and data types, can serve as valuable resources. Although primarily used for studies focused on prediction tasks, where methods are evaluated directly on the real datasets, these repositories can also be used for real-data-based simulation studies with other aims (e.g., \citealp{stolte2024simulation}, albeit in a meta-scientific context). 
In addition to data repositories that are fully open to the public, we also consider data repositories that are broadly accessible to the research community under controlled conditions to fulfill {\dsrepro}.  Examples include clinical research data-sharing platforms such as Vivli (\url{https://vivli.org}; \citealp{Bierer2016Vivli}), the Yale Open Data Access (YODA) Project (\url{https://yoda.yale.edu/}; \citealp{ross2018YODAProject}), and the Virtual International Stroke Trials Archive (VISTA; \url{https://www.virtualtrialsarchives.org/vista/}; \citealp{ali2007VISTA}).\\
Public data repositories may offer many datasets eligible under requirement {\dsextract} (i.e.\ providing sufficient information for inference and eligibility assessment), as researchers have direct access to the dataset. 
However, finding a repository with enough datasets to fulfill requirement {\dsaim} (i.e.\ adequately representing the domain of interest) can be more challenging. An alternative approach is to reconstruct datasets from tables and figures in research publications---such as journal articles and reports---that present aggregated or visualized data.
For example, survival data can be reconstructed from digitized survival curves \citep[e.g.,][]{guyot2012enhanced}, as employed by \citet{trinquart2016comparison}, \citet{royston2019combined}, \citet{dormuth2022which}, and, in the simulation context, \citet{thurow2024how}. In these cases, possible databases may be collections of publications, such as PubMed (\url{https://pubmed.ncbi.nlm.nih.gov/}), as referenced by \citet{dormuth2022which} and \citet{thurow2024how}, or specific journals \citep{royston2019combined,trinquart2016comparison}. \\
Another alternative specifically for parametric DGMs is to rely on aggregated data instead of raw datasets. This approach can still satisfy requirement {\dsextract}, provided that all necessary information can be extracted without direct access to the full datasets. Although the underlying datasets are not directly accessible (and also not reconstructed), we continue to use the notation $\ds$ to refer to the respective dataset used in the publication, which is implicitly represented through the extracted information.
Relying on aggregated data can, for example, be feasible in {\exordinal}, where the information needed to estimate outcome probabilities---the number of individuals in each ordinal category per group---is often available in tables or figures included in publications. In other cases, parameter inference may have already been conducted within the publication itself, meaning that the reported values can be directly adopted. 
For instance, in the context of {\exmeta}, both meta-analysis-level parameters (e.g., the [estimated] overall effect) and study-level parameters (e.g., group means) are commonly reported in published meta-analyses, thus providing the necessary information for both the meta-analysis dataset and the individual study datasets without requiring access to raw data. The approach of using aggregated information from published meta-analyses was implemented in the simulation study by \citet{langan2019comparison}.
When using aggregated data, suitable databases again include PubMed and specific journals but also systematic reviews summarizing multiple publications (the latter serving as a database in contexts other than {\exmeta}). 
Given the vast number of publications available in collections like PubMed or individual journals, it is often reasonable to define the database as all publications within the collection, restricted to a specific time frame. \\
Depending on the chosen database, several potential limitations may arise that researchers should be aware of, address transparently, and, if possible, mitigate through additional eligibility criteria (see \hyperref[sec:3.3.2_rdb_select_criteria]{Section~\ref*{sec:3.3.2_rdb_select_criteria}}). For public repositories containing existing datasets, similar concerns to those typically discussed in the context of real-data studies apply (see, for example, the discussions by \citealp{strobl2024against}, \citealp{boulesteix2015framework}, and \citealp{friedrich2024ontherole}). A key concern is that the true DGMs of real datasets donated to repositories may fail to adequately represent the domain of interest to which the simulation study’s results and recommendations are intended to apply. For instance, the real datasets in public repositories might focus heavily on specific subpopulations of DGMs (e.g., a particular cancer type). Furthermore, the quality of the data may be low depending on how it was collected and curated, with issues such as poor documentation or missing values. \\
For databases consisting of collections of publications, similar issues may arise. Specifically, publication bias can result in datasets with true DGMs that are not representative of the domain of interest (e.g., if only results based on datasets with large true effect sizes are published), and the quality and methodological rigor of how the underlying datasets were collected and reported may also vary widely. However, when using publication collections as databases, additional challenges arise, because the original datasets are rarely directly accessible. For reconstruction procedures, the quality and accuracy of reconstructed datasets depend on several factors, including the validity of assumptions underlying the reconstruction algorithm. 
For instance, the previously mentioned algorithm by \citet{guyot2012enhanced} for reconstructing survival data relies on the assumption that censoring occurs at a constant rate within each time interval, which may not hold in all cases. 
When aggregated dataset information is used, inconsistencies may emerge if different publications apply varying aggregation methods. Moreover, publication bias may not only introduce issues with representativeness but also lead to optimistically biased parameter estimates by incentivizing practices like $p$-hacking or selective reporting. In addition, relying on publication collections as databases may demand more time and effort than using open repositories with readily available datasets. This is because reconstructing or extracting relevant information from publications involves additional steps, and assessing publications for eligibility criteria often requires considerably more time.\\
Finally, a practical issue that arises independently of the chosen database is that the database itself only determines the maximum possible number of datasets that could be selected but not how many will actually meet all eligibility criteria. 
Accordingly, after applying these criteria, researchers may end up with too few datasets, which would risk inadequate representation of the domain of interest, or too many datasets, which would make it impractical to reasonably process them in subsequent simulation steps. As a pragmatic approach, we suggest that researchers specify a minimum and maximum number of datasets to be selected and, after applying the eligibility criteria, check whether the number of selected datasets falls within this range. If the number is too low, the database should be expanded (e.g., by including additional publication years or related repositories). If the number is too high, we recommend using random selection to reduce the number of datasets to the specified maximum.
While we cannot provide a general recommendation for an appropriate minimum and maximum, as these values depend on the specific simulation study, we argue that a minimum of only one or two datasets---which is common in practice---is typically insufficient.

\subsection{Eligibility criteria}\label{sec:3.3.2_rdb_select_criteria}
Once the database has been specified, datasets can be selected based on eligibility criteria, which may be formulated as inclusion or exclusion criteria. In general, assessing their fulfillment may involve a combination of automated methods (e.g., filtering datasets in a repository or applying search strings for journal databases) and manual review (e.g., screening data documentation or publications).\\
Ideally, all eligibility criteria would be specified before dataset selection to prevent bias from post hoc modifications. However, this is often not feasible, as criteria may need to be refined or extended during the assessment process to account for unforeseen challenges or inconsistencies. In any case, to meet requirement {\dsrepro} (i.e.\ ensuring the selection process is transparent and reproducible), it is essential to clearly and comprehensively report the final criteria applied during the selection process. \\
The criteria themselves can be categorized into those addressing {\dsaim} (i.e.\ ensuring adequate representation of the domain of interest) and those related to {\dsextract} (i.e.\ providing sufficient information for inference and eligibility assessment), and researchers should explicitly indicate which requirement each criterion is intended to fulfill.\\
Defining eligibility criteria to fulfill {\dsaim} involves translating the domain of interest---i.e.\ the population of true DGMs to which the simulation study’s results and recommendations are intended to apply---into concrete criteria. As stated in \hyperref[sec:2.2.1_spec_diff]{Section~\ref*{sec:2.2.1_spec_diff}}, the specifications that define the domain of interest consist of researcher-specified components of interest (e.g., the number of treatment groups, the violation of a specific assumption) and constraints imposed on real-data-based components (e.g., restricting parameter values to those observed in a specific type of disease). In contrast, researcher-specified components of convenience are not incorporated into the eligibility criteria, meaning that datasets differing in these aspects may still be eligible. 
In general, defining {\dsaim}-related criteria can be a challenging task, in part because researchers typically do not distinguish explicitly between components of interest and components of convenience when planning simulations. Note that for databases consisting of collections of publications, an effective strategy for defining a search string to identify an initial set of relevant publications is to include the names of methods commonly used to analyze datasets from this domain as keywords (e.g., the methods compared in the simulation study). \\
Beyond the challenge of specifying criteria related to {\dsaim}, assessing whether these criteria are met presents additional difficulties. 
The domain of interest is defined by the \textit{true} DGMs, meaning that, in an ideal scenario, datasets would be selected based directly on these underlying true DGMs. However, in practice, only the observed datasets (or their aggregated versions) are available for assessment. Accordingly, if a criterion refers to parts of the model structure or parameters whose true form or value in the true DGM underlying a real dataset cannot be known with certainty (i.e.\ unknown components, as defined in \hyperref[sec:2.2.2_know_diff]{Section~\ref*{sec:2.2.2_know_diff}}), its fulfillment cannot be determined with certainty either. This applies, for instance, when only datasets in which a specific assumption does not hold or those with a large true effect size should be selected. In principle, this introduces the same difficulties discussed in \hyperref[sec:3_implementation]{Section~\ref*{sec:3_implementation}}, with the key difference that, in those cases, inference was performed on an already selected set of datasets, $\dsall$, whereas here, inference is required as part of the process of identifying $\dsall$ in the first place. As a consequence, when a criterion involves a component whose true form or value in the true DGM underlying a dataset is unknown, the criterion should also specify how fulfillment should be assessed  (e.g., by defining a statistical test or other procedure).\\
As stated above, in contrast to {\dsaim}, the criteria addressing {\dsextract} do not relate to the study's domain of interest but rather ensure that the selected datasets contain the necessary information for meaningful parameter inference and the assessment of eligibility. Such criteria may, for instance, address the sample size of the dataset. 
As already mentioned in \hyperref[sec:3.1.1_rdb_param_direct]{Section~\ref*{sec:3.1.1_rdb_param_direct}}, this could include requiring a minimum total sample size $n$ to ensure stable parameter estimation or, in the case of {\exordinal}, ensuring that there are more than zero observations per group and ordinal category. 
Note that, in this case, $n$ and other parameters related to sample size cannot be real-data-based, and since their specific values are typically not of direct interest, they would be considered researcher-specified components of convenience (and represent examples of this component category that neither have negligible impact nor are expected to hold by default but also cannot be reasonably inferred; see \hyperref[sec:2.2.1_spec_diff]{Section~\ref*{sec:2.2.1_spec_diff}}).
Other criteria may address the quality of the selected datasets, which, as noted in the previous section, can be a concern. Here, criteria could specify that certain metadata must be available, or they could, for example, restrict inclusion to publications from reputable journals. \\
While the criteria mentioned above primarily relate to the \enquote{meaningfully} aspect of {\dsextract}, other criteria---particularly when databases consist of collections of publications---ensure that parameter inference and the assessment of eligibility are possible in the first place. 
For example, when reconstructing datasets for {\exsurv} with the algorithm by \citet{guyot2012enhanced}, relevant criteria may require that the number at risk is reported and that survival curves are presented in high resolution \citep{dormuth2022which,thurow2024how}. Similarly, when using aggregated information for {\exordinal}, a relevant criterion is that the number of patients in both treatment groups and ordinal categories is clearly reported in tables or figures.\\
Importantly, the criteria specified so far determine whether a dataset as a whole should be selected or not. We refer to these as dataset-level criteria. 
However, some of these criteria also inherently specify which subsets of the data within an included dataset are used. For example, requiring datasets to include specific types of outcome variables already restricts the usable subset of each dataset. Beyond such implicit subset specification, additional explicit subset-level criteria may be applied after dataset selection is complete. These criteria refine the selection of specific elements (e.g., outcome variables, treatment groups, or covariates) within an included dataset to ensure consistency across the selected datasets. Additionally, considering subset-level criteria allows for broader dataset inclusion by enabling the use of relevant subsets within datasets that would otherwise be excluded. 
For instance, if a simulation study focuses on two treatment groups, rather than excluding all datasets that do not match this criterion exactly, one could exclude only those with a single treatment group while keeping those with more than two groups, subsequently selecting the two treatment groups with the largest sample sizes. This subset-level criterion would relate both to {\dsaim} (ensuring the selection of the two treatment groups) and to {\dsextract} (ensuring an adequate sample size).

%%%%%%%%%%%%%%%%%%%%%%%%%%%%%%%%%%%%%%%%%%%%%%%%%%%%
%%%%%%%%%%%%%%%%%%%% Section 5 %%%%%%%%%%%%%%%%%%%%%
%%%%%%%%%%%%%%%%%%%%%%%%%%%%%%%%%%%%%%%%%%%%%%%%%%%%
\section{Example illustrations}
\label{sec:4_results}
To empirically illustrate the implementation of parametric DGMs based on a systematically selected set of real datasets, we conduct two simulation studies that build on two of the examples discussed in the previous sections ({\exordinal} and {\exde}). We also compare the systematic parameter inference from multiple datasets with other approaches by additionally considering purely researcher-specified parameters in the first illustration (\hyperref[sec:ill_ordinal]{Section~\ref*{sec:ill_ordinal}}) and parameters inferred from a single real dataset in the second (\hyperref[sec:ill_de]{Section~\ref*{sec:ill_de}}). Importantly, the simulations are not intended as comprehensive simulation studies with in-depth analyses---each of which could warrant a dedicated paper---but rather as illustrative examples demonstrating the implementation of parametric DGMs based on a systematically selected set of datasets and their impact on results. The simulations and analyses are conducted in the software environment \texttt{R} \citep{RCore2023}, and the code to reproduce all results is available at \url{https://github.com/NiesslC/realdata_simulations}.

\subsection{Two-arm randomized controlled trial with an ordinal outcome}\label{sec:ill_ordinal}
\subsubsection{Design}
\paragraph{ADEMP structure} 
An overview of the ADEMP structure for this simulation can be found in \hyperref[tab:ademp_ordinal]{Table~\ref*{tab:ademp_ordinal}}. The \enquote{A}, \enquote{D}, and \enquote{E} aspects correspond to those listed for {\exordinal} in \hyperref[sec:2.4_examples]{Section~\ref*{sec:2.4_examples}}, at least for the aspects already specified there. Accordingly, we consider a two-arm (i.e.\ $\groupnum = 2$) randomized controlled trial, and we aim to evaluate the ability of methods to detect a true treatment effect between groups.

%%%%% Begin Table: tab_ADEMP_ordinal %%%%%-----------------------
\begin{table}[h]
    \caption{ADEMP structure for the example illustration on hypothesis testing in the context of a two-arm randomized controlled trial with an ordinal outcome. 
    Either all parameters are researcher-specified, or all parameters except the outcome probabilities $(\bm{\pi}_\groupone, \bm{\pi}_\grouptwo)$ are researcher-specified, with the latter being real-data-based.
    Accordingly, in the first case, 
    $\lambdaall = \{\groupnum, \ordnum, \obsnum, \bm{\pi}_\groupone, \bm{\pi}_\grouptwo\}$, while in the second case, $\thetaall = \{\bm{\pi}_\groupone, \bm{\pi}_\grouptwo\}$  and $\lambdaall = \{\groupnum, \ordnum, \obsnum\}$. 
    }
    \centering
    \small
   {\renewcommand{\arraystretch}{1.3}
   \begin{tabularx}{1\textwidth}{|W{2cm}|X|}\hline
%%%%%%%%%%%%%%%%%%%%%%%%%%%%%%%%%%%%%%%%%%%%%%%%%%%%%%%%%%%%%%%%%%%%%%%%%%%%%
       \raggedright Aim &   Evaluation of methods testing the null hypothesis $H_0$ of no treatment differences in two-arm (i.e.\ $\groupnum = 2$) randomized controlled trials with ordinal outcomes having $\ordnum$ categories, in settings where $H_0$ is false \\ \hline
%%%%%%%%%%%%%%%%%%%%%%%%%%%%%%%%%%%%%%%%%%%%%%%%%%%%%%%%%%%%%%%%%%%%%%%%%%%%%
       \raggedright Data-generating mechanisms (DGMs) &\textbf{Model structure $\struc$}  \vspace{0.2cm}
       \begin{tabitemize}
       \item Ordinal outcome of individual $\obsind$, $\obsind = 1,\dots,\obsnum$, in group $\groupind \in \{\groupone, \grouptwo\}$ (equal group sizes) is drawn from $\operatorname{Multinomial}(1, \bm{\pi}_{\groupind})$, where $\bm{\pi}_{\groupind} = (\pi_{1,{\groupind}},\dots,\pi_{\ordnum,\groupind})$,  $\pi_{\ordind,\groupind} = P(Y = \ordind \mid X = \groupind)$, $\sum_{\ordind=1}^\ordnum \pi_{\ordind,\groupind} = 1$, and $\pi_{\ordind,\groupone} \neq \pi_{\ordind,\grouptwo}$ for at least one $\ordind  \in \{1, \dots, \ordnum\}$.
       \end{tabitemize} 
       \vspace{0.4cm}
       \textbf{Parameters $\lambdaall$ and $\thetaall$ (varied fully factorially)} \vspace{0.2cm}
       \begin{tabitemize}
            \item $\lambdaall$: $\groupnum = 2$; $\ordnum = 7$; $\obsnum\in\{60,120,200,300,600\}$
            \item \vspace{-0.25\baselineskip}\begin{tabular}[t]{@{}l@{}}$\lambdaall$/$\thetaall$: $(\bm{\pi}_{\groupone},\bm{\pi}_{\grouptwo})$ (4 researcher-specified and 15 real-data-based pairs of outcome\\[-0.12cm]\quad\;\;\;\; probabilities, see \hyperref[tab:ordinal_parameters_user]{Table~\ref*{tab:ordinal_parameters_user}} and \hyperref[tab:ordinal_datasets_probs_rdb]{Table~\ref*{tab:ordinal_datasets_probs_rdb}})\end{tabular}\vspace{-0.25\baselineskip}
        \end{tabitemize}
        \vspace{0.4cm}
        \textBF{Number of repetitions per DGM}: $\nrep=10{,}000$  \\ \hline
% %%%%%%%%%%%%%%%%%%%%%%%%%%%%%%%%%%%%%%%%%%%%%%%%%%%%%%%%%%%%%%%%%%%%%%%%%%%%%
       \raggedright  Estimand / Target  &    The null hypothesis $H_0: \pi_{\ordind,\groupone} = \pi_{\ordind,\grouptwo}$ for all $\ordind \in \{1, \dots, \ordnum\}$\\ 
       \hline
% %%%%%%%%%%%%%%%%%%%%%%%%%%%%%%%%%%%%%%%%%%%%%%%%%%%%%%%%%%%%%%%%%%%%%%%%%%%%
       \raggedright Methods  &   4 methods:   Chi-square test,  Fisher's exact test, Wilcoxon rank-sum test, proportional odds ordinal logistic regression  \\ \hline 
% %%%%%%%%%%%%%%%%%%%%%%%%%%%%%%%%%%%%%%%%%%%%%%%%%%%%%%%%%%%%%%%%%%%%%%%%%%%%%
       \raggedright Performance measure  & Power, estimated as $\frac{1}{\nrep}\sum_{s=1}^{\nrep} \mathbbm{1}(p_s \leq 0.05)$, with $p_s$ being the $p$-value from repetition $s$, $s=1,\dots,\nrep$ \\ \hline 
    \end{tabularx}
    }
    \label{tab:ademp_ordinal}
\end{table}
%%%%% End Table: tab_ADEMP_ordinal %%%%%-------------------------
In {\exordinal}, the remaining parameters to specify are the number of ordinal categories, $\ordnum$, the sample size $\obsnum$, and the outcome probabilities $\bm{\pi}_\groupone$ and $\bm{\pi}_\grouptwo$. To simplify the illustration, we focus on real-data-based inference for the outcome probabilities while fixing the number of ordinal categories to $\ordnum = 7$ and setting the sample size manually to five predefined values ($\obsnum \in \{60,120,200,300,600\}$).
For the outcome probabilities $(\bm{\pi}_\groupone, \bm{\pi}_\grouptwo)$, which are specified jointly for both groups and treated as a single factor, we consider 4 researcher-specified outcome probabilities (\hyperref[tab:ordinal_parameters_user]{Table~\ref*{tab:ordinal_parameters_user}}) and 15 real-data-based  outcome probabilities (\hyperref[tab:ordinal_datasets_probs_rdb]{Table~\ref*{tab:ordinal_datasets_probs_rdb}}).
The process of generating the real-data-based outcome probabilities, which were inferred after defining the researcher-specified outcome probabilities, will be detailed below. The different sample sizes and outcome probabilities are combined using a fully factorial design, resulting in 20 DGMs when $(\bm{\pi}_\groupone, \bm{\pi}_\grouptwo)$ is researcher-specified ($5 \times 4 = 20$) and 75 DGMs when it is real-data-based ($5 \times 15 = 75$).\\ 
Regarding the remaining ADEMP aspects---methods and performance measures---we evaluate four statistical tests: Chi-square test, Fisher's exact test, Wilcoxon rank-sum test, and proportional odds (PO) ordinal logistic regression, which are all tests that may be used in this context \citep{selman2024statistical}. For a given DGM, the performance of the methods is assessed by their power to reject the null hypothesis of no treatment difference, which is estimated using the proportion of rejected null hypotheses at a nominal significance level of $\alpha = 0.05$.\\
The number of repetitions (i.e.\ simulated datasets) per DGM, denoted as $\nrep$, is set to 10{,}000, ensuring that the Monte Carlo standard error (MCSE) remains below $0.5\%$ for a worst-case rejection proportion of 0.5 (the probability at which MCSE is maximized; \citealp{morris2019using}). 
However, we only run the methods on simulated datasets where all seven ordinal categories are observed, which reduces the number of analyzed datasets for some DGMs. To still ensure stable power estimates, we exclude all DGMs for which the number of repetitions where all seven ordinal categories are observed is lower than 8{,}000.

\paragraph{Dataset selection and parameter inference} 
As stated in \hyperref[sec:3.3.1_rdb_select_database]{Section~\ref*{sec:3.3.1_rdb_select_database}}, the information needed to estimate $\bm{\pi}_\groupone$ and $\bm{\pi}_\grouptwo$ is often reported in publications on corresponding trials, meaning access to the raw datasets is not necessarily required.
Accordingly, one may use a collection of publications as a database, which in this illustration is specified as all research publications in \textit{The New England Journal of Medicine} (\textit{NEJM}), with publication years restricted to 2017--2022. The dataset selection process from this database is summarized below, with full details provided in \hyperref[sec:S1.2_dataset_selection]{Supplementary Section~\ref*{sec:S1.2_dataset_selection}}.\\
To construct the search string for identifying relevant publications, we include the terms \enquote{randomized} and \enquote{ordinal}, along with variations of the names of the considered methods. This search yields 270 publications, which are manually screened according to 11 dataset-level eligibility criteria. Of these criteria, eight relate to {\dsaim}, including conditions such as requiring at least one ordinal outcome, excluding studies where the ordinal outcome has fewer or more than $\ordnum = 7$ categories, and excluding those in which participants were randomized in groups or clusters rather than individually. The remaining three criteria are associated with {\dsextract}, two of which ensure that the relevant information is clearly reported and that all ordinal categories contain at least one observation, while the third addresses cases where two publications use the same dataset. Applying these criteria results in $\dsnum=15$ eligible publications and corresponding datasets (see \hyperref[tab:ordinal_datasets_probs_rdb]{Table~\ref*{tab:ordinal_datasets_probs_rdb}}). On the subset level, we specify the following criteria for the 15 datasets: When a publication includes more than two treatment groups, we select the two with the largest sample sizes. Similarly, if multiple ordinal outcomes are available, we prioritize the outcome considered most important; if no clear priority is established, we select the outcome with the highest sample size.\\
Notably, we do not impose any criteria on minimum sample size or whether the null hypothesis of no treatment effect, $H_0$, is false (both of which were discussed as potential criteria throughout Sections~\hyperref[sec:3_implementation]{\ref*{sec:3_implementation}} and \hyperref[sec:3.3_selecting_datasets]{\ref*{sec:3.3_selecting_datasets}}). The latter implies that $H_0$ may, in fact, be true in some selected datasets. This decision is intentional to illustrate the impact of not applying such a criterion. However, in a formal simulation study, it may be reasonable to include such restrictions.\\
The outcome probabilities are estimated using simple maximum likelihood estimation, where $\hat{\pi}_{\ordind, \groupind}$ represents the proportion of individuals in treatment group $\groupind$ who fall into ordinal category $\ordind$.

\subsubsection{Results} 
\begin{figure}[ht]
    \centering
    \includegraphics[width=0.8\linewidth]{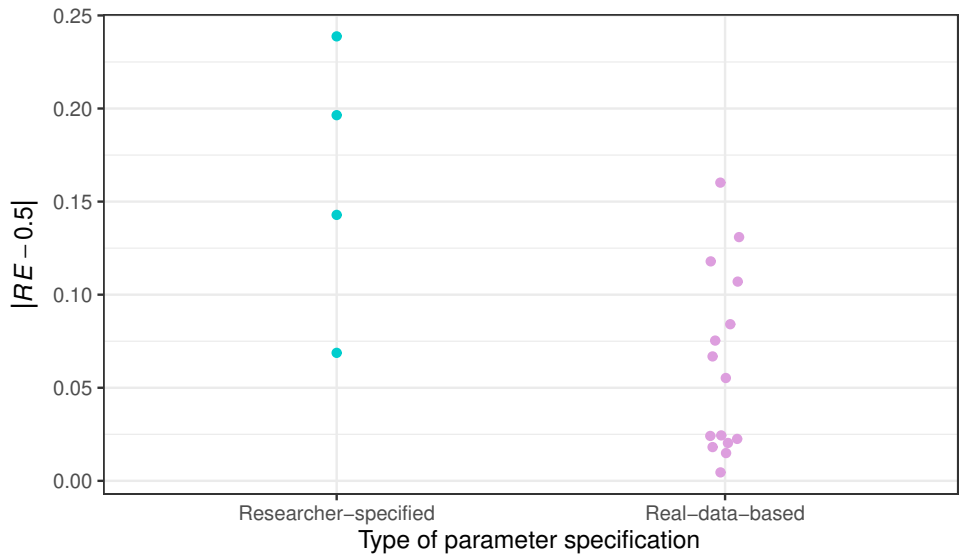} 
    \caption{Absolute deviation from 0.5 in the relative effect ($|\releff - 0.5|$) for 4 researcher-specified and 15 real-data-based outcome probabilities $(\bm{\pi}_\groupone,\bm{\pi}_\grouptwo)$. The more the relative effect deviates from 0.5, the greater the difference between the two treatment groups.}
    \label{fig:charac_ordinal}
\end{figure}

\paragraph{Parameter characteristics}
To systematically compare the researcher-specified and real-data-based outcome probabilities, directly examining their individual values is impractical due to their multi-dimensional nature (but see \hyperref[fig:ordinal_bsp]{Figure~\ref*{fig:ordinal_bsp}} for an example of a researcher-specified and a real-data-based set of outcome probabilities). Instead, one or several summary measures are needed to characterize them. For simplicity, we focus here on the relative effect, which has also been considered by \citet{funatogawa2023comparison} in their similar simulation study.
The relative effect, denoted as $\releff$, is defined as $P(Y_\groupone > Y_\grouptwo) + \frac{1}{2} P(Y_\groupone = Y_\grouptwo)$, where $Y_\groupone$ and $Y_\grouptwo$ denote the ordinal outcome variables in the two treatment groups. 
A relative effect of 0.5 indicates no systematic difference between the groups, whereas values greater or smaller than 0.5 suggest that observations in group $\groupone$ tend to be larger or smaller, respectively, relative to those in group $\grouptwo$ \citep{brunner2021win,agresti2010analysis}. \hyperref[fig:charac_ordinal]{Figure~\ref*{fig:charac_ordinal}} presents the relative effects for the researcher-specified and real-data-based outcome probabilities. Since both values below and above 0.5 indicate differences between groups, we consider the absolute deviation from 0.5, $|\releff - 0.5|$, where 0 indicates no difference, and larger values correspond to greater differences between the two groups. 

As shown in \hyperref[fig:charac_ordinal]{Figure~\ref{fig:charac_ordinal}}, the real-data-based outcome probabilities generally imply smaller group differences than the researcher-specified probabilities, illustrating that the two different approaches to parameter specification can lead to systematic differences in effect sizes. In the specific case considered here, as discussed in \hyperref[sec:3.1.1_rdb_param_direct]{Section~\ref*{sec:3.1.1_rdb_param_direct}}, this may be due to the fact that, for some of the true DGMs underlying the 15 real datasets, $H_0$ of no treatment effect holds. 
However, even if this applies to some datasets, a systematic difference remains between the two types of parameter specification.

\begin{figure}
    \centering
    \includegraphics[width=0.9\linewidth]{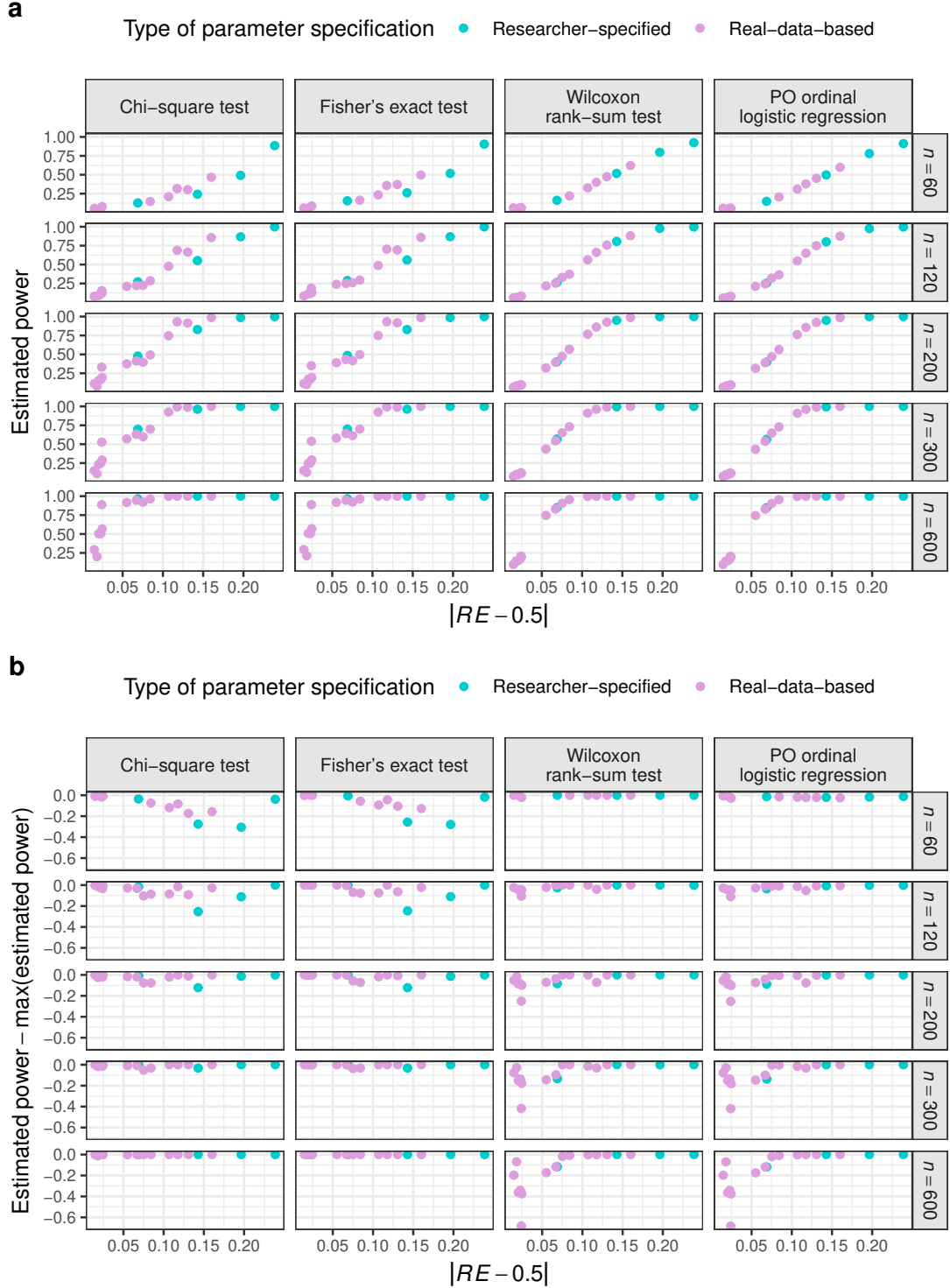}
    \caption{Absolute (panel a) and relative performance (panel b) of four statistical tests in relation to the absolute deviation of the relative effect from 0.5 ($|\releff - 0.5|$), calculated from the outcome probabilities $(\bm{\pi}_\groupone,\bm{\pi}_\grouptwo)$, across different sample sizes $\obsnum$. Panel a shows the absolute estimated power. Panel b shows the difference between the estimated power and the highest estimated power within each DGM. In both panels, the DGMs for all 4 researcher-specified and 14 of the 15 real-data-based outcome probabilities are included for $\obsnum \in \{ 200, 300, 600\}$, while for $\obsnum \in \{60,120\}$, some of these 14 real-data-based outcome probabilities are not included due to an insufficient number of repetitions where all seven ordinal categories are observed. The remaining one of the 15 real-data-based probabilities is not included at all for the same reason.}
    \label{fig:performance_ordinal}
\end{figure}

\paragraph{Method performance} 
\hyperref[fig:performance_ordinal]{Figure~\ref*{fig:performance_ordinal}} presents both the absolute performance (panel a: estimated power) and the relative performance (panel b: difference to the best, i.e.\ highest power) in relation to the relative effect deviation from 0.5 ($|\releff - 0.5|$) for different sample sizes $\obsnum$. 
Note that in both panels, not all 19 ($4+15$) outcome probabilities are included. As stated before, we exclude DGMs for which the number of repetitions where all seven ordinal categories are observed is lower than $8{,}000$. For one of the 15 real-data-based outcome probabilities, namely the probabilities extracted from \citet{perkins2018randomized}, this was the case for all five DGMs, i.e.\ for all five possible values for $\obsnum$. For the other 14 real-data-based outcome probabilities, only DGMs with smaller sample size values ($\obsnum \in \{60,120\}$) fell short of the $8{,}000$ repetitions. Specifically, fewer than $8{,}000$ datasets with observations in all seven ordinal categories were simulated for six real-data-based DGMs with $\obsnum = 60$ and one real-data-based DGM with $\obsnum =120$, leading to their exclusion from the respective rows of the panels.\\
As seen in \hyperref[fig:performance_ordinal]{Figure~\ref*{fig:performance_ordinal}a}, the estimated power generally increases with the relative effect deviation from 0.5. For the Wilcoxon rank-sum test and PO ordinal logistic regression, the estimated power follows a nearly deterministic monotonic increase with the relative effect deviation, and both tests yield highly similar results (see \hyperref[sec:S1.3_additional_infos_alignment]{Supplementary Section~\ref*{sec:S1.3_additional_infos_alignment}} for a brief explanation of the theoretical basis for this alignment).
For the Chi-square test and Fisher's exact test, the relationship between estimated power and the relative effect deviation is less clear. A more detailed investigation into additional characteristics (e.g., asymmetry in outcome probabilities or the expected number of observations per category) would be necessary in a more comprehensive simulation study but is beyond the scope of this illustration.\\
With regard to the comparison between researcher-specified and real-data-based probabilities, the systematic differences in relative effect deviation (\hyperref[fig:charac_ordinal]{Figure~\ref{fig:charac_ordinal}}) are reflected in corresponding differences in estimated power (\hyperref[fig:performance_ordinal]{Figure~\ref{fig:performance_ordinal}a}).
Specifically, all tests tend to yield higher estimated power for the researcher-specified probabilities. Additionally, \hyperref[fig:performance_ordinal]{Figure~\ref*{fig:performance_ordinal}b} highlights that the relative performance of the four tests varies across the two types of parameter specification. For Chi-square and Fisher’s exact test, the difference in power between the best-performing test and these tests is larger for researcher-specified parameters at smaller sample sizes. Conversely, for the Wilcoxon rank-sum test and PO ordinal logistic regression, the difference in power is more pronounced for real-data-based probabilities at larger sample sizes. Accordingly, if only one type of parameter specification had been considered, the conclusions regarding the relative performance of the tests would have differed, depending on whether the parameters were specified by the researcher or based on real data.

\subsection{Differential gene expression analysis}\label{sec:ill_de}
\subsubsection{Design}
\paragraph{ADEMP structure}  
For the second illustrative simulation study, we adopt the simulation design from \citet{baik2020benchmarking} (hereafter referred to as {\baik}) on differential gene expression analysis, extending it only by the number of real datasets used for parameter inference, as detailed below. Since {\baik}'s simulation is highly comprehensive, we include only a subset of the original DGMs (see \hyperref[sec:S2.1_excluded_DGMs]{Supplementary Section~\ref*{sec:S2.1_excluded_DGMs}} for details on excluded DGMs) and focus on a single performance measure. An overview of the resulting ADEMP structure is provided in \hyperref[tab:ademp_de]{Table~\ref*{tab:ademp_de}}.

%%%%% Begin Table: tab_ADEMP_deanalysis %%%%%--------------------
\begin{table}[ht]
    \caption{ADEMP structure for the example illustration on methods for differential gene expression analysis. Parameters $\lambdaall = \{\groupnum, \varnum, \lambda_{\text{FC}}, minFC, \propupreg, \propDE, \obsnum\}$ are researcher-specified, and parameters $\thetaall = \{\boldsymbol{\mu},\boldsymbol{\phi}\}$ are real-data-based.}
    \small
    \centering
    {\renewcommand{\arraystretch}{1.3}
    \begin{tabularx}{1\textwidth}{|W{2cm}|X|}\hline
%%%%%%%%%%%%%%%%%%%%%%%%%%%%%%%%%%%%%%%%%%%%%%%%%%%%%%%%%%%%%%%%%%%%%%%%%%%%%
       \raggedright Aim &   Evaluation of methods for differential gene expression analysis, i.e.\ methods that identify genes with differences in their RNA-Seq expression levels, in a two-group (i.e.\ $\groupnum = 2$) setting (e.g., cancer vs.\ normal) \\ \hline
%%%%%%%%%%%%%%%%%%%%%%%%%%%%%%%%%%%%%%%%%%%%%%%%%%%%%%%%%%%%%%%%%%%%%%%%%%%%%
       \raggedright Data-generating mechanisms (DGMs) &  
       \textbf{Model structure $\struc$}  \vspace{0.2cm}
       \begin{tabitemize}
            \item Read count $r_{\obsind,\varind}$ of gene $\varind$, $\varind = 1,\dots,\varnum$, and sample $\obsind$, $\obsind = 1,\dots,\obsnum$, in group $\groupind \in \{\groupone, \grouptwo\}$ (equal group sizes) is simulated as $R_{\obsind,\varind} \sim \operatorname{NB}(\mu_{\varind} \cdot \operatorname{FC}_{\varind}, \phi_{\varind})$, $\mu_{\varind}, \phi_{\varind} \geq 0$, if $x_{\obsind} = \groupone$, and $R_{\obsind,\varind} \sim \operatorname{NB}(\mu_{\varind}, \phi_{\varind})$, $\mu_{\varind}, \phi_{\varind} \geq 0$, if $x_{\obsind} = \grouptwo$. Among $\varnum$ genes, proportion $\propDE$ are differentially expressed (DE) with $\operatorname{FC}_{\varind} \neq 1$. Among DE genes, proportion $\propupreg$ are upregulated ($\operatorname{FC}_{\varind} > 1$), rest are downregulated ($\operatorname{FC}_{\varind} < 1$). $\operatorname{FC}_\varind$ is defined as \newline
            $\operatorname{FC}_{\varind} =\begin{cases}
                (minFC + randFC_\varind) & \text{if gene } \varind \text{ is DE and upregulated,}\\
                (minFC + randFC_\varind)^{-1} & \text{if gene }  \varind \text{ is DE and downregulated},\end{cases}$\newline 
            where $randFC_\varind$ is drawn from $\operatorname{Exp}(\lambda_{\text{FC}})$.
        \end{tabitemize} 
        \vspace{0.4cm}
        \textbf{Parameters $\lambdaall$ and $\thetaall$ (varied fully factorially)} \vspace{0.2cm}
        \begin{tabitemize}
            \item \vspace{-0.25\baselineskip}\begin{tabular}[t]{@{}l@{}}$\lambdaall$: $\groupnum = 2$; $\varnum = 10{,}000$; $\propupreg = 0.5 $; $\propDE\in \{0.05, 0.10, 0.30, 0.60\}$; $\lambda_{\text{FC}}=1$;\\[-0.12cm]\quad\, $minFC=1.5$ for $\obsnum = 6$ and $minFC=1.2$ for $\obsnum = 20$; $\obsnum \in \{6,20\}$\end{tabular}\vspace{-0.25\baselineskip}
            \item $\thetaall$: $(\boldsymbol{\mu},\boldsymbol{\phi}) = ((\mu_{1}, \dots, \mu_{\varnum}),(\phi_{1}, \dots, \phi_{\varnum})) $ (14 pairs of values, see \hyperref[tab:deanalysis_datasets]{Table~\ref*{tab:deanalysis_datasets}} for datasets) 
        \end{tabitemize}
        %\vspace{0.4cm}
        \textBF{Number of repetitions per DGM}: $\nrep =50$ \\ \hline
% %%%%%%%%%%%%%%%%%%%%%%%%%%%%%%%%%%%%%%%%%%%%%%%%%%%%%%%%%%%%%%%%%%%%%%%%%%%%%
       \raggedright  Estimand / Target  &    The null hypothesis $H_0: \operatorname{FC}_{\varind} = 1$ for all $\varind \in \{1, \dots, \varnum\}$ \\ \hline
% %%%%%%%%%%%%%%%%%%%%%%%%%%%%%%%%%%%%%%%%%%%%%%%%%%%%%%%%%%%%%%%%%%%%%%%%%%%%
       \raggedright Methods  &   11  methods: edgeR, edgeR.ql, edgeR.rb, DESeq.pc, DESeq2, voom.tmm, voom.qn, voom.sw, ROTS, baySeq, PoissonSeq  \\ \hline
% %%%%%%%%%%%%%%%%%%%%%%%%%%%%%%%%%%%%%%%%%%%%%%%%%%%%%%%%%%%%%%%%%%%%%%%%%%%%%
        \raggedright Performance measure  & Area under the receiver operating characteristic curve (AUC) \\ \hline
    \end{tabularx} 
    }
    \label{tab:ademp_de}
\end{table}
%%%%% End Table: tab_ADEMP_deanalysis %%%%%----------------------
Similar to the first illustration, the \enquote{A}, \enquote{D}, and \enquote{E} aspects largely align with those specified for {\exde} (see \hyperref[sec:2.4_examples]{Section~\ref*{sec:2.4_examples}}), which was itself based on {\baik}'s study. That is, we aim to evaluate methods for identifying DE genes from RNA-Seq data in a two-group (i.e.\ $\groupnum = 2$) setting with $\obsnum$ samples and $\varnum$ genes. The only difference is that in {\exde}, the fold changes of DE genes were directly assigned, whereas in the actual study by {\baik}, they are specified in a more refined manner, incorporating additional parameters and a stochastic component. Our simulation study follows this more detailed formulation by {\baik} (see \hyperref[tab:ademp_de]{Table~\ref*{tab:ademp_de}}).\\
In the subset of DGMs considered in our simulation, all parameters except for the gene-wise mean expression $\boldsymbol{\mu} = (\mu_1, \dots, \mu_\varnum)$ and dispersion $\boldsymbol{\phi} = (\phi_1, \dots, \phi_\varnum)$ are researcher-specified (by {\baik}), and only the number of samples and the proportion of DE genes are varied ($\obsnum \in \{6,20\}$ and $\propDE \in \{0.05, 0.10, 0.30, 0.60\}$).  
{\baik} estimate the values of $\boldsymbol{\mu}$ and $\boldsymbol{\phi}$ from the Kidney Renal Clear Cell Carcinoma (KIRC) RNA-Seq dataset from the dataset collection of TCGA (mentioned earlier in \hyperref[sec:3.3.1_rdb_select_database]{Section~\ref*{sec:3.3.1_rdb_select_database}}), resulting in a single vector for both $\boldsymbol{\mu}$ and $\boldsymbol{\phi}$. As an extension, we consider 13 additional TCGA datasets, leading to 14 $(\boldsymbol{\mu}, \boldsymbol{\phi})$ pairs in total, which we consider jointly. Details on dataset selection and parameter inference are provided below.
{\baik} employ a fully factorial design, which we also adopt. When using only one TCGA dataset to infer $(\boldsymbol{\mu},\boldsymbol{\phi})$, this results in $2 \times 4 = 8$ DGMs. When all 14 (eligible) TCGA datasets are considered, this results in $8 \times 14 = 112$ DGMs. Following {\baik}, the number of simulated datasets per DGM is set to $\nrep = 50$.\\
{\baik} evaluate 12 methods, including, e.g., edgeR \citep{Rpkg_edgeR2010}, DESeq2 \citep{Rpkg_DESeq22014}, and their variants (see {\baik} for details). However, in our simulation, we are only able to consider 11 methods, as SAMseq \citep{samseq2013} is excluded due to persistent execution errors that prevent it from running.  
For performance evaluation, we consider only the area under the receiver operating characteristic curve (AUC), which is the primary measure used by {\baik}, although {\baik} also consider the true positive rate and the false discovery rate.\\
We conduct the simulation using the \texttt{compareDEtools} \texttt{R} package, which accompanies {\baik}'s study.  

\paragraph{Dataset selection and parameter inference} 
Since the real dataset used by {\baik} (KIRC) is from the dataset collection of TCGA, we consider TCGA as the database, which contains RNA-Seq datasets for 33 cancer types. To determine dataset eligibility, we follow the approach that {\baik} used for the KIRC dataset, whereby only tumor and normal tissues are considered and only paired samples (i.e.\ those in which tumor and normal tissue originate from the same patient) are included. Specifically, we define one {\dsaim} criterion, which excludes datasets that do not contain both tumor and normal samples, and one {\dsextract} criterion, which excludes unmatched samples and, as an additional requirement for our study, datasets with fewer than 10 matched sample pairs.
After applying both criteria, $\dsnum = 14$ datasets remain. 
Details on the datasets and the selection process are provided in \hyperref[sec:S2.2_dataset_selection]{Supplementary Section~\ref*{sec:S2.2_dataset_selection}}.\\
For each TCGA dataset, the parameters $\boldsymbol{\mu}$ and $\boldsymbol{\phi}$ are inferred following the same approach used by {\baik} for the KIRC dataset. Specifically, for each gene in a given TCGA dataset, the mean expression is computed as the average across all samples, while dispersion is estimated in a more complex manner: first, RNA-Seq counts are normalized to account for differences in sequencing depth (i.e.\ variations in the total number of reads per sample). Then, an empirical Bayes approach is applied to improve the stability of gene-wise dispersion estimates (see the \texttt{compareDEtools} package for more details). Since each TCGA dataset contains 20{,}501 genes, this process results in 20{,}501 mean and dispersion values per dataset. 
From these values, those corresponding to genes with a mean expression of less than 10 are excluded. Finally, $\varnum = 10{,}000$ genes (the number used in the simulation) are randomly selected, and their corresponding mean and dispersion estimates constitute the resulting vectors $\boldsymbol{\mu}$ and $\boldsymbol{\phi}$. Although these exclusion steps may appear to be subset-level criteria (see \hyperref[sec:3.3.2_rdb_select_criteria]{ Section~\ref*{sec:3.3.2_rdb_select_criteria}}), the excluded genes still contribute to parameter inference as they are used for dispersion estimation, making this better viewed as a filtering procedure.
Moreover, note that we follow the implementation by \baik, who draw new genes for each repetition.

\subsubsection{Results}
\begin{figure}[ht]
    \centering
    \includegraphics[width=0.8\linewidth]{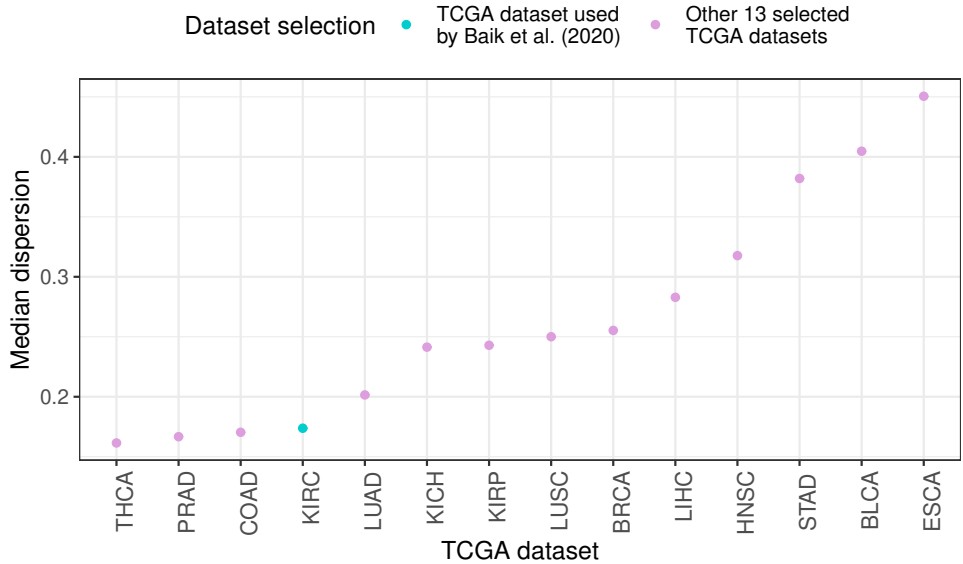} 
    \caption{Median dispersion values across the 14 selected TCGA datasets, including the KIRC dataset used by \citet{baik2020benchmarking}, sorted in ascending order. Median dispersion is calculated based on genes with a mean expression greater than 10, which serve as the basis for drawing dispersion values in the simulation. Dataset abbreviations are detailed in \hyperref[tab:deanalysis_datasets]{Table~\ref*{tab:deanalysis_datasets}}. }
    \label{fig:charac_de}
\end{figure}

\paragraph{Parameter characteristics}
Similar to the outcome probabilities in the first illustration, the multi-dimensional nature of the mean and dispersion values $(\boldsymbol{\mu},\boldsymbol{\phi})$ makes a direct comparison across the 14 TCGA datasets impractical. Instead, one or more summary measures are needed to characterize their differences.
For further analysis, we focus on the median dispersion of each dataset, calculated from the set of dispersion estimates that serve as the basis for drawing $\boldsymbol{\phi}$ in the simulation---that is, the dispersion values from all genes with a mean expression greater than 10. 
The median dispersion across datasets ranges from 0.161 to 0.451, with the median values for each dataset shown in \hyperref[fig:charac_de]{Figure~\ref*{fig:charac_de}}. Notably, the KIRC dataset, which is used in the original simulation by {\baik}, has one of the lowest median dispersion values at 0.174.

\paragraph{Method performance} 
\begin{figure}
    \centering
    \includegraphics[width=0.9\linewidth]{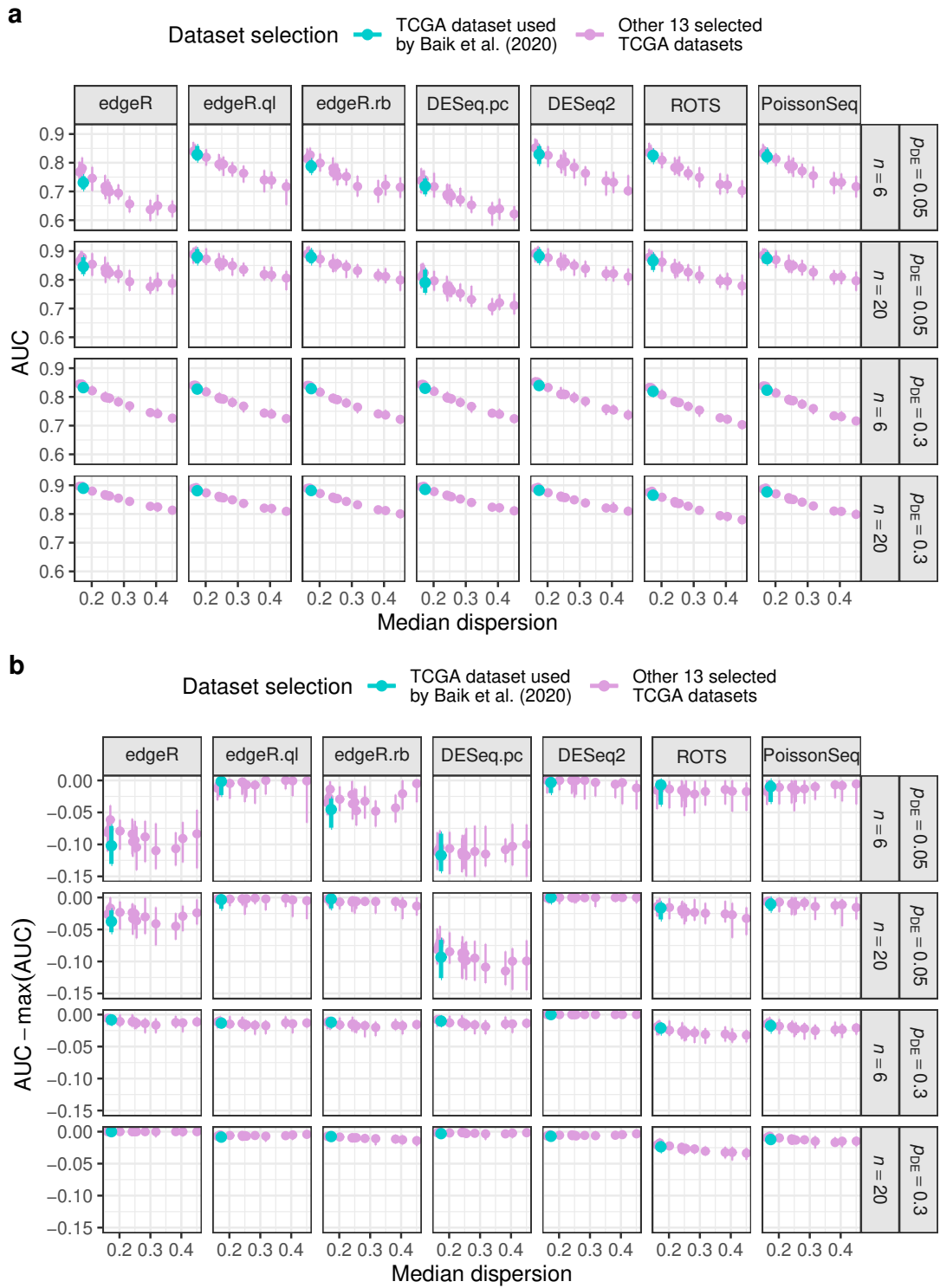} 
    \caption{Absolute (panel a) and relative (panel b) performance of six differential expression analysis methods in relation to the median dispersion (averaged across all genes in the real datasets after filtering), across different sample sizes $\obsnum$ and proportions of DE genes, $\propDE$, comparing results based on the KIRC dataset used by \citet{baik2020benchmarking} and the results based on 13 other selected TCGA datasets. Panel a displays the median and range of absolute AUC values, while panel b presents the median and range of the difference between the AUC and the highest AUC observed within each DGM.}
    \label{fig:performance_de}
\end{figure}
In \hyperref[fig:performance_de]{Figure~\ref*{fig:performance_de}}, the absolute performance (panel a: AUC) and the relative performance (panel b: difference to the best, i.e.\ the highest AUC within each repetition) are shown in relation to the median dispersion estimated from each TCGA dataset. To improve clarity, only 7 of the 11 considered methods are displayed---those recommended by {\baik} for the DGMs under consideration (see {\baik}, Table 2). 
Similarly, \hyperref[fig:performance_de]{Figure~\ref*{fig:performance_de}} only includes results for two of the four specified values of the proportion of DE genes ($\propDE \in \{0.05, 0.30\}$), as these values are used by {\baik} to differentiate recommendations in their summary table. The results for all methods and $\propDE$ values are provided in \hyperref[sec:S2.3_additional results]{Supplementary Section~\ref*{sec:S2.3_additional results}} but lead to similar conclusions as those discussed below.\\
As shown in \hyperref[fig:performance_de]{Figure~\ref*{fig:performance_de}a}, regardless of the number of samples and the proportion of DE genes, the AUC of all methods decreases as median dispersion increases. Since the KIRC dataset is among the datasets with the smallest median dispersion, relying solely on this specific dataset for simulation may overestimate the ability of all methods to identify DE genes. With regard to relative performance (\hyperref[fig:performance_de]{Figure~\ref*{fig:performance_de}b}), if for a given $\obsnum$ and $\propDE$ a method is among the top-performing methods when considering only the KIRC dataset (i.e.\ its AUC difference is close to zero), then its relative performance tends to remain stable across the remaining 13 datasets. However, if a method is not among the best (particularly edgeR, edgeR.rb, and DESeq.pc under DGMs with $\propDE = 0.05$), its performance ranking may change more substantially when additional datasets are considered. While expanding the analysis beyond the KIRC dataset does not lead to different conclusions for all methods, we argue that confirming the stability of relative performance across datasets also provides valuable insight.

%%%%%%%%%%%%%%%%%%%%%%%%%%%%%%%%%%%%%%%%%%%%%%%%%%%%
%%%%%%%%%%%%%%%%%%%% Section 6 %%%%%%%%%%%%%%%%%%%%%
%%%%%%%%%%%%%%%%%%%%%%%%%%%%%%%%%%%%%%%%%%%%%%%%%%%%
\section{Suggested workflow}\label{sec:5_workflow}
\hyperref[tab:workflow]{Table~\ref*{tab:workflow}} provides a structured workflow for constructing parametric DGMs based on a systematically selected set of real datasets, summarizing the considerations discussed in Sections~\ref{sec:3_implementation} and \ref{sec:3.3_selecting_datasets}. Once the real-data-based DGMs are fully specified, the simulation study can proceed as usual. When reporting on this process, the dataset selection and inference steps should be documented in detail, including the database used, eligibility criteria applied, inference methods employed, the mapping of inferred parameter values to DGMs, and any additional information necessary for reproducibility. Finally, as generally recommended in simulation studies, an overview of the resulting DGMs should be provided.

%%%%%%%%%%%%%%%%%%%%%%%%%%%%%%%%%%%%%%%%%%%%%%%%%%%%
%%%%%%%%%%%%%%%%%%%% Section 7 %%%%%%%%%%%%%%%%%%%%%
%%%%%%%%%%%%%%%%%%%%%%%%%%%%%%%%%%%%%%%%%%%%%%%%%%%%
\section{Conclusion}
\label{sec:6_conclusion}
%%%%% Begin Table: tab_workflow %%%%%----------------------------
\begin{table}[p]
\small
\centering
\caption{Structured workflow for constructing parametric DGMs based on a systematically selected set of real datasets.}
\renewcommand{\arraystretch}{1.3} 
% set widths to later calculate the width of the \multicolumn for the bold main steps
\newlength{\tablewidth}
\setlength{\tablewidth}{0.945\textwidth}
\newlength{\lastcolumnwidth}
\setlength{\lastcolumnwidth}{1cm}
% set widths to adjust what is \labelsep in itemize and automatically adjust hanging indentation, so you only need to change \spaceaftersubstepnumber
\newlength{\spaceaftersubstepnumber}
\setlength{\spaceaftersubstepnumber}{0.5em}
\newlength{\hangingindent}
\setlength{\hangingindent}{1.60em + \spaceaftersubstepnumber - 0.3333em}
\begin{tabulary}{\tablewidth}{L>{\hangindent=\hangingindent}LR{\lastcolumnwidth}}
\hline
\multicolumn{3}{r}{\textbf{Section}}\\
\hline
\multicolumn{3}{p{\tablewidth - 1.5\lastcolumnwidth}}{\textbf{Step 1: Apply specification-based and knowledge-based differentiation and plan for variation of researcher-specified components.}} \\ 
& 1.1\hspace{\spaceaftersubstepnumber}Determine which components are researcher-specified components of interest, researcher-specified components of convenience, and real-data-based components. & \ref{sec:2.2.1_spec_diff} \\ 
& 1.2\hspace{\spaceaftersubstepnumber}For researcher-specified components of interest and real-data-based components, clarify whether their true form or value in the true DGM is known, as unknown components introduce uncertainty for dataset selection and inference, respectively. & \ref{sec:2.2.2_know_diff} \\ 
& 1.3\hspace{\spaceaftersubstepnumber}Decide whether multiple options/values should be considered for any researcher-specified component of interest. If so, repeat steps 2--7 for each relevant combination. For researcher-specified components of convenience, multiple options/values can be considered without repeating the full process; only the combination with the inferred real-data-based components needs to be specified. & \ref{sec:2.3_notation} \\ \hline
\multicolumn{3}{p{\tablewidth - 1.5\lastcolumnwidth}}{\textbf{Step 2: Specify researcher-specified components and additional constraints.}} \\ 
& 2.1\hspace{\spaceaftersubstepnumber}Specify the researcher-specified components of interest (one option/value per component; see step 1.3), the researcher-specified components of convenience, and any explicit or implicit constraints for the real-data-based components. & \ref{sec:2.2.1_spec_diff}
\\ \hline
\multicolumn{3}{p{\tablewidth - 1.5\lastcolumnwidth}}{\textbf{Step 3: Specify inference procedures for real-data-based DGM components.}} \\ 
& 3.1\hspace{\spaceaftersubstepnumber}Specify the inference method for each real-data-based parameter or part of the model structure, ensuring that it accounts for potential misalignment between the considered and the true model structure as well as for sampling variability. & \ref{sec:3.1_inferring_params},~\ref{sec:3.2_rdb_ms} \\ 
& 3.2\hspace{\spaceaftersubstepnumber}For parameters, specify for each one whether its values will be directly inferred or aggregately inferred  and specify how the sets of inferred values for the individual parameters will be mapped to the set of considered parameter vectors, acknowledging the trade-off between realism and greater control or practical feasibility, as implied by the one-to-one approach versus the deviations from it. & \ref{sec:3.1_inferring_params} \\
\hline
\multicolumn{3}{p{\tablewidth - 1.5\lastcolumnwidth}}{\textbf{Step 4: Specify the systematic selection of real datasets.}} \\ 
& 4.1\hspace{\spaceaftersubstepnumber}Specify eligibility criteria addressing {\dsaim} by translating the researcher-specified components of interest and constraints imposed on real-data-based components (step~2.1) into concrete criteria.  Where not directly evident, also specify how the fulfillment of these criteria should be assessed. & \ref{sec:3.3.2_rdb_select_criteria} \\
& 4.2\hspace{\spaceaftersubstepnumber}Specify eligibility criteria addressing {\dsextract} based on the inference procedure specified in step~3.1 and the criteria specified for {\dsaim} in step~4.1. & \ref{sec:3.3.2_rdb_select_criteria} \\
& 4.3\hspace{\spaceaftersubstepnumber}Specify a minimum and maximum number of datasets to be selected. & \ref{sec:3.3.1_rdb_select_database} \\
& 4.4\hspace{\spaceaftersubstepnumber}Select a database that meets accessibility requirements to fulfill {\dsrepro} and is likely to contain datasets eligible with respect to the criteria specified in steps~4.1 and 4.2. & \ref{sec:3.3.1_rdb_select_database} \\
\hline
\multicolumn{3}{p{\tablewidth - 1.5\lastcolumnwidth}}{\textbf{Step 5: Conduct dataset selection and adjust criteria if necessary.}} \\
& 5.1\hspace{\spaceaftersubstepnumber}Apply the eligibility criteria to the datasets in the database, refining or extending the criteria during the assessment as needed. & \ref{sec:3.3.2_rdb_select_criteria} \\
\hline
\multicolumn{3}{p{\tablewidth - 1.5\lastcolumnwidth}}{\textbf{Step 6: Check the number of selected datasets.}} \\
& 6.1\hspace{\spaceaftersubstepnumber}If the number of selected datasets falls within the predefined range, proceed with step~7; if it is too low, expand the database (e.g., more publication years); if it is too high, randomly draw the specified maximum number of datasets. & \ref{sec:3.3.1_rdb_select_database} \\
\hline
\multicolumn{3}{p{\tablewidth - 1.5\lastcolumnwidth}}{\textbf{Step 7: Infer real-data-based DGM components.}} \\
& 7.1\hspace{\spaceaftersubstepnumber}If only parameters are inferred, apply the specified inference procedure to the $\dsnum$ selected datasets.  If parts of the model structure are also inferred, determine them first, then infer parameters separately for each subset of datasets corresponding to each resulting model structure derived from the inference of real-data-based parts. & \ref{sec:3.1_inferring_params},~\ref{sec:3.2_rdb_ms} \\
& 7.2\hspace{\spaceaftersubstepnumber}Check the resulting DGMs for plausibility. & \ref{sec:3.1_inferring_params}\\
\hline
\end{tabulary}
\label{tab:workflow}
\end{table}
%%%%% End Table: tab_workflow %%%%%------------------------------
Basing parametric simulations on real data can be a viable approach to improve the practical relevance of DGMs, ideally achieving alignment with real-world DGMs in the study’s domain of interest. 
However, current implementations often lack a systematic approach---both in determining which components of the DGM should be based on real data and how they should be inferred, as well as in selecting the real datasets. In particular, the rationale behind dataset selection and the domain of interest they are meant to represent is often unclear, and the number of utilized datasets is typically very limited.
As a result, the findings of simulation studies using real-data-based DGMs may not necessarily generalize better to practical applications than those based on fully researcher-specified DGMs, despite potentially creating that impression.\\
To address these issues, this paper provided a detailed discussion on the construction of real-data-based parametric DGMs, aiming to support researchers in assessing the possibilities and implications of inferring specific DGM components from real datasets and in making dataset selection more systematic. In addition to the formal discussion, we conducted two simulation studies demonstrating the implementation of parametric DGMs based on a systematically selected set of datasets and illustrating that they may lead to different conclusions than fully researcher-specified DGMs or DGMs based on a single real dataset.\\
Importantly, throughout the paper, several practical and conceptual limitations associated with constructing real-data-based parametric DGMs as we suggest have emerged.
One such limitation is finding an adequate database. 
It is true that parametric DGMs offer more potential database options than, for example, semi-parametric DGMs, which require access to complete datasets, and certain issues---such as dataset quality---can be mitigated through carefully chosen eligibility criteria. However, even with the most meticulous planning, one major limitation remains: some datasets relevant to the domain of interest may not be included in the database at all (e.g., if the database primarily contains datasets from a specific subpopulation of DGMs, a restriction that is not always immediately apparent).
Additionally, it is often unclear whether the selected datasets truly belong to the domain of interest, as some eligibility criteria may refer to DGM components whose true form or values in the underlying real-world DGM cannot be determined solely from the real dataset itself. As discussed in detail, this issue arises not only when datasets are selected based on DGM components that cannot be fully known but also when these components are inferred from the selected datasets to construct the DGM. Specifically, in the case of parameters, even the one-to-one inference approach does not guarantee that the inferred values are close to the true ones. Similarly, if parts of the model structure are also intended to be based on real data, the specified set of possible options (e.g., distributions) may already deviate substantially from the true underlying structure (which may not even have a closed-form representation).\\
As a consequence, there is no guarantee that the suggested approach will lead to practically relevant parametric DGMs---and, at the same time, it demands substantially more effort than simply relying on a single convenience dataset or entirely researcher-specified DGMs. 
However, we propose the following considerations.
First, the increased effort is worthwhile in itself because it encourages researchers to think more deeply about the simulation design.
Moreover, reporting how and why the real datasets were selected and how the DGMs were constructed based on them enhances transparency and ideally provides a clearer understanding of the study’s domain of interest.
In principle, this follows a similar line of reasoning as other (complementary) approaches aimed at improving thoroughness and transparency, such as the writing (and potential preregistration) of research protocols for simulation studies \citep{siepe2024simulation}, which, despite seeming like an unnecessary burden, already improves study quality by requiring careful consideration and documentation of decisions.\\
Regarding the extent to which practical relevance can be achieved, we argue that pursuing it should not be abandoned just because it cannot be fully attained---especially since, also due to our certainly idealized definition, it may never be completely achievable in the first place. Moreover, our proposed approach would likely lead to more practically relevant parametric DGMs than the alternatives currently used in practice, i.e.\ fully researcher-specified parametric DGMs or DGMs based on one or two convenience datasets. Of course, a rigorous discussion of the limitations that hinder the achievement of practical relevance remains essential. This is particularly important to prevent the mere use of a larger-than-usual number of real datasets from being misinterpreted---by readers or even the researchers conducting the study---as a guarantee of practical relevance.\\
Finally, it is worth reiterating that while this paper focused on parametric DGMs, researchers aiming for practically relevant DGMs may also consider semi-parametric DGMs, for which the use of a systematic selection of real datasets is also uncommon, just as it is for parametric DGMs. Although our discussion is partially applicable to semi-parametric DGMs, a more detailed investigation specifically tailored to these cases would be valuable for future research.
Moreover, our approach for constructing real-data-based parametric DGMs does not address all potential pitfalls in simulation studies that may contribute to overgeneralization and misinterpretation---for example, biased post hoc selection of performance measures or considered methods \citep{pawel2024pitfalls}. While these issues fall outside the scope of our approach, they underscore the need for careful study design beyond just the choice of DGM. 
Still, we hope that the proposed approach to constructing real-data-based parametric DGMs is a step toward simulation studies that yield more well-founded recommendations, ultimately helping applied researchers make more informed choices when selecting statistical methods.

\FloatBarrier

\section*{Funding information}
This work was supported by the German Research Foundation (individual grants BO3139/7-2 and BO3139/9-1 to ALB for the work of CS and project 352692197 for the work of MT).
%This work was supported by the German Research Foundation (BO3139/7-2, BO3139/9-1) to ALB and project 352692197.
%The work of Maria Thurow was supported by the Deutsche Forschungsgemeinschaft (DFG, German Research Foundation) - project 352692197. 
The authors of this work take full responsibility for its content.

\section*{Acknowledgments}
We thank Luzia Hanßum for assistance with supporting tasks related to the manuscript.

\section*{Conflicts of interest}
The authors have declared no conflicts of interest for this article.

\FloatBarrier

\printbibliography

\newpage

%%%%%%%%%%%%%%%%%%%%%%%%%%%%%%%%%%%%%%%%%%%%%%%%%%%%
%%%%%%%%%%%%%%%%%%%% Supplement %%%%%%%%%%%%%%%%%%%%
%%%%%%%%%%%%%%%%%%%%%%%%%%%%%%%%%%%%%%%%%%%%%%%%%%%%
\appendix
\beginsupplement
\section*{Supplementary material}

%%%%%%%%%%%%%%%%%%%%%%%%%%%%%%%%%%%%%%%%%%%%%%%%%%%%
%%%%%%%%%%%%%%%%%%%% Appendix A %%%%%%%%%%%%%%%%%%%%
%%%%%%%%%%%%%%%%%%%%%%%%%%%%%%%%%%%%%%%%%%%%%%%%%%%%
\section{Example illustration 1: Two-arm randomized controlled trial with an ordinal outcome}
\label{sec:S1_supp_ordinal}

\subsection{Specification of researcher-specified parameters}

%%%%% Begin Table: tabappendix_ordinal_probuser %%%%%------------
\begin{table}[ht]
\centering
{\setlength\tabcolsep{4.8pt}
\small
\centering
\caption{The 4 researcher-specified pairs of outcome probabilities $(\bm{\pi}_\groupone,\bm{\pi}_\grouptwo)$ considered in the example illustration on hypothesis testing in the context of a two-arm randomized controlled trial with an ordinal outcome (see \hyperref[sec:ill_ordinal]{Section~\ref*{sec:ill_ordinal}}).}
\begin{tabular}{lcccccccccccccc}
\hline
\begin{tabular}[]{@{}l@{}} Outcome pro-\\babilities ID\end{tabular} & $\pi_{1,\groupone}$& $\pi_{2,\groupone}$ & $\pi_{3,\groupone}$ & $\pi_{4,\groupone}$ & $\pi_{5,\groupone}$ & $\pi_{6,\groupone}$ & $\pi_{7,\groupone}$ & $\pi_{1,\grouptwo}$ & $\pi_{2,\grouptwo}$ & $\pi_{3,\grouptwo}$ & $\pi_{4,\grouptwo}$ & $\pi_{5,\grouptwo}$ & $\pi_{6,\grouptwo}$ & $\pi_{7,\grouptwo}$ \\ 
\hline
k7\_id1 & 0.04 & 0.07 & 0.11 & 0.14 & 0.18 & 0.21 & 0.25 & 0.14 & 0.14 & 0.14 & 0.14 & 0.14 & 0.14 & 0.14 \\ 
k7\_id2 & 0.14 & 0.14 & 0.14 & 0.14 & 0.14 & 0.14 & 0.14 & 0.05 & 0.05 & 0.07 & 0.10 & 0.10 & 0.28 & 0.35 \\ 
k7\_id3 & 0.05 & 0.05 & 0.07 & 0.10 & 0.10 & 0.28 & 0.35 & 0.05 & 0.10 & 0.20 & 0.30 & 0.20 & 0.10 & 0.05 \\ 
k7\_id4 & 0.05 & 0.05 & 0.20 & 0.20 & 0.30 & 0.10 & 0.10 & 0.05 & 0.10 & 0.20 & 0.30 & 0.20 & 0.10 & 0.05 \\ 
\hline
\end{tabular}
\label{tab:ordinal_parameters_user}
}
\end{table}
%%%%% End Table: tabappendix_ordinal_probuser %%%%%--------------

\subsection{Specification of real-data-based parameters}

\subsubsection{Dataset selection}\label{sec:S1.2_dataset_selection}
\paragraph{Database} As a database, we considered all research publications published in \textit{The New England Journal of Medicine} (\url{https://www.nejm.org/}) between 2017 and 2022, which corresponds to the journal's volumes 376--387.
\paragraph{Search string} We first identified 270 articles for screening using this search string:
\texttt{fulltext:"randomized" AND (fulltext:"ordinal" OR fulltext:"proportional-odds" OR fulltext:"Mann-Whitney U" OR  fulltext:"Mann-Whitney-Wilcoxon" OR  fulltext:"Wilcoxon-Mann-Whitney" OR fulltext:"Wilcoxon rank-sum" OR fulltext:"Chi-Square" OR fulltext:"Fisher") AND (startDate:2017-01-01 AND endDate:2022-12-31) AND (articleCategory:"research")}. 

\paragraph{Dataset-level criteria for screening}
The inclusion and exclusion criteria on the dataset level are listed below, followed by some notes to clarify how we defined ordinal outcomes for our assessment of these criteria. Before each criterion, it is indicated to which of the three proposed requirements for real datasets used to construct real-data-based DGMs (see \hyperref[sec:3.3_selecting_datasets]{Section~\ref*{sec:3.3_selecting_datasets}}) the criterion is related.

\begin{itemize}
    \item Inclusion criteria 
    \begin{itemize}
        \item {\dsaim} Randomized controlled trials
        \item {\dsaim} At least one ordinal outcome% 
    \end{itemize}
        \item Trial exclusion criteria
    \begin{itemize}
        \item {\dsaim} Trials where individuals were not randomized individually but in groups or clusters, for example
        \item {\dsextract} Trials whose data overlaps with another trial considered at this stage, with preference given to the trial with the larger sample size
    \end{itemize}
\enlargethispage{-\baselineskip}

    \item Outcome exclusion criteria
    \begin{itemize}
        \item {\dsaim} Ordinal outcomes that are non-efficacy outcomes (e.g., safety, procedural, treatment adherence, or health economics outcomes)
        \item {\dsaim} Ordinal outcomes that are patient-reported outcomes 
        \item {\dsaim} Ordinal outcomes analyzed according to anything other than the intention-to-treat principle
        \item {\dsaim} Ordinal outcomes that were not analyzed beyond the presentation of frequencies, not analyzed as ordinal variables (e.g., if an ordinal outcome was dichotomized for the analysis), or analyzed with methods inappropriate for ordinal data (e.g., methods for continuous data)
        \item {\dsextract} Ordinal outcomes for which the data was not clearly reported, either in tables or figures (in the main/full text or supplement), for all categories
        \item {\dsaim} Ordinal outcomes with more or fewer than 7 categories
        \item {\dsextract} Ordinal outcomes with empty categories
    \end{itemize}
    \item Details on the definition of ordinal outcomes we applied when assessing articles with respect to the criteria above: 
    \begin{itemize}
        \item Ordinal outcomes must be explicitly declared as trial outcomes either in the main/full text, supplement, or study protocol to be considered.
        \item We considered an outcome variable ordinal if it was a categorical variable with ordered categories that are mutually exclusive and explicitly labeled. 
        \item If a reported distribution contained a category labeled \enquote{could not be evaluated} or \enquote{unknown} in addition to otherwise ordinal categories, we did not consider the variable ordinal. 
        \item Both ordinal outcomes based on an ordinal scale and ordinal outcomes defined by categorizing continuous measures were considered suitable for inclusion.        
        \item Non-ordinal outcomes involving an ordinal scale/measure, i.e.\ binary or continuous outcomes based on ordinal scales/scores, such as dichotomized ordinal variables or continuous variables reflecting the change in an ordinal scale/score, were not considered, even if there was data available for the involved ordinal scale/score.
    \end{itemize}

\end{itemize}

\begin{figure}[h]
    \centering
    \includegraphics[width=1\linewidth]{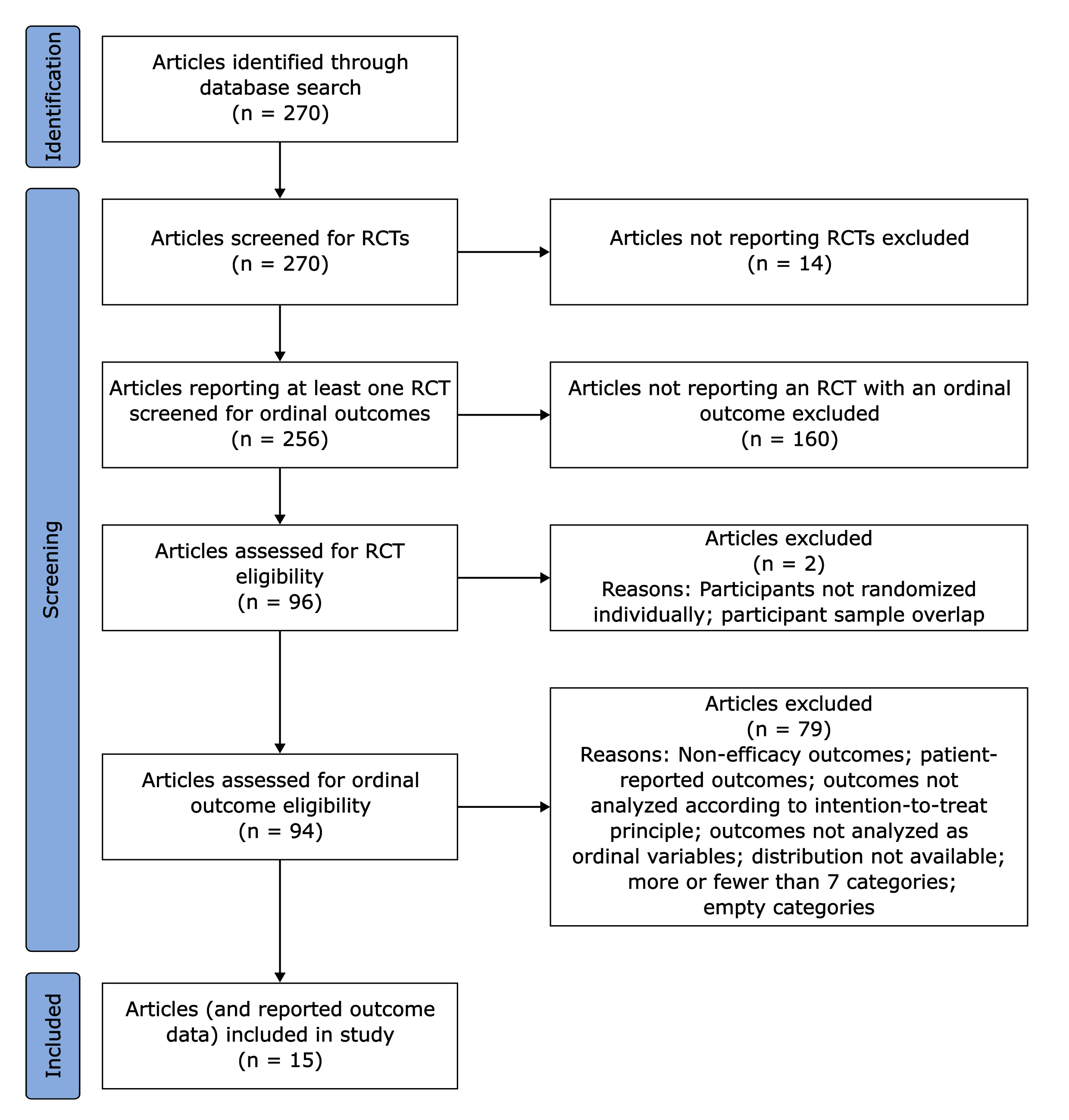}
    \caption{PRISMA flow diagram \citep{page2021PRISMA2020} to illustrate the dataset-level screening process for the example illustration on hypothesis testing in the context of a two-arm randomized controlled trial with an ordinal outcome (see \hyperref[sec:ill_ordinal]{Section~\ref*{sec:ill_ordinal}}).}
    \label{fig:prisma_ordinal}
\end{figure}

We first assessed for each of the 270 articles identified from the search whether or not it met the two inclusion criteria. 174 articles failed to meet the inclusion criteria, resulting in 96 remaining articles. These were then assessed with respect to the trial exclusion criteria. For two articles, the reported randomized controlled trials met trial exclusion criteria, leaving 94 articles with randomized controlled trials with ordinal outcomes to be assessed with respect to the outcome exclusion criteria. Out of these articles, 79 only had ordinal outcomes that met at least one of the outcome exclusion criteria, resulting in a final number of 15 articles with eligible ordinal outcomes. The screening process of the 270 publications is illustrated in \hyperref[fig:prisma_ordinal]{Figure~\ref*{fig:prisma_ordinal}} and the spreadsheet documenting the eligibility assessment can be found at \url{https://github.com/NiesslC/realdata_simulations}. 
Note that when assessing the eligibility of trials/outcomes, we did not factor in specifics of the randomization procedure (as long as individuals were randomized individually), treatment of missing values, or small details regarding the conducted analysis (e.g., covariates or random effects in regression models).
\vspace{-1em}

\paragraph{Subset-level criteria}
\enlargethispage{\baselineskip}
For the 15 selected articles, or more specifically, their underlying datasets, we applied the following subset criteria.
    \begin{itemize}
        \item {\dsaim}/{\dsextract} If there are two or more suitable ordinal outcomes in an article/trial, include the outcome that is considered most important in the trial (e.g., prefer primary outcomes to secondary outcomes and prefer secondary outcomes to tertiary/exploratory/additional outcomes). If such a distinction is not possible, include the outcome that has the highest sample size.
        \item {\dsaim}/{\dsextract} If more than two groups are compared in a trial with a suitable outcome, include the figures for the two groups with the highest sample sizes. 
    \end{itemize}

\subsubsection{Resulting parameters}
\hyperref[tab:ordinal_datasets_probs_rdb]{Table~\ref*{tab:ordinal_datasets_probs_rdb}} shows the 15 resulting real-data-based pairs of outcome probabilities. Note that we extracted the data for ordinal outcomes as it was presented in the article, which means that we did not change the order of the categories and extracted the distributions across the categories either in absolute terms (counts) or in relative terms (proportions), whichever was reported. Moreover, some papers reporting the distributions in relative terms included statements such as ``percentages may not total 100 because of rounding''. If that was the case, we scaled the resulting probabilities to 1.

\subsection{Additional results}
\label{sec:S1.3_additional_infos_alignment}
\begin{figure}[h]
    \centering
    \includegraphics[width=0.7\linewidth]{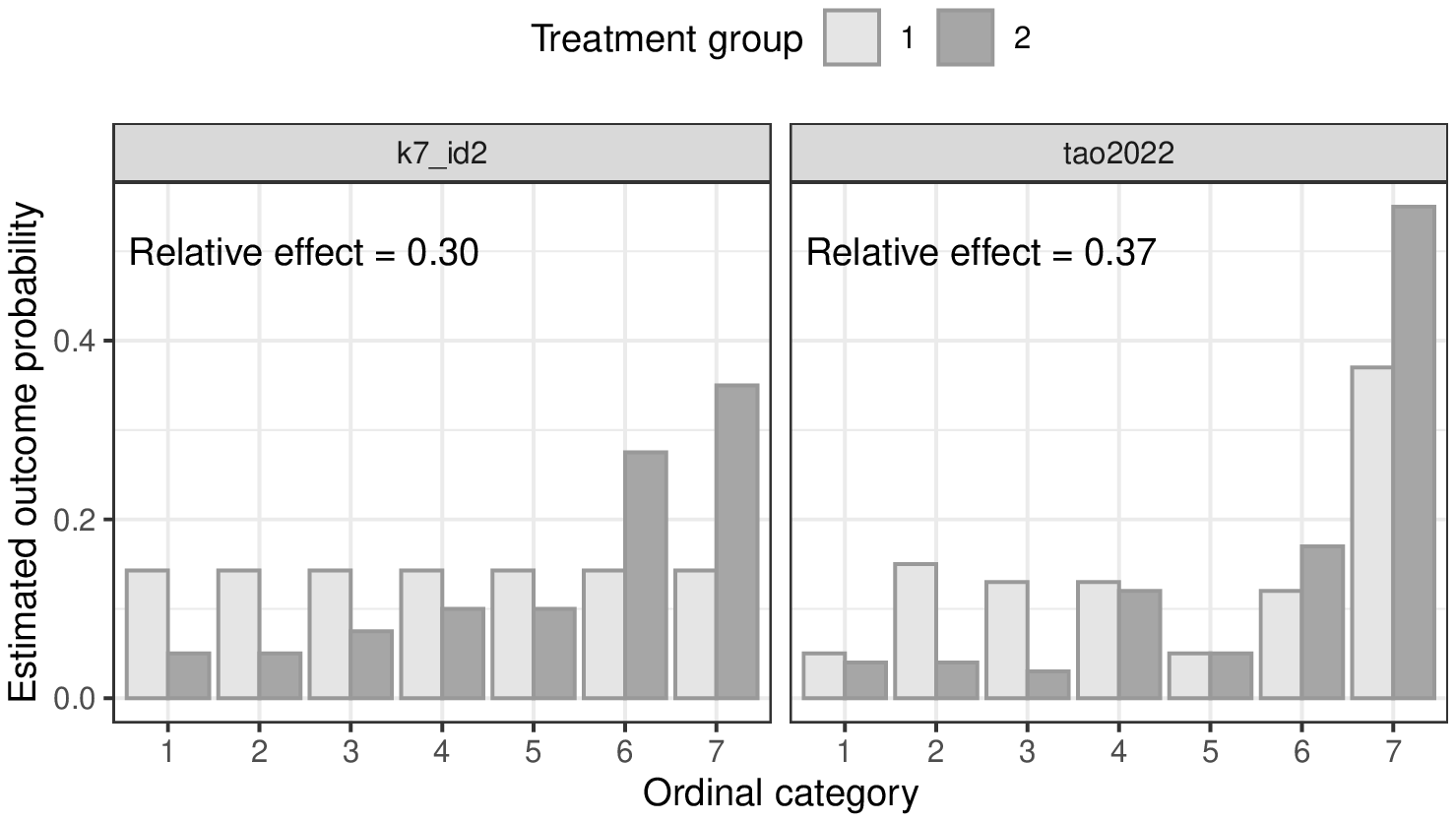} 
    \caption{Two of the sets of outcome probabilities $(\bm{\pi}_\groupone,\bm{\pi}_\grouptwo)$ considered in the example illustration on hypothesis testing in the context of a two-arm randomized controlled trial with an ordinal outcome (see \hyperref[sec:ill_ordinal]{Section~\ref*{sec:ill_ordinal}}), one researcher-specified (left) and one real-data-based (right), as well as the corresponding relative effect for each set. The shown real-data-based probabilities are the estimates published by \citet{tao2022trial}. }
    \label{fig:ordinal_bsp}
\end{figure}

\paragraph{Parameter characteristics} 
\hyperref[fig:ordinal_bsp]{Figure~\ref*{fig:ordinal_bsp}} shows an example of a researcher-specified and an example of a real-data-based set of outcome probabilities $(\bm{\pi}_\groupone,\bm{\pi}_\grouptwo)$, selected from the four researcher-specified and 15 real-data-based outcome probabilities. As expected, the researcher-specified outcome probabilities are more structured and systematically chosen, whereas those based on real data appear less uniform and more irregular. 
\vspace{-1em}

\paragraph{Method performance} The strong alignment between the Wilcoxon rank-sum test and the relative effect deviation from 0.5 arises because the Wilcoxon test statistic is based on rank-based comparisons, which are inherently linked to the relative effect. However, this relationship is not necessarily deterministic in all DGMs beyond those considered in this simulation (e.g., for smaller sample sizes; \citealp{thas2010comparing}). 
Similarly, the close agreement between the Wilcoxon rank-sum test and PO ordinal logistic regression can be attributed to the fact that the score test for the treatment effect in the PO ordinal logistic regression model is asymptotically equivalent
%%%%% Begin Table: tabappendix_ordinal_probs_rdb %%%%%-----------
\begin{landscape}
\begin{table}[p]
{\setlength\tabcolsep{3.6pt}
\renewcommand{\arraystretch}{1.1}%
\small
\caption{The 15 real-data-based pairs of outcome probabilities $(\bm{\pi}_\groupone,\bm{\pi}_\grouptwo)$ considered in the example illustration on hypothesis testing in the context of a two-arm randomized controlled trial with an ordinal outcome (see \hyperref[sec:ill_ordinal]{Section~\ref*{sec:ill_ordinal}}) as well as additional information about the corresponding trials.}
\label{tab:ordinal_datasets_probs_rdb}
\begin{tabular}{llllcccccccccccccccc}
\hline 
Dataset ID    & Publication & \begin{tabular}[]{@{}l@{}}Treated\\ condition\end{tabular}  & Measure                          & $n_\groupone$   & $n_\grouptwo$ & $\pi_{1,\groupone}$& $\pi_{2,\groupone}$ & $\pi_{3,\groupone}$ & $\pi_{4,\groupone}$ & $\pi_{5,\groupone}$ & $\pi_{6,\groupone}$ & $\pi_{7,\groupone}$ & $\pi_{1,\grouptwo}$ & $\pi_{2,\grouptwo}$ & $\pi_{3,\grouptwo}$ & $\pi_{4,\grouptwo}$ & $\pi_{5,\grouptwo}$ & $\pi_{6,\grouptwo}$ & $\pi_{7,\grouptwo}$       \\
\hline
albers2018     & \citet{albers2018thrombectomy}  & Stroke             & mRS                                                   
& 92   & 90  & 0.10 & 0.16 & 0.18 & 0.15 & 0.18 & 0.08 & 0.14 & 0.08 & 0.04 & 0.04 & 0.16 & 0.27 & 0.16 & 0.26 \\
campbell2018   & \citet{campbell2018tenectplase} & Stroke             & mRS                                                   
& 101  & 101 & 0.28 & 0.21 & 0.14 & 0.14 & 0.08 & 0.06 & 0.10 & 0.18 & 0.23 & 0.09 & 0.12 & 0.14 & 0.07 & 0.18 \\
cavalcanti2020 & \citet{cavalcanti2020hydro}     & COVID-19           & Other       
& 159  & 173 & 0.64 & 0.17 & 0.08 & 0.04 & 0.01 & 0.03 & 0.03 & 0.68 & 0.17 & 0.05 & 0.03 & 0.01 & 0.04 & 0.03 \\
goldman2020    & \citet{goldman2020remdesivir}   & COVID-19           & Other                  
& 200  & 197 & 0.08 & 0.08 & 0.04 & 0.10 & 0.06 & 0.04 & 0.60 & 0.11 & 0.17 & 0.05 & 0.07 & 0.07 & 0.02 & 0.52 \\
hutchinson2020 & \citet{hutchinson2020trial}     & \begin{tabular}[]{@{}l@{}}Subdural\\ hematoma\end{tabular}  & mRS                                                 
& 341  & 339  & 0.48 & 0.14 & 0.04 & 0.18 & 0.03 & 0.04 & 0.09 & 0.48 & 0.16 & 0.06 & 0.19 & 0.03 & 0.02 & 0.05 \\
jovin2022      & \citet{jovin2022trial}          & Stroke             & mRS                                                   
& 110  & 107 & 0.06 & 0.18 & 0.15 & 0.07 & 0.09 & 0.14 & 0.31 & 0.01 & 0.06 & 0.07 & 0.10 & 0.19 & 0.15 & 0.42 \\
lecouffe2021   & \citet{lecouffe2021randomized}  & Stroke             & mRS                                                   
& 273  & 266  & 0.04 & 0.12 & 0.33 & 0.10 & 0.10 & 0.11 & 0.21 & 0.06 & 0.09 & 0.36 & 0.09 & 0.14 & 0.09 & 0.16 \\
ma2019         & \citet{ma2019thrombolysis}      & Stroke             & mRS                                                   
& 113  & 112 & 0.12 & 0.23 & 0.14 & 0.13 & 0.13 & 0.12 & 0.12 & 0.11 & 0.19 & 0.13 & 0.14 & 0.21 & 0.12 & 0.09 \\
martins2020    & \citet{martins2020thrombectomy} & Stroke             & mRS                                                   
& 111  & 110  & 0.08 & 0.12 & 0.15 & 0.22 & 0.13 & 0.06 & 0.24 & 0.03 & 0.06 & 0.12 & 0.15 & 0.19 & 0.16 & 0.30 \\
perkins2018    & \citet{perkins2018randomized}   & \begin{tabular}[]{@{}l@{}}Cardiac\\ arrest\end{tabular}     & mRS                                               
& 4007 & 3994 & 0.00 & 0.00 & 0.01 & 0.01 & 0.00 & 0.01 & 0.97 & 0.00 & 0.00 & 0.01 & 0.01 & 0.00 & 0.00 & 0.98 \\
rosas2021      & \citet{rosas2021tocilizumab}    & \begin{tabular}[]{@{}l@{}} COVID-19\\ pneumonia\end{tabular} & Other  
& 294  & 144  & 0.56 & 0.02 & 0.05 & 0.02 & 0.09 & 0.06 & 0.20 & 0.49 & 0.06 & 0.03 & 0.07 & 0.10 & 0.06 & 0.19 \\
tao2022        & \citet{tao2022trial}            & Stroke             & mRS                                                   
& 226  & 114  & 0.05 & 0.15 & 0.13 & 0.13 & 0.05 & 0.12 & 0.37 & 0.04 & 0.04 & 0.03 & 0.12 & 0.05 & 0.17 & 0.55 \\
thomalla2018   & \citet{thomalla2018MRI-guided}  & Stroke             & mRS                                                   
& 254  & 249 & 0.21 & 0.32 & 0.21 & 0.12 & 0.07 & 0.02 & 0.04 & 0.15 & 0.27 & 0.23 & 0.17 & 0.13 & 0.04 & 0.01 \\
vandenberg2017 & \citet{vandenberg2017two-year}  & Stroke             & mRS                                                   
& 194  & 197 & 0.03 & 0.05 & 0.30 & 0.18 & 0.06 & 0.08 & 0.30 & 0.01 & 0.05 & 0.18 & 0.17 & 0.10 & 0.11 & 0.39 \\
yang2020       & \citet{yang2022endovascular}    & Stroke             & mRS                                                   
& 326  & 328 & 0.13 & 0.11 & 0.12 & 0.19 & 0.11 & 0.15 & 0.18 & 0.14 & 0.09 & 0.14 & 0.15 & 0.12 & 0.18 & 0.19 \\ 
\hline \\[-1.5ex]
\end{tabular}
{\raggedright Note: Due to rounding to two decimal places, seven non-zero probability values (ranging between $0.0020$ and $0.0042$) are shown as $0.00$. \\ Abbreviations: COVID-19, Coronavirus disease 2019; mRS, modified Rankin Scale. \par}
}
\end{table}
\end{landscape}
%%%%% End Table: tabappendix_ordinal_probs_rdb %%%%%-------------
to the Wilcoxon rank-sum test under the PO assumption. That is, when the treatment effect results in a constant shift in the log-odds of higher categories, the two tests behave similarly \citep{agresti2010analysis}, which appears to hold for the outcome probabilities considered here.

%%%%%%%%%%%%%%%%%%%%%%%%%%%%%%%%%%%%%%%%%%%%%%%%%%%%
%%%%%%%%%%%%%%%%%%%% Appendix B %%%%%%%%%%%%%%%%%%%%
%%%%%%%%%%%%%%%%%%%%%%%%%%%%%%%%%%%%%%%%%%%%%%%%%%%%
\section{Example illustration 2: Differential gene expression analysis}
\label{sec:S2_supp_de}

\subsection{Excluded DGMs}
\label{sec:S2.1_excluded_DGMs}
From the DGMs investigated by \citet{baik2020benchmarking}, we adopted or excluded DGMs as specified below.
\begin{itemize}
    \item We adopted the DGMs with independent samples within groups; thus, we excluded those with genetically identical replicates within groups.
    \item We adopted the DGMs where the proportion of DE genes ($\propDE$) is greater than zero; thus, we excluded those with $\propDE = 0$.
    \item We adopted the DGMs whose results are presented as figures in the main text; thus, we excluded those found only in the supplement.
    \item We adopted the DGMs representing the default mode (D) with respect to outliers; thus, we excluded those with random outlier counts (R), where 5\% of counts are turned into outliers, and those with outlying dispersion samples (OS), where one third of the samples in each group have their dispersions increased fivefold to simulate low-quality samples.
\end{itemize}

\subsection{Dataset selection}
\label{sec:S2.2_dataset_selection}
\paragraph{Database} As a database, we use The Cancer Genome Atlas (TCGA) program (\url{https://www.cancer.gov/tcga}), which contains RNA-Seq datasets for 33 different cancer types. The datasets are accessed via the \texttt{R} package \texttt{curatedTCGAData} \citep{Rpkg_curatedTCGAData2020,Rpkg_MultiAssayExperiment2017}.

\paragraph{Dataset-level criteria}
The following exclusion criteria are applied: 
\begin{itemize}
    \item {\dsaim} Exclude datasets that do not contain samples of both type \enquote{01-Primary Solid Tumor} and type \enquote{11-Solid Tissue Normal}.
    \item {\dsextract} Exclude datasets with fewer than 10 matched sample pairs across the two groups.
\end{itemize}
The first criterion excludes 10 datasets, and the second excludes an additional 9 datasets (7 with fewer than 10 samples in total and 2 with fewer than 10 samples in both groups when considering only paired samples). After applying both criteria, 14 datasets remain (see \hyperref[tab:deanalysis_datasets]{Table~\ref*{tab:deanalysis_datasets}}).

%%%%% Begin Table %%%%%------------------------------------------
\begin{table}[ht]
\centering
\caption{Information about the 14 TCGA datasets considered in the example illustration on methods for differential gene expression analysis, including the KIRC dataset used by \citet{baik2020benchmarking}. Each dataset contains 20{,}501 genes. More information about the data, cancers, and studies can be found at \url{https://www.cancer.gov/ccg/research/genome-sequencing/tcga/studied-cancers}.}
\begin{tabular}{llr}
\hline
\begin{tabular}[c]{@{}l@{}}Study\\ abbreviation\end{tabular} & Study name & $n$ \\
\hline
BLCA & Bladder urothelial carcinoma          & 38        \\
BRCA & Breast invasive carcinoma             & 224       \\
COAD & Colon adenocarcinoma                  & 52        \\
ESCA & Esophageal carcinoma                  & 22        \\
HNSC & Head and neck squamous cell carcinoma & 86        \\
KICH & Kidney chromophobe                    & 50        \\
KIRC & Kidney renal clear cell carcinoma     & 144        \\
KIRP & Kidney renal papillary cell carcinoma & 64        \\
LIHC & Liver hepatocellular carcinoma        & 100        \\
LUAD & Lung adenocarcinoma                   & 116        \\
LUSC & Lung squamous cell carcinoma          & 102        \\
PRAD & Prostate adenocarcinoma               & 104        \\
STAD & Stomach adenocarcinoma                & 64        \\
THCA & Thyroid carcinoma                     & 118        \\
\hline
\end{tabular}
\label{tab:deanalysis_datasets}
\end{table}
%%%%% End Table %%%%%--------------------------------------------

\clearpage

\subsection{Additional results}
\label{sec:S2.3_additional results}
\vfill{}
\begin{figure}[h]
    \centering
    \includegraphics[width=0.9\linewidth]{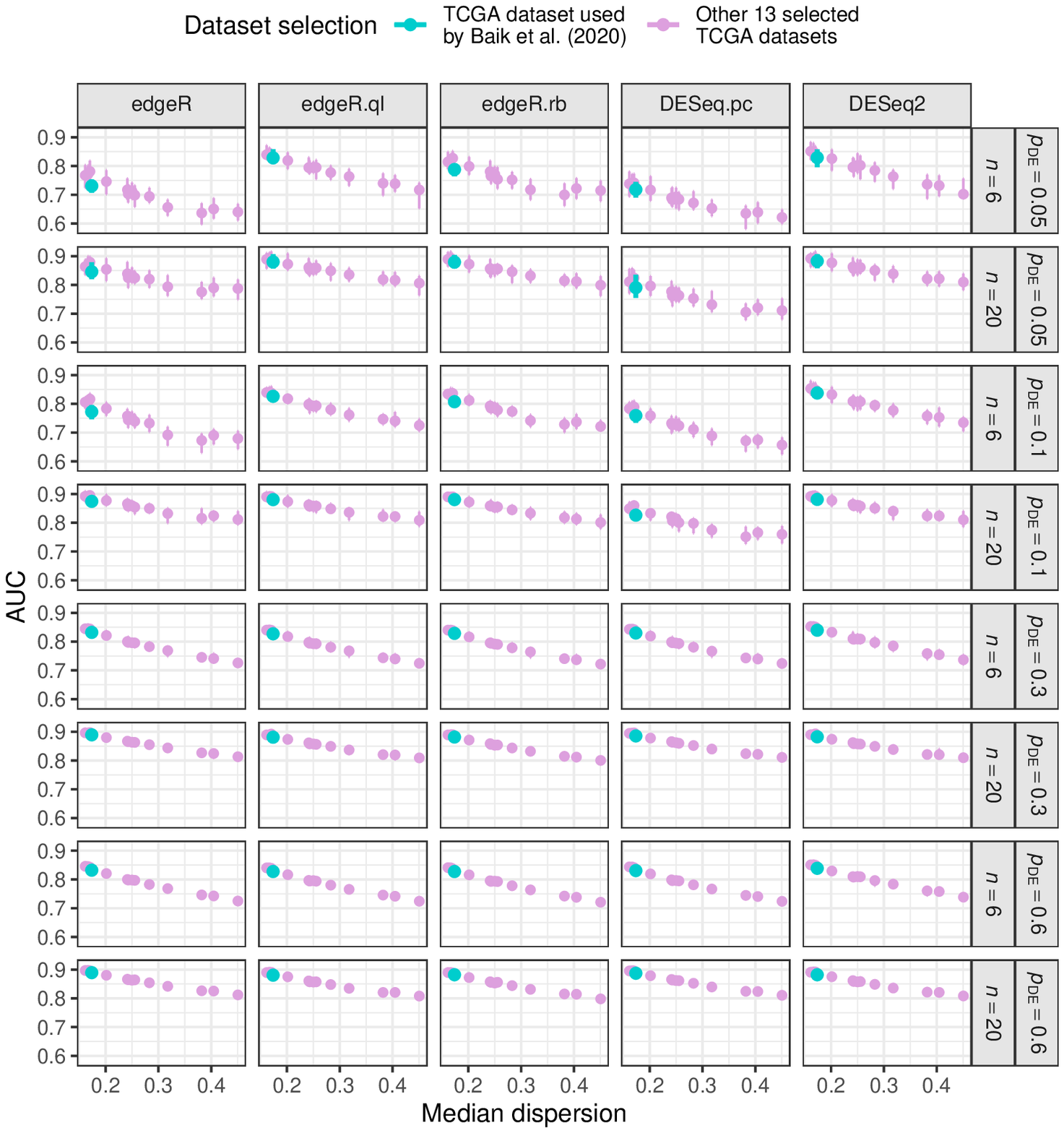}
    \caption{Absolute performance of edgeR, edgeR.ql, edgeR.rb, DESeq.pc, and DESeq2 in relation to the median dispersion (averaged across all genes in the real datasets after filtering), across all considered sample sizes ($\obsnum \in \{6,20\}$) and proportions of DE genes ($\propDE \in \{0.05,0.1,0.3,0.6\}$), comparing results based on the KIRC dataset used by \citet{baik2020benchmarking} and the results based on 13 other selected TCGA datasets. Each panel displays the median and range of absolute AUC values.}
\end{figure}
\vfill{}
\begin{figure}
    \centering
    \includegraphics[width=0.9\linewidth]{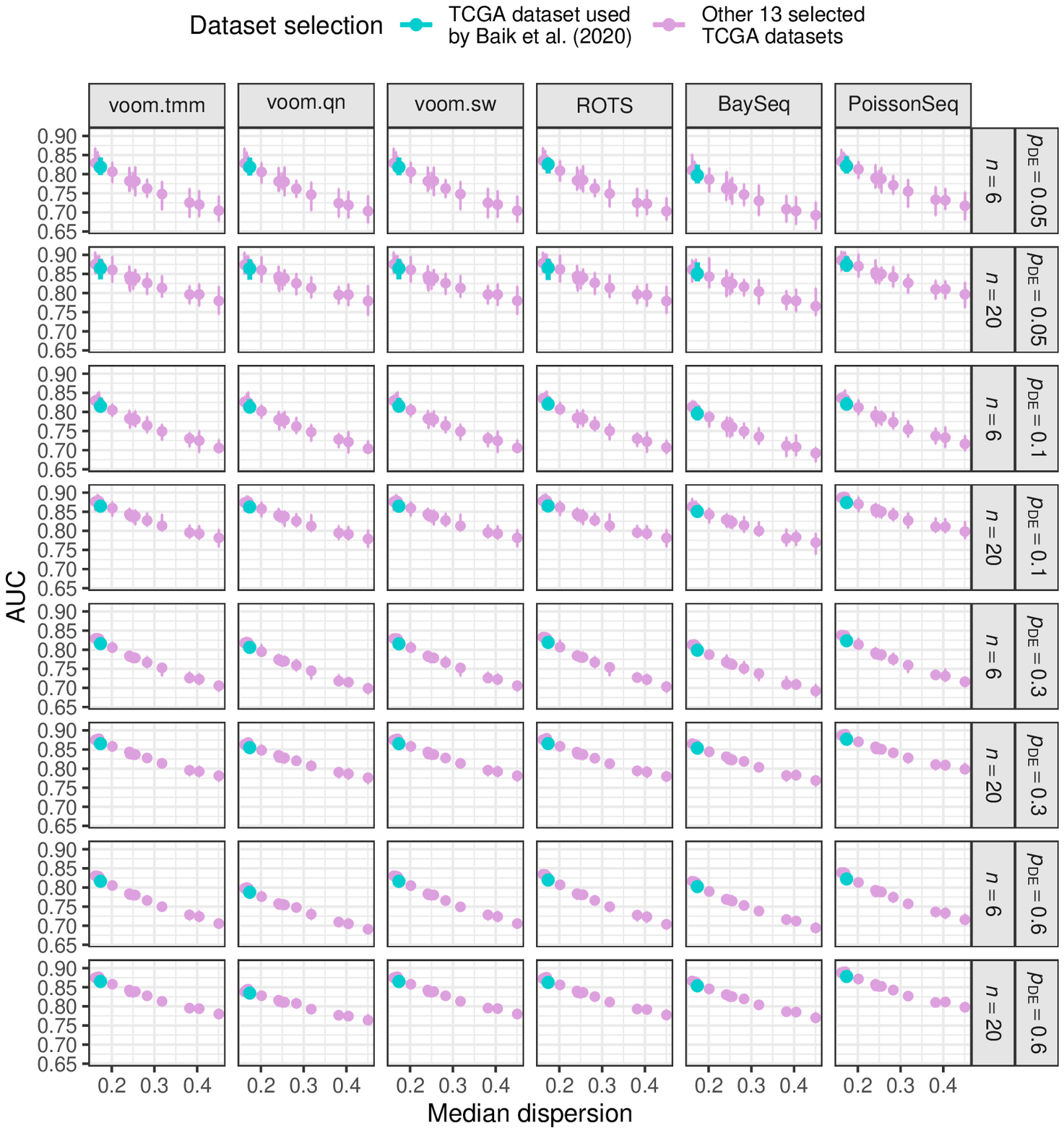}
    \caption{Absolute performance of voom.tmm, voom.qn, voom.sw, ROTS, BaySeq, and PoissonSeq in relation to the median dispersion (averaged across all genes in the real datasets after filtering), across all considered sample sizes ($\obsnum \in \{6,20\}$) and proportions of DE genes ($\propDE \in \{0.05,0.1,0.3,0.6\}$), comparing results based on the KIRC dataset used by \citet{baik2020benchmarking} and the results based on 13 other selected TCGA datasets. Each panel displays the median and range of absolute AUC values.}
\end{figure}
\begin{figure}
    \centering
    \includegraphics[width=0.9\linewidth]{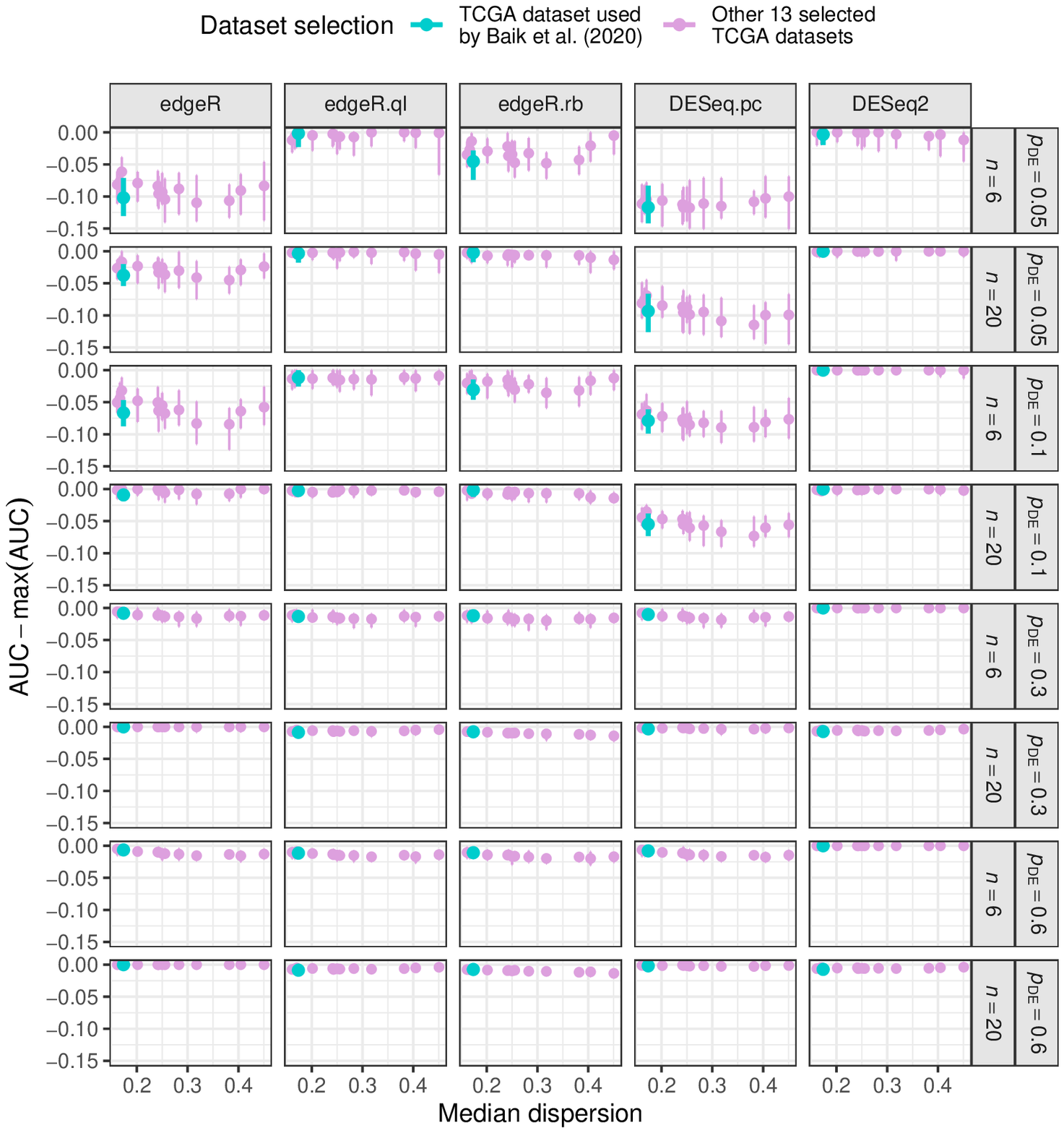}
    \caption{Relative performance of edgeR, edgeR.ql, edgeR.rb, DESeq.pc, and DESeq2 in relation to the median dispersion (averaged across all genes in the real datasets after filtering), across all considered sample sizes ($\obsnum \in \{6,20\}$) and proportions of DE genes ($\propDE \in \{0.05,0.1,0.3,0.6\}$), comparing results based on the KIRC dataset used by \citet{baik2020benchmarking} and the results based on 13 other selected TCGA datasets. Each panel displays the median and range of the difference between the AUC and the highest AUC observed within each DGM.}
\end{figure}
\begin{figure}
    \centering
    \includegraphics[width=0.9\linewidth]{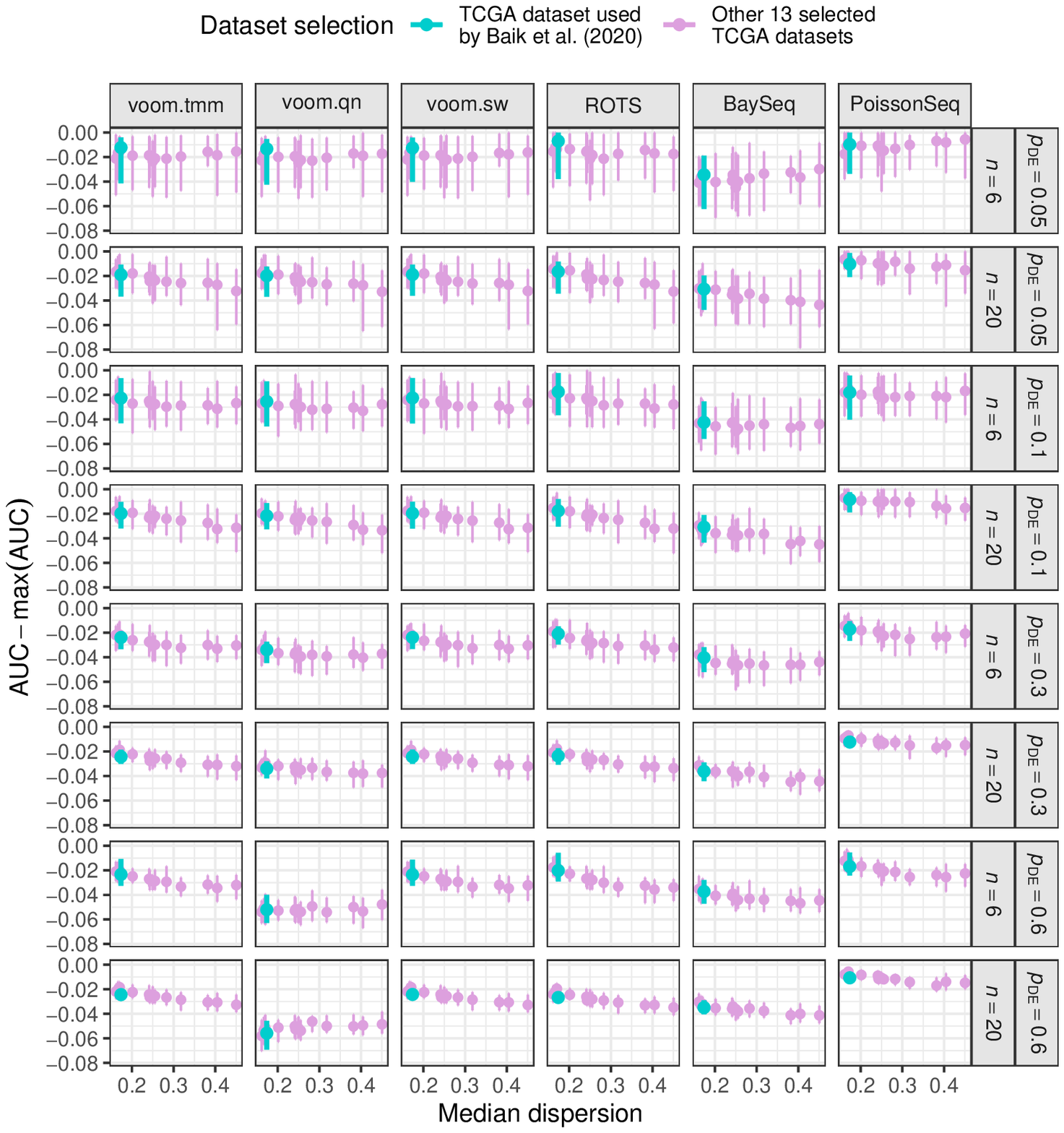}
    \caption{Relative performance of voom.tmm, voom.qn, voom.sw, ROTS, BaySeq, and PoissonSeq in relation to the median dispersion (averaged across all genes in the real datasets after filtering), across all considered sample sizes ($\obsnum \in \{6,20\}$) and proportions of DE genes ($\propDE \in \{0.05,0.1,0.3,0.6\}$), comparing results based on the KIRC dataset used by \citet{baik2020benchmarking} and the results based on 13 other selected TCGA datasets. Each panel displays the median and range of the difference between the AUC and the highest AUC observed within each DGM.}
\end{figure}

\end{document}